\documentclass[rmp,aps,twocolumn,floatfix]{revtex4-1}

\usepackage{graphics}
\usepackage{epsfig}

\def\inbar{\,\vrule height1.5ex width.4pt depth0pt}
\def\IR{\relax{\rm I\kern-.18em R}}
\def\IC{\relax\hbox{$\inbar\kern-.3em{\rm C}$}}

\begin{document}
\title{Ultrahigh Energy Cosmic Rays}

\author{Antoine Letessier-Selvon}
\email{Antoine.Letessier-Selvon@in2p3.fr}
\affiliation{LPNHE, University of Paris UPMC, CNRS/IN2P3, 4, place Jussieu\\
 75252 Paris Cedex 05 - France}
\author{Todor Stanev}
\email{stanev@bartol.udel.edu}
\affiliation{Bartol Research Institute, Department of Physics and Astronomy,
University of Delaware, Newark, DE 19716, USA}

\begin{abstract}  
 This is a review of the most resent results from the investigation of 
 the Ultrahigh Energy Cosmic Rays, particles of energy exceeding
 10$^{18}$ eV. After a general introduction to the topic and a brief
 review of the lower energy cosmic rays and the detection methods,
 the two most recent experiments, the High Resolution Fly's Eye (HiRes)
 and the Southern Auger Observatory are described.
 We then concentrate on the results from these two experiments on the
 cosmic ray energy spectrum, the chemical composition of these cosmic
 rays and on the searches for their sources.
 We conclude with a brief analysis of the controversies in these
 results and the projects in development and construction that can
 help solve the remaining problems with these particles.   
\end{abstract}

\maketitle
\tableofcontents

\section{INTRODUCTION}
\label{sec:intro}
  Cosmic rays are defined as charged nuclei that originate 
 outside the solar system. Such nuclei of total energies 
 between one GeV to above 10$^{11}$ GeV have been detected.
 Below energies of several GeV cosmic rays are usually
 studied in terms of kinetic energy $E_k \; = \; E_{tot} - m c^2$.
 In such terms the cosmic ray energy spectrum extends
 more than 14 orders of magnitude, from 10$^6$~eV to above
 10$^{20}$~eV. 

  The exploration of cosmic rays began as a mixture of physics and
 environmental studies almost a hundred years ago. After the discovery of
 radioactivity it was noticed that between 10 and 20 ions were
 generated per cubic centimeter of air every second. The main question
 was if this ionization was a product of the natural radioactivity of
 the Earth. The agent of this radioactivity was assumed to be 
 $\gamma$-rays because the two other types of radioactive rays:
 $\alpha$-rays (ionized He nuclei) and $\beta$-rays (electrons)
 were easily shielded. To prove that natural radioactivity is
 the culprit physicists started measurements of the ionization
 at different heights above the surface. Such measurements were
 done at the Eiffel tower.

  Just before the First World War Victor Hess started measuring
 the ionization on balloons. In 1912 he flew a balloon from Austria to
 an altitude of 5 km and to everybody's surprise the ionization
 increased by a factor of two rather than decrease. Werner Kohlh\"{o}rster 
 flew balloons to altitudes exceeding 9 km in Germany and
 measured even higher ionization level of the 
 {\em H\"{o}henstrahlung} (high altitude radiation)
 as the cosmic rays were called 
 by the first explorers. The term {\it cosmic rays} 
 was put together by Robert Millikan, who was trying to prove
 that cosmic rays are 10 to 100 MeV $\gamma$-rays from 
 nucleosynthesis of the common C and O elements.

  Kohlh\"{o}rster continued his cosmic ray research during 1930s.
 In collaboration with Walther Bothe he proved that cosmic rays can
 penetrate through heavy absorbers.
 Bruno Rossi shielded his detectors with one meter of lead
 and saw some cosmic rays still penetrating.  Many expeditions were
 organized at high mountains to study the interactions of cosmic
 rays with the geomagnetic field. Arthur Compton organized
 expeditions at different geomagnetic latitudes which proved 
 that cosmic rays are positively charged particles. More of them 
 come from the West than from the East because the geomagnetic
 field bends positively
 charged particles coming from the West towards the surface of the
 Earth and those from the East away from it.

 Cosmic ray research was
 the basis for the development of the QED and the electromagnetic
 cascade theory. Towards the end of the decade Pierre Auger and
 collaborators made several experiments at high mountain altitude
 where they ran in coincidence Geiger-M\"{u}ller tubes at
 large distances from each other.
 They concluded that primary cosmic
 rays generate showers in the atmosphere.
 Kohlh\"{o}rster and Rossi ran similar experiments even earlier
 but of smaller dimensions.
 Auger estimated that the showers that were
 detected came from a primary cosmic ray of energy up to 10$^6$ GeV.
 The term `shower' is an English translation by Patrick Blackett
 of the italian expression {\em sciami} that Rossi used in
 conversations with Beppo Occhialini. The knowledge accumulated
 in the 1930s was published in the magnificent article of 
 \cite{RG41} ``Cosmic Ray Theory''.
 This is the beginning of the investigations of the high energy
 cosmic rays, of their energy spectrum and composition. 
      
\begin{figure}[thb]
\centerline{\includegraphics[width=8cm]{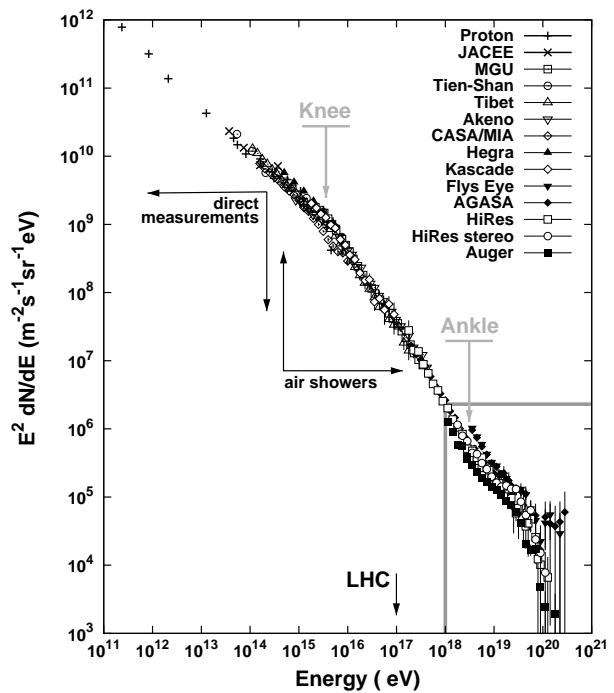}}
\caption{Differential energy spectrum of cosmic rays of energy
 above 10$^{11}$~eV multiplied by E$^2$.
 The positions of the cosmic rays knee and ankle are indicated
 with gray arrows. The experiments that contribute data to this graph
 are shown. The equivalent laboratory energy of the Large Hadron
 Collider is also shown. 
 }
\label{spe_200}
\end{figure}

  Figure~\ref{spe_200} shows the energy spectrum of cosmic rays 
 with energy above 10$^{11}$~eV. Note that lower energy cosmic ray
 spectrum at Earth is affected by the magnetic fields of the heliosphere
 and the geomagnetic field.
 The cosmic ray flux as a function
 of energy is multiplied by E$^2$ to emphasize the spectral shape
 and to indicate the amount of energy carried by cosmic rays of
 different energy. 
 This is a smooth power law spectrum that 
 contains three general features: the cosmic ray {\em knee}
 above 10$^{15}$~eV, the cosmic ray {\em ankle} at about
 3$\times$10$^{18}$~eV (3 EeV),
 and the {\em cut-off} above 3$\times$10$^{19}$~eV. The approximate
 positions of the knee and ankle are indicated with arrows above
 them. The cosmic ray spectrum below the knee is a power law 
 E$^{-\alpha}$ with spectral index $\alpha$ = 2.7. Above the knee
 the spectral index increases with $\Delta \alpha$ = 0.3.
 Above the ankle the power law spectrum becomes flatter and 
 similar to that before the knee.    
 
  The values of the spectral indices show that below the knee 
 the flux decreases by a factor of 50 when the energy increases by
 an order of magnitude. Above the knee the decrease is by a factor
 of 100. Because of the decrease, cosmic rays of energy above 
 10$^{14}$~eV are difficult to measure by direct experiments
 performed on balloons and satellites. The flux of such cosmic rays
 is about 3 particles per hour per steradian in one square meter
 detector. Particles above 10$^{15}$~eV can only be measured by
 air shower arrays of areas more than 10$^4$~m$^2$. Various
 air shower experiments obviously have different energy 
 assignments that lead to the inconsistencies in the presented
 spectra.

  The standard thinking in the field of cosmic rays is that
 particles of energy below and around the knee are accelerated 
 at galactic astrophysical objects, mainly at supernova
 remnants and possibly at powerful binary systems. The knee
 itself is probably a result of reaching the maximum energy
 of such accelerators. Particles above the ankle are believed
 to be of extragalactic origin. They may be accelerated at
 active galactic nuclei (AGN), at radio galaxies, in gamma-ray
 bursts (GRB), or in other powerful astrophysical systems.
 It is not obvious where the particles above the
 knee and below the ankle are accelerated, possibly at some
 special, very efficient galactic accelerators.   

 In this article we will concentrate on the cosmic rays of 
 energy above 10$^{18}$~eV, in the lower right hand of the graph.
 The search for such high energy cosmic rays started in the
 1950s by the MIT group led by B.~Rossi. The first announcement 
 of a cosmic ray shower of energy above 10$^{19}$~eV came from
 the Volcano Ranch air shower array in New Mexico (\cite{LSR61})
 that had an area of about 8 km$^2$. Two years later John Linsley
 reported on the detection of an event of energy
 10$^{20}$~eV (\cite{JL63}).
 The discoveries continued during the next 50 years  with larger
 and larger arrays but the total world statistics is still small.

 These are the ultrahigh energy cosmic rays (UHECR) at least a
 part of  which are of extragalactic origin. We will discuss the 
 requirements for acceleration of such particles that carry
 more than seven orders of magnitude more energy than
 the LHC beam 
 and their propagation in the intergalactic space from
 their sources to us. We will introduce the UHECR detection methods
 and detectors and the results on the cosmic ray spectrum and
 composition.
 We concentrate on the new results presented by the HiRes experiment
 and the Auger Southern Observatory to which we will often
 refer as HiRes and Auger. Please consult the excellent review of
 \cite{NaganoW} for the older experiments and results and that of
 \cite{Cronin-1999} for the importance of the research in this field.
 Some more information could be found in the reviews of
 \cite{Bluemer-2009} and \cite{BW2009}.
 We will conclude with a discussion of the 
 remaining problems and description of possible future 
 experiments. 

\section{EXTENSIVE AIR SHOWERS}
\label{sec:EAS}
Extensive Air Showers (EAS) are the particle cascades following the interaction
of a cosmic ray  with an atom of the atmosphere. After this first 
interaction the atmosphere acts like a calorimeter of variable
 density with a vertical thickness of more than 11 interaction lengths
 and 26 radiation length. 

 A 10$^{19}$ eV (10 EeV, 1 EeV = 10$^{18}$ eV) proton striking vertically
 the top of the atmosphere
 produces at sea level (atmospheric thickness of 1033~g/cm$^2$)
 about $3\times 10^{10}$ particles (with energy in excess of 200 KeV). 
\begin{figure}[htbp] 
   \centering
   \includegraphics[width=7cm]{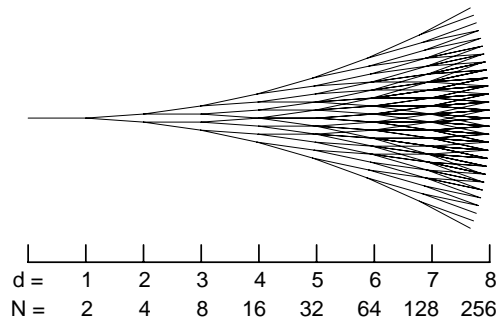} 
   \caption{Heitler's schematic evolution of an electromagnetic cascade.
 At each stage of the cascade the number of particle is multiplied
 by two, either through pair creation or single photon bremsstrahlung.
 The evolution stops when individual particle energy fall below the
 critical energy, about 80 MeV in Air.}
   \label{fig:Heitler-a1}
\end{figure}
 99\% of these are photons and electrons/positrons
 (referred simply as electrons in the following) in a ratio
 of about 6 to 1. Their energy is mostly in the range 1 to 10~MeV and they
 transport 85\% of the total energy. The remaining
 particles are either muons with an average energy of about 1~GeV
 (carrying about 10\% of the total energy),
 few GeV pions (about 4\% of the total energy) and,
 in smaller proportions, neutrinos and baryons.
 The shower footprint (more than 1 muon per m$^2$) on the ground
 extends over a few km$^2$.

 The basic properties of the development of the cascade can be
 extracted from a simplified model due to Heitler. It describes
 the evolution of a pure electromagnetic cascade (\cite{Heitler-a1}). 

\subsection{Heitler's model of electromagnetic showers}
 In his model, Heitler described the evolution of electromagnetic
 cascades  as a perfect binary tree 
 (see Fig~\ref{fig:Heitler-a1}).
 At each step all particles interact and produce two
 secondaries of equal energy.  This description assumes that at each step 
electrons split their energy in half via bremsstrahlung emission
 of a single photon while photons 
 produce an electron/positron pair of equal energy.
 In this simplified approach, all the processes 
 cross sections are taken as independent of energy and collision
 energy losses are ignored. 

 The interaction step length $d$ in the cascade is therefore given
 by the radiation length of the medium $\lambda_r$ ($\lambda_r= 37$
 g/cm$^2$ in air)  as $d = \lambda_r\ln{2}$. After $n$ steps the
 particle number is $N_n=2^n$ and their individual energy is $E_0/N_n$.
 This development continues until the individual energy drops below a
 critical value where the rate of energy loss by electrons via
 bremsstrahlung is equal to the rate of energy loss by ionization.
 This energy is about $E^\gamma_{c} = 80$ MeV in Air. At this point
 of development the electromagnetic cascade has reached a maximum
 and the number of particles is given by the ratio of the original
 energy to the critical one.

 Although very simplified, Heitler's model reproduces correctly
 three properties of electromagnetic cascades :\\
1) The number of particles at the maximum of the cascade development
 is proportional to the incoming primary cosmic ray energy :
 \begin{equation} N_{max} = E_{0}/E^\gamma_c\end{equation}\\
2) The evolution of the depth of maximum of the shower
 (measured in g/cm$^2$) is logarithmic with energy:
 \begin{equation}\label{eq:xmax}X_{max} = X_0+ \lambda_r\ln({E_0/E^\gamma_c})
 \end{equation} where $X_0$ is the position of the start of the cascade.\\
3) The rate of evolution of $X_{max}$ with energy, the elongation
 rate,  defined as
 \begin{equation}\label{eq:Lambda} D_{10}
 \equiv \frac{dX_{max}}{d\log_{10}{E_0}} = 2.3\lambda_r\end{equation}
 is given by the radiation length of the medium. This elongation rate 
 is  about 85 g/cm$^2$ in air.

 Extensive simulations of electromagnetic cascades confirm
 these properties although the particle number at maximum
 is overestimated by about a factor 2 to 3. Moreover,
 Heitler's model predicts a ratio of electrons to photons
 of  2  while simulations and direct cascade measurements
 in Air show a ratio of the order of 1/6th.
 This is in particular due to the facts that multiple photons
 are emitted during bremsstrahlung and that electrons lose
 energy much faster than photons do.

\subsection{Extension to hadronic showers}
\begin{figure}[htbp] 
   \centering
   \includegraphics[width=6.5cm, angle=-90]{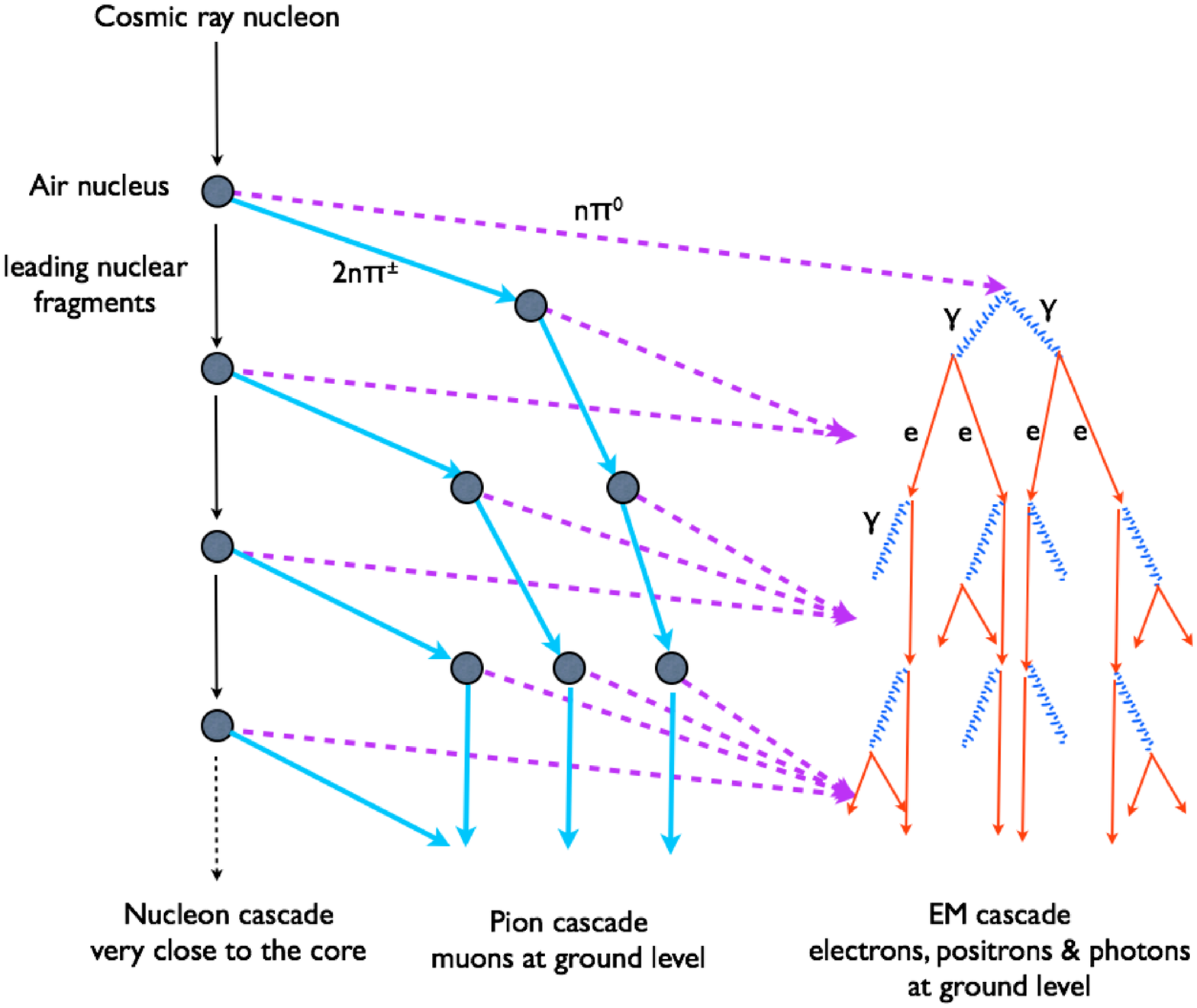} 
   \caption{Schematic evolution of an hadronic cascade. At each step roughly 1/3rd of the energy is transferred from the hadronic cascade to the electromagnetic one.}
   \label{fig:hadcas}
\end{figure}

 Heitler's model can be adapted  to describe hadronic showers
 (\cite{Matthews-2004,Stanev-book}).
 In this case the relevant parameter is the hadronic interaction
 length $\lambda_I$. At each step of thickness $\lambda_I\ln{2}$
 it is assumed that hadronic interactions produce $2N_\pi$ charged
 pions and $N_\pi$ neutral ones. While $\pi^0$ decay immediately
 and feed the electromagnetic part of the shower, $\pi^+$ and $\pi^-$
 interact further. The hadronic cascade continues to grow,
 feeding the electromagnetic part at each step, until charged pions
 reach an energy where decay is more likely than a new
 interaction. A schematic of an hadronic cascade is shown in Fig.~\ref{fig:hadcas}. 
 The interaction length and the pion multiplicity ($3N_\pi$) are
 energy independent in the model. The energy is
 equally shared by the secondary pions. For pion energy between
 1 GeV and 10 TeV a charged multiplicity of 10 ($N_\pi$ = 5)
 is an appropriate number.
 
 One third of the available energy goes into the
 electromagnetic component while the remaining 2/3rd continues
 as hadrons. Therefore the longer it takes for pions to reach
 the critical energy $E^\pi_c$ (20 GeV in air, below which they
 will decay into muons), the larger will be the electromagnetic
 component. Consequently in long developing showers  the
 energy of the muons from decaying pion will be smaller. 
 In addition, because of the density profile of the atmosphere,
 $E^\pi_c$ is larger high above ground than at see level
 and deep showers will produce fewer muons. 

 This positive correlation  introduces a link between the primary
 cosmic ray interaction cross section with Air and the muon
 content at ground. According to those principles
 primaries with higher cross
 sections will have a larger muon to electron ratio at ground.  

 To obtain the number of muons in the shower one  simply assumes
 that all pions decay into muons when they reach the critical energy.
 $N_\mu = (2N_\pi)^{n_c}$ where $n_c = \ln({E_0/E_c^\pi})/\ln{3N_\pi}$
 is the number of steps needed for the pions to reach
 $E^\pi_c$. Introducing $\beta=\ln{2N_\pi}/\ln{3N_\pi}$
 (0.85 for $N_\pi=5$) we have:
 \begin{equation}\label{eq:nmu} N_\mu = (E_0/E^\pi_c)^\beta\end{equation} 
 Unlike the electron number, the muon multiplicity does not grow
 linearly with the primary energy but at a slower rate.
 The precise value of $\beta$ depends on the average pion
 multiplicity used.
 It also depends on the inelasticity of the hadronic interactions.
 Assuming that only half of the available energy goes into the pions
 at each step (rather than all of it as done above) would lead
 to $\beta=0.93$. Detailed simulations give values of $\beta$ in
 the range 0.9 to 0.95 (\cite{Alvarez-Muñiz-2002}).

 The determination of the position of shower maximum is more
 complex in the case of hadronic shower than in the case of a pure
 electromagnetic one. The larger cross section and the larger
 multiplicity at each step will reduce the value of $X_{max}$
 while the energy evolution of those quantities will modify the
rate of change of $X_{max}$ with energy - a quantity known as 
the elongation rate. In addition the inelasticity of the interaction
 will also modify both the position of the maximum and the elongation
 rate. A proper account for the energy transfer from the hadronic
 component to the electromagnetic one at each step together with 
 a correct superposition of each electromagnetic sub-showers to
 compute $X_{max}$ is beyond the scope of a simple model but can be
 successfully done in a simulation. 
 An approximation based on the sole evolution of the EM
 cascade initiated by the first interaction falls short
 of the full simulation value by about 100 $g/cm^2$ (\cite{Matthews-2004}). 

 A good approximation of the elongation rate can be obtained
 when introducing the cross-section and multiplicity energy dependance.
 Using a proton Air cross section of 550 mb at 10$^{18}$ eV and a rate
 of change of about 50 mb per decade of energy (\cite{Ulrich-2009})
 one obtains:
\begin{equation}\label{lambda-i}
 \lambda_I \simeq 90 - 9 \log{(E_0/EeV)}\mbox{ g/cm$^2$} 
\end{equation}
 Assuming, as in~\cite{Matthews-2004}, that the first interaction
 initiates $2N_\pi$ EM cascades of energy $E_0/6N_\pi$ with
 $N_\pi\propto (E_0/PeV)^{1/5}$ for the evolution of the first interaction
 multiplicity with energy, one can calculate the elongation rate: 
\begin{equation}
 D_{10}^ p=\frac{dX_{max}}{d\log{E_0}}= \frac{d(\lambda_I\ln{2}+\lambda_r \ln{[E_0/(6N_\pi E_c^\gamma)}]}{d\log{E_0}}
\end{equation}
or
\begin{equation}
 D_{10}^p=\frac{4}{5}D_{10}^\gamma - 9\ln{2} \simeq 62\mbox{ g/cm$^2$}
\end{equation} 
 This result is quite robust as it only depends on the cross section
 and multiplicity evolution with energy. It is in good agreement with
 simulation codes (\cite{Alvarez-Muñiz-2002}).

 The fast rate of the energy transfer in hadronic showers was noted
 long ago by~\cite{Linsley-1977} who introduced the {\it elongation
 rate theorem} that stipulates that the elongation rate for
 electromagnetic showers ($D_{10}^\gamma$) is an upper limit to the
 elongation rate of hadronic showers. This is of course a direct
 consequence of the larger hadronic multiplicity which increases the
 rate of conversion of the primary energy into secondary particles. 

\begin{figure}[htbp] 
   \centering
   \includegraphics[width=6cm, angle=-90]{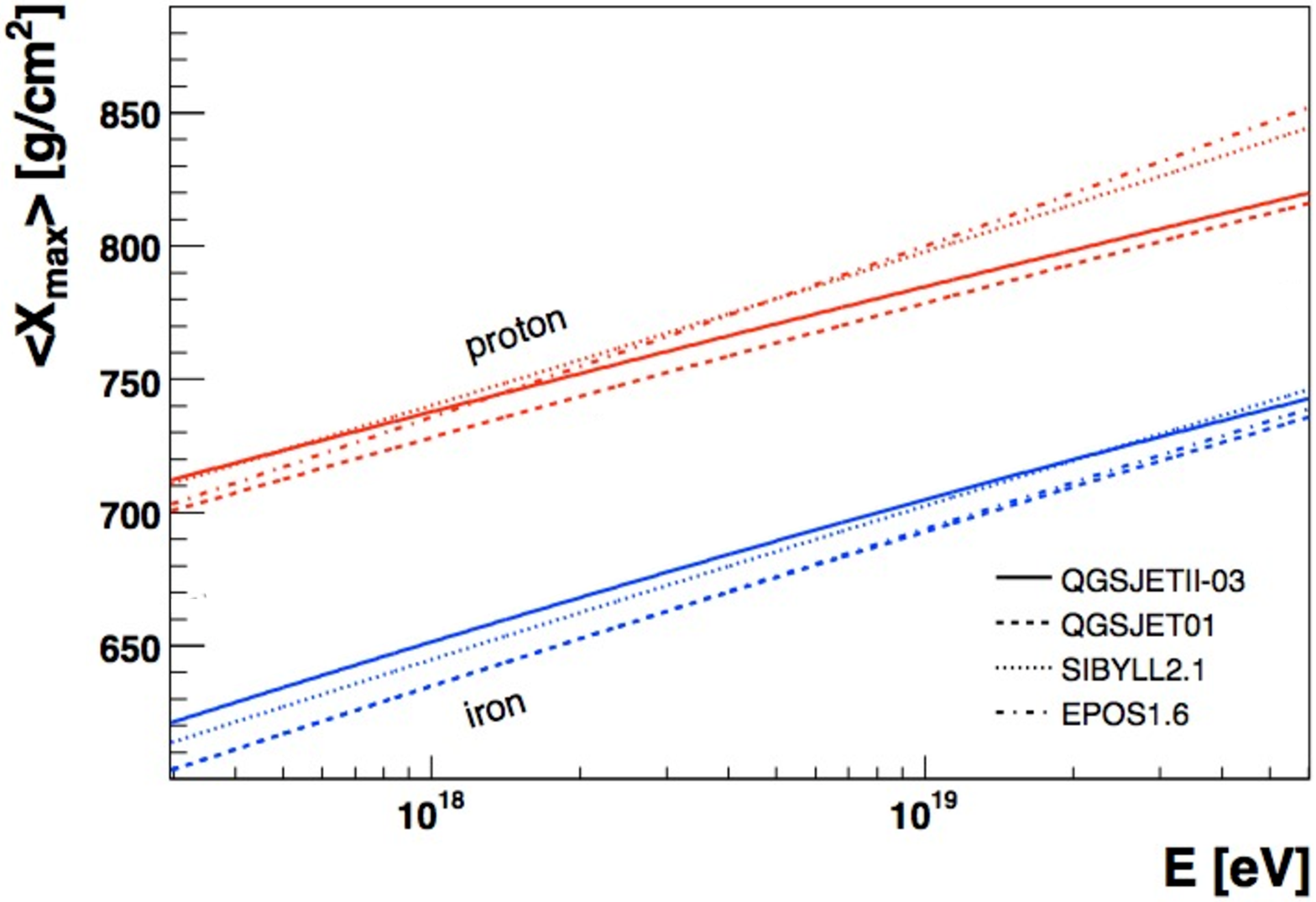} 
       \vspace*{-0.3cm}  \caption{ Evolution of the position of X$_{max}$ as a function of energy (elongation rate) 
   for iron and proton induced air showers. Elongation rate of different nuclear species are
   with nearly constant slope and almost parallel to each other. Shown here are the results of 
   detailed simulation performed by the Auger collaboration using various interaction models.}
   \label{fig:ElRate}
\end{figure}

 Extension of this description to nuclear primaries can finally be done
 using the superposition model. In this framework the nuclear interaction
 of a nucleus with atomic number $A$ is simply viewed as the
 superposition of the interactions of $A$ nucleons of individual energy
 $E_0/A$.
 Showers from heavy nuclei will therefore develop higher, faster and
 with less shower to shower fluctuations than showers initiated by
 lighter nuclei. The faster development implies that pions in the
 hadronic cascade will reach their critical energy (where they decay
 rather than interact) sooner and therefore augment the relative
 number of muons with respect to the electromagnetic component. 
 From this simple assumptions one can directly see that:\\ 
1) Shower induced by nuclei with atomic number $A$ will
 develop higher in the atmosphere.
 The offset with respect to proton showers is simply :
 \begin{equation} X_{max}^A = X_{max}^p - \lambda_r\ln{A}\end{equation}
2) Showers initiated by nuclei
 with atomic number $A$ will have a larger muon number :
 \begin{equation} N_\mu^A = N_\mu^p A^{1-\beta} \end{equation}
3) The evolution of the primary cross section and multiplicity with
 energy for nuclei is the same as for protons. Different nuclei
 will have identical elongation rates and will show up as parallel
 lines in an $X_{max}$ vs energy plot. See figure~\ref{fig:ElRate}.\\ 
4) The fluctuation of the position of $X_{max}$ from one shower to
 another is smaller for heavy nuclei than for light ones. 

 All the above results and properties are qualitatively confirmed by
 detailed simulations. All interaction models share those basic
 principles and they all predict that  iron showers have a smaller
 average $X_{max}$, less fluctuations on $X_{max}$ and a larger muon
 to electron ratio at ground than proton ones.  In particular the
 offset in $X_{max}$ from iron to proton showers is more than 100~g/cm$^2$
 and iron showers carry about 1.8 times as many muons as proton
 showers of the same energy.  Of course in quantitative terms
 there are differences but all the basic trends regarding the evolution
 of  $X_{max}$ and $N_\mu$ with energy and atomic number are reproduced.
 This is of particular importance in the attempts to relate experimentally
 measured quantities to mass composition.     

\subsection{Main features used for composition studies}
 On a shower to shower basis, composition studies are particularly
 difficult because of the intrinsic shower to shower fluctuation of
 $X_{max}$ and $N_\mu$. Those fluctuations come from the random
 nature of the interaction processes (in particular the position of
 the first interaction) and from the large spacing and limited sampling
 size of the detectors.
 Nevertheless, due to the difference in their cross section with Air,
 showers originating from different primaries can, at least statistically,
 be distinguished.

 In a real situation, where the composition evolves with energy,
 one observes changes in the elongation rate that are not compatible
 with a single species because the rate of change is either too large,
 when composition evolves from heavy to light (violating the
 {\it elongation rate theorem}) or too small going from light to heavy.

 From the superposition principle we have seen that distinct primaries
 will show up as parallel lines of constant slopes in an elongation rate
 plot. Detailed simulations qualitatively confirm this principle although
 the lines are neither totally parallel nor exactly of constant slope.
 Those features are model dependent  as they depend on the inelasticity
 treatment of the cross section and on the leading (the fastest)
 particle effect
 together with the evolution of the rate of change of the cross section
 with energy which is not measured at the highest energies
 (above 10$^{17}$\ eV). Nevertheless for a detector that can measure 
 in each individual cosmic ray event the position of $X_{max}$ with
 decent precision (a few tens of g/cm$^2$ or less) the elongation rate
 plot provides information about the evolution of cosmic ray composition
 with energy. It is important to note that both the absolute
 value of $X_{max}$ and its rate of change with energy are not the same
 for the various interaction models that have been
 developed (\cite{Sybill, QGSJet, EPOS}). Therefore going from the
 experimental average value of $X_{max}$ at a given energy to an average
 atomic number is strongly model dependent. 

 Beside the average value of X$_{max}$ at a given energy another
 statistical observable which distinguishes composition and
 is less model dependent is given by the width of the X$_{max}$
 distribution. In a simple approach
 the shower to shower (of the same energy) fluctuations of X$_{max}$
 are dominated by the fluctuations of the first interaction
 point X$_0$. X$_{0}$ follows an exponential distribution. As in
 Heitler's model, we take the 50\% energy loss distance:
 $\lambda_I\ln{2}$ for the characteristic length in the distribution
 of X$_0$.  At 10$^{18}$ eV this approach gives, according to
 equation~(\ref{lambda-i}), a fluctuation of X$_0$ for protons of
 60 g/cm$^2$ which is in good agreement with simulation results.
 For an iron nucleus  of the same energy,
 the lower energy of individual nucleons and the strict application
 of the superposition principle give RMS[X$_0$] =
 $\lambda_I(10^{18} {\rm eV} /56)\ln{2}/\sqrt{56} \simeq 14$~g/cm$^2$.
 This is in reasonable agreement with detailed simulations which give
 a value between 20 and 24 g/cm$^2$. Leading fragment effect
 (violating the superposition principle) and fluctuation in subsequent
 interactions also play a role in the case of heavy nuclei.
 Here the fluctuations of X$_0$ give only a lower bound to the fluctuations
 of X$_{max}$.

 On an individual shower basis, the identification of the primary cosmic
 ray is experimentally more challenging but not totally hopeless.
 Due to the fast rate of growth of the particle number in the cascade
 and to the large phase space available for secondaries at each interaction,
 the particle fluxes converge rapidly towards distributions that are
 independent of the primary particle type. This is especially true 
 around the maximum development of the shower.  Electromagnetic and
 muons fluxes are adequately described by a Gaisser-Hillas
 (\cite{Gaisser-Hillas}) type function (as in Eq.~\ref{GH})
 whose "age" parameter describing their stage of development is
 derived from X$_{max}$ and whose normalizations are given by the
 primary energy and the muon fraction. This universality property 
 was recently discussed in~\cite{Chou-2005} and studied
 in~\cite{Giller-2004, Nerling-2006, Lipari-2009,
 Schmidt-2007,Schmidt-2008,Apel-2008}. 

 Air shower universality states that the longitudinal development 
 of the electromagnetic component of nuclei-induced air showers
 can be completely described in terms of two parameters: the primary
 nucleus energy and the shower age.
 Shower universality tells us that all information about the primary
 particle can in principle be recovered from the measurements of
 3 parameters. While, due to fluctuations, it is insufficient to measure
 only X$_{max}$ and E$_0$, efficient separation can be achieved if
 the muonic content of the shower is also measured.  
 Additional information on the first interaction cross section can
 also be retrieved by fitting an exponential to the right hand side
 (deeper side) of the X$_{max}$ distribution in fixed energy bins.

 Unlike the electromagnetic component of EAS muons reaching ground level
 still carry information about their production point along the shower
 axis. Because they mostly travel in straight lines without much scattering
 they dominate the early part of the signal at ground. Therefore detectors
 with good timing capabilities can construct composition sensitive
 parameters based on the signal shape either in individual detectors
 (rise time parameters) or comparing signals in several detectors
 (asymmetries and curvature parameters). Even when the absolute muon
 content cannot be retrieved these shape parameters provide valuable
 information characterizing the primary composition.

\subsection{Detection methods}

 Above $10^{15}$~eV the cosmic ray flux drops below a few tens of
 particle per m$^2$ and per year. It is no longer possible to detect
 the incident particles above the atmosphere before they interact.
 Detectors flying in balloons or satellites that are less than a
 few m$^2$ in size must be replaced by ground based instruments that
 cover up to several thousands of km$^2$. 

 From the direct measurement of the incident particle properties, energy,
 mass, charge, etc., one must revert to the indirect measurement of the
 EAS produced by the interaction in Air. The atmosphere acts as a
 calorimeter and becomes part of the detection system. As this is not a
 fully controlled environment, atmospheric conditions must be carefully
 monitored and recorded along with the air shower data.   
 All experiments aim at measuring, as accurately as possible, the primary
 direction (by the relative times of the signals), the primary
 energy (inferred from the integrated signals densities), and the primary
 nature or mass (extracted from the signals shapes).

 With the exception of fluorescence light from the nitrogen molecules
 excited along the shower trajectory and the possible microwave
 emission (\cite{PGetal2008}, most radiations emitted from EAS are
 concentrated in the forward direction and cannot be detected far away
 from the shower axis. Hence the original, and most frequent technique,
 used to detect UHECR is to build an array of sensors (scintillators,
 water Cherenkov tanks, muon detectors, Cherenkov telescopes, ...)
 spread over a large area. When a cosmic ray event falls within the array
 boundary, the sub-sample of detectors placed near enough to the shower
 axis will observe the radiation reaching ground. The surface area of the
 array is chosen according to the incident flux, i.e. the energy range 
 one wants to explore. 

 Ignoring the remaining fragments of the hadronic cascade which are
 concentrated very near the shower core
 electrons and photons from the EM cascade, muons and forward beamed
 Cherenkov light propagate along the shower axis.
 Particles reaching ground from the EM cascade are the result of a
 long chain of interactions, they are constantly regenerated and
 progress in a diffusive way. Those observed at ground are produced in
 the vicinity (a couple of Moli\`{e}re radii or a few 100~m) of the ground
 detector that measures them. Their time profiles carry little
 information on the shower development itself but their density gives
 information on the primary energy. This radiation is concentrated
 around the shower axis, but at the highest energies, above 10$^{18}$ eV,
 particles  can be observed up to a couple of kilometers away with detectors
 of about 10 m$^2$ in size. Like the EM cascade, muons and (direct)
 Cherenkov light are concentrated around the shower axis. However,
 they reach ground essentially  unaltered.  Their time profile carries
 the memory of their production point along the shower axis and can be
 used to construct composition sensitive parameters.   

 Fluorescence light is emitted isotropically and hence can be detected
 with appropriate telescopes tens of kilometers away from the shower axis. 
 The light is emitted proportionally to the number
 of electrons in the EM cascade and reaches the telescopes essentially
 unaltered (we neglect the losses due to diffusion). The time profile
 will then reflect the evolution of the electromagnetic cascade and
 allows for direct measurement of composition sensitive parameters such
 as X$_{max}$. Moreover, because the radiation can be  observed far away,
 a clear lateral view of the shower profile is possible, unlike in the
 case of a detection close to the shower axis.
  
\subsubsection{Air shower arrays}
 Air shower arrays are networks of particles detectors. They cover a surface
 area in direct proportion to the CR flux in  the region of the spectra one
 wishes to study. A few thousands m$^2$ is enough for the knee region
 around 10$^{15}$~eV, while thousands of km$^2$ are necessary for studies
 near the spectral cutoff at energies above 10$^{19}$~eV.  The spacing
 of the detector is also a function of the energy range of interest.
 For cosmic rays of energy 10$^{18}$ eV and above spacing is of the order of 1~km. 

 The array of detectors counts the number of secondary particles which
 cross them as a function of time. They  sample the part of the shower
 which reaches the ground. The incident cosmic ray direction and
 energy are measured assuming that the shower has an axial symmetry
 in the transverse shower plane. This assumption is valid for zenith
 angles up to about  $60^{\circ}$. At larger zenith
 angles the EM part of the cascade is largely absorbed and the muons
 start to be bent by the geomagnetic field. Above 75$^\circ$, the ground
 pattern shows a clear butterfly shape characteristic of the geomagnetic
 field effect. 

 The pioneer work of J. Linsley at Volcano Ranch (\cite{LSR61})
 used an array of  3~m$^2$ scintillators 900 meters apart
 covering a total surface area of about 8 km$^2$.  It is with this
 detector that the first event in the 10$^{20}$~eV range was
 detected (\cite{JL63}). 
 Scintillator arrays are usually made of m$^2$ flat pieces of
 plastic scintillators laid on the ground and connected by cables.
 They are
 equally sensitive to all charged particles, thus measure 
 mostly the EM component of the cascade.
 Particularly simple to use and deploy they have been quite
 popular for studies at the highest
 energies (\cite{LSR61, Efimov-1991, Chiba-1992}).
The aperture of flat scintillator arrays drops quickly with zenith
 angle because of the decrease of their effective
 surface and because of the absorbtion of the E.M. component. For accurate measurements data of scintillators array is
 usually restricted to zenith angles below 45$^\circ$.

 In principle the measurement of the EM cascade allows for a
 calorimetric and essentially mass independent measure of the
 primary cosmic ray energy. However, detector arrays sample the
 particle densities at a fixed atmospheric depth which varies
 from shower to shower because of the variations of the position
 of X$_{max}$. This introduces a mass dependent bias in the energy
 estimates. In practice the energy calibration of scintillator arrays
 relies on simulations. This has always been the major difficulty
 of the technique.

 Water Cherenkov tanks have also been successfully used in large
 cosmic ray arrays. The Haverah Park array, made of Cherenkov tanks
 of various sizes spread over about 12 km$^2$ took data for almost
 20 years (\cite{Lawrence-1991}). 
 Heavy, requiring extra pure water with excellent protection against
 contamination, water Cherenkov datectors
 are not as easy to deploy as scintillators.
 However, since the Cherenkov light generated in the water is proportional
 to the pathlength of the particle, water tanks are sensitive to both
 the numerous electrons and photons, and the shower muons.
 On the average, depending on the exact detector geometry, a muon will
 deposit about 10 times more light than a single 20 MeV electron.
 Because of their height, water tanks also offer a non zero effective
 surface for horizontal showers. Together with the muon sensitivities
 this extends the aperture of such arrays to nearly horizontal showers.

 Reconstruction of the primary particle parameters is based on timing
 for the geometry and on the distribution of signal densities as a
 function of the lateral distance to the shower axis for the energy. 

 From the position of the different detectors and from the onset of the
 shower front signal recorded in each of them, one can reconstruct the
 shower axis and hence the original cosmic ray direction.  Precision
 of one to three degrees are usually obtained given the large base line
 of the detector spacing (1 km). For charged cosmic rays this precision
 is sufficient given the deflection expected from the galactic coherent
 and random magnetic field. 

 For the energy, the detector position are projected onto the plane
 transverse to the shower axis and  a "lateral distribution function"
 (LDF) is adjusted to the measured  signals. \cite{Hillas-1970} 
 proposed to use the signal at an optimal distance
 $r_{opt}$ depending on the energy range and the array spacing. 
 At $r_{opt}$  the sum of the fluctuations from shower to shower
 and of the statistical fluctuations from
 particle counting are minimum.

\begin{figure}[htbp] 
   \centering
   \includegraphics[width=6cm, angle=-90]{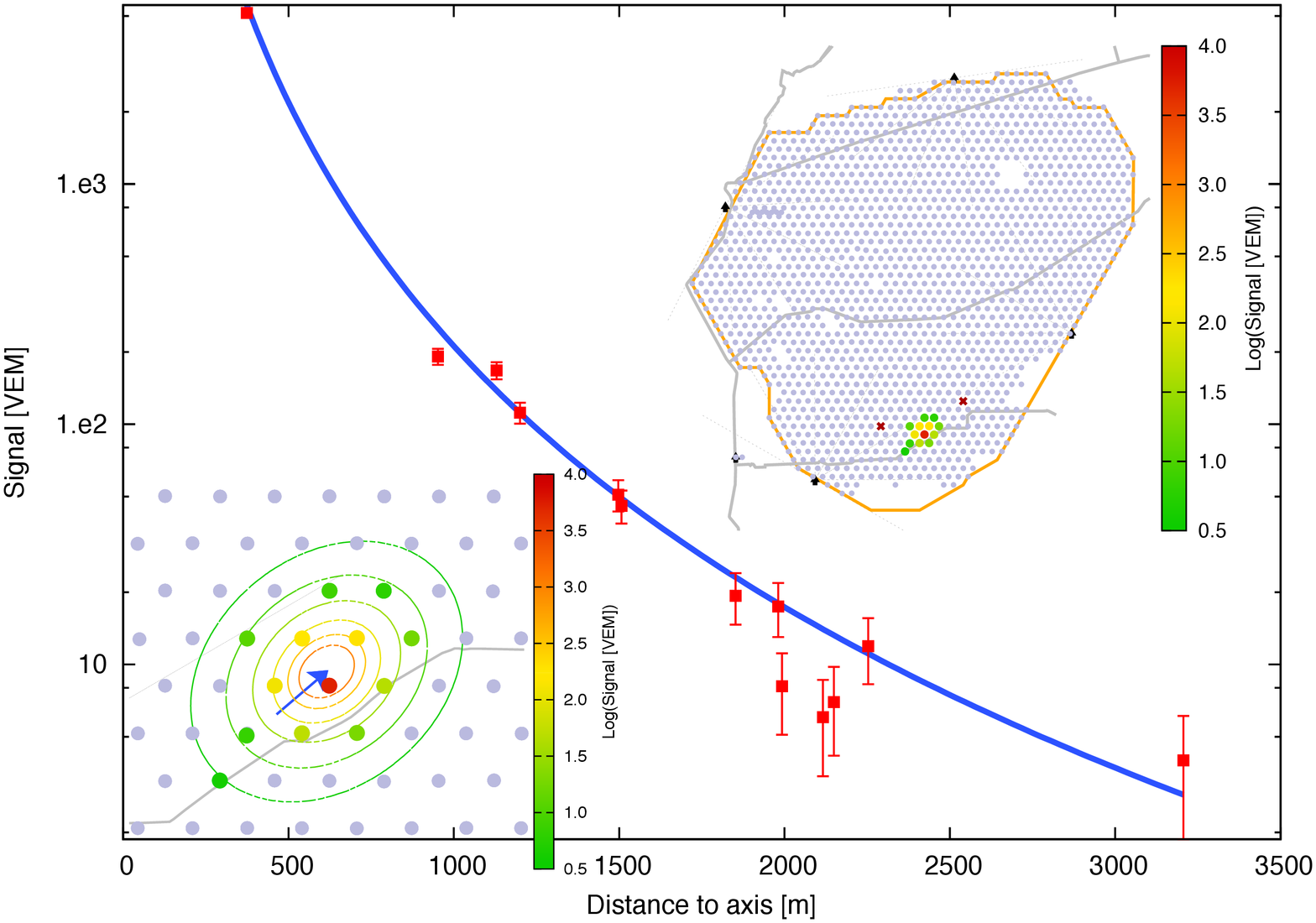} 
  \vspace*{-0.3cm}  \caption{ Example of detection using a surface array. The upper right insert shows the whole
   auger surface array and the footprint of the shower, each dot represented a detector the spacing between them is 1.5~km. 
   The lower insert  shows details of this footprint with the estimated
 contours of the particle density levels. The curve represent the adjusted
 LDF (lateral distribution function) and the red point the measured densities
 as a function of the distance to the shower core. (extracted from the
 public event display of the Auger collaboration).}
   \label{fig:ED}
\end{figure}

 Several LDFs have been used to represent the lateral signal distribution.
 The Haverah Park experiment used as LDF the function:
\begin{equation}
S(r,\theta,E)=kr^{-[\beta(\theta,E)+r/4000]}
\end{equation}
 for distances less than 1 km from the shower core. Here $r$ is in
 meters, $\theta$ is the zenith angle and $\beta$ can be expressed as:
\begin{equation}
\beta(\theta,E)=a+b\sec\theta
\end{equation}
 The value of  $\beta$ is derived from Monte Carlo simulations and
 can also incorporate a slow logarithmic evolution with energy. 
 Roughly $a\simeq 3.5$ and $b\simeq -1.2$ for a vertical value of 
 $\beta$ under 3. At larger distances (and higher
 energies), this function has to be modified to take into account
 a change in the rate at which the densities decrease with distance.
 This is due to the increasing dominance of muons over
 electrons at large core distance, since muons have a flatter
 LDF. A more
 complicated form was used by the AGASA group (\cite{AGASA}).
 However, the principle remains the same. On figure~\ref{fig:ED} an example, taken from the Auger collaboration
 public event display~\footnote{http://auger.colostate.edu/ED/},  of the footprint of an air shower on the ground 
 together with the reconstruciton of the LDF is shown. 

 Once the attenuation of the signal due to the zenith angle is
 accounted for, an estimator of the energy 
 is obtained  from the corrected density at $r_{opt}$ in the form :
\begin{equation}
E=kS(r_{opt},\theta_{ref})^\alpha
\end{equation}
 where $\alpha$ is a parameter of order 1. 

 To reconstruct the primary parameters, a  minimum  of three detectors
 with signal is necessary. The spacing between those detectors will
 determine the array energy threshold. For a vertical shower the
 500~m spacing of the trigger stations in Haverah Park corresponds
 to a threshold of a few $10^{16}$~eV, while the 1.5~km separation
 of  the Auger Observatory stations gives a few 10$^{18}$~eV.

 Ground arrays do not have a direct access to the position of the
 shower maximum and this is a strong limitation to this technique
 for primary identification. Muon counting can be done with buried
 detectors or, in favorable conditions, when the EM to muon signal
 ratio is not too large, by counting muon spikes in the recorded traces
 of water Cherenkov tanks. Additionally, when again the EM to muon
 signal  ratio is not too large, the early part of the traces is
 dominated by the muon signal and its time evolution carries information
 on the position of the shower maximum. This is always less
 sensitive than a direct measurement of $X_{max}$ as done
 by the fluorescence technique. Only a combination of both
 measurements, $X_{max}$ and $N_\mu$ can give a shower by shower
 composition indication.

 Alternative techniques trying to exploit the emission of EAS in the radio
 band have also been explored. Between 1967 and 1973 extensive studies took
 place in the 10-100 MHz band. However the technique was judged unworthy 
 in particular due to the strong beaming of the emission in the forward direction 
 and to the poor signal to noise ratio achieved at the time~\cite{Allan-1977}. 
 With the progress in fast digital electronics and low noise amplifiers the interest for the technique
 was revived.  These new efforts(\cite{LOPES, CODALEMA}). 
 aim at replacing ground array detectors with radio antennas which
 are both less expensive and easy to deploy. In addition, the radio signal
 which propagates essentially non altered from its source to the detector
 carries information on the shower evolution. Important progress has
 been made and the radio signal in the VHF band has been showed to be
 dominated by the geo-synchrotron emission of the  electrons and positrons
 of the EM cascade. However, detection of transient signal in those
 frequencies is still a challenging task, even more so given the very tight
 lateral distribution of the radio signal. 
 Recent measurements confirm the strong concentration of the signal in the forward region with an
 exponential decrease from the core with a characteristic distance of
 order 150-200~m.  However they demonstrated the possibility to 
 reconstruct the CR direction with reasonable accuracy~(\cite{CODALEMA-2009}).
 Progress regarding this technique and its exploitation at the highest energies
 are expected in the coming years as important R\&D efforts are being
 pursued (\cite{AERA}).

\subsubsection{Cherenkov light }
 According to Brennan and Chudakov (\cite{Brennan-1958,Chudakov-1960}) 
the Cherenkov light emission from the charged particle component of 
an air shower can provide an integrated measurement of the longitudinal 
development. The Cherenkov intensity is proportional to the primary energy,
while the slope of the lateral distribution is related to the depth
of maximum shower development. Thus, if one samples
the Cherenkov lateral distribution, i.e. the photon density
as a function of the distance from the air shower core, it is possible to
estimate both the primary 
energy and composition.

 From air shower simulations, it was shown (\cite{Patterson-Hillas-1983}) 
 that at distances larger than about  $\sim 150$ m from the shower core, 
 the density of Cherenkov light is proportional to the primary energy but 
 essentially independent of its nature. The light profile close
 to the core is sensitive to the depth of penetration of the shower
 in the atmosphere which correlates 
 with the primary cross section.

  Experimental setups exploiting the Cherenkov light for EAS are
 usually associated to standard particle detector
 arrays (\cite{Efimov-1991,Cassidy-1997, Dickinson-1999,
 Arqueros-2000,Swordy-2001, Chernov-2006}).
 Cherenkov light is also used in CR observation at other energies.
 For a complete overview and history of Cherenkov detection of cosmic
 rays see~\cite{Lidvansky-2005}.

 In~\cite{Cassidy-1997},~\cite{Chernov-2006}
 and~\cite{Arqueros-2000}  the experimental setups consist of 
 open photomultipliers fitted with Winston cones and looking upward in the sky. 
 The largest array composed of 150 of such sensors distributed about every 40~m 
 was installed on the Fly's Eye site in Dugway (Utah, USA) together with
 the CASA
 and CASA-MIA (\cite{CASA}) detector arrays. 
 Near the core, the  lateral distribution of Cherenkov light was
 shown to be
 exponential as in~\cite{Patterson-Hillas-1983}. The CASA-BLANCA
 group (\cite{Fowler-2001}) 
 used a two component function 
 which matches both their real and simulated data.  The function is exponential
 in the range 30\,m--120\,m from the shower core and a power law from
 120\,m--350\,m.  It has three
 parameters: a normalization $C_{120}$, the exponential ``inner slope''
 $s$, and the power law index $\beta$:

\begin{equation}
C(r) = \left\{
   \begin{array}{ll}
   C_{120}\   e^{s(120{\rm m}-r)},     & 30\mbox{m}<r\leq 120\mbox{m} \\
   C_{120}\   (r/120{\rm m})^{-\beta}, & 120\mbox{m}<r\leq
   350\mbox{m} \\
   \end{array}  \right.
\label{eq.CherenkovLDF}
\end{equation}

 The primary energy depends primarily on
 $C_{120}$, the Cherenkov intensity 120\,m from the core. 
 Detailed simulation of the shower and of the detector can be
 used to derive the relation between those two quantities.
 Hadronic models predict that $C_{120}$ grows approximately
 as $E^{1.07}$, because in hadronic cascade the fraction of 
 primary energy directed  into the electromagnetic component 
 increases with energy.

 Similarly the slope of the exponential can be related to the
 shower maximum using simulations. The relation between the two
 is essentially linear (\cite{Arqueros-2000,Fowler-2001}).
 
 The low duty cycle (Cherenkov detector can only be operated on clear
 dark nights), the short core distance up to which the inner slope
 parameter can be used to estimate $X_{max}$ and consequently the small
 spacing within units made this technique inappropriate to study
 EAS beyond an energy of about 10$^{17}$~eV.
 The success of the fluorescence detection technique contributed to
 the decline in the interest for this technique at the highest energies. 

\subsubsection{Fluorescent light}
 The charged secondary particles in EAS produce ultraviolet light
 through nitrogen fluorescence. Nitrogen molecules,  excited by a
 passing shower, emit photons isotropically into several spectral bands
 between 300 and 420 nm. As discussed above, a much larger fraction
 of UV light is emitted as Cherenkov photons. But this emission is
 strongly beamed along the shower axis and usually considered as a
 background to fluorescence detection. 

\begin{figure}[htbp] 
   \centering
   \includegraphics[width=5.5cm, angle=-90]{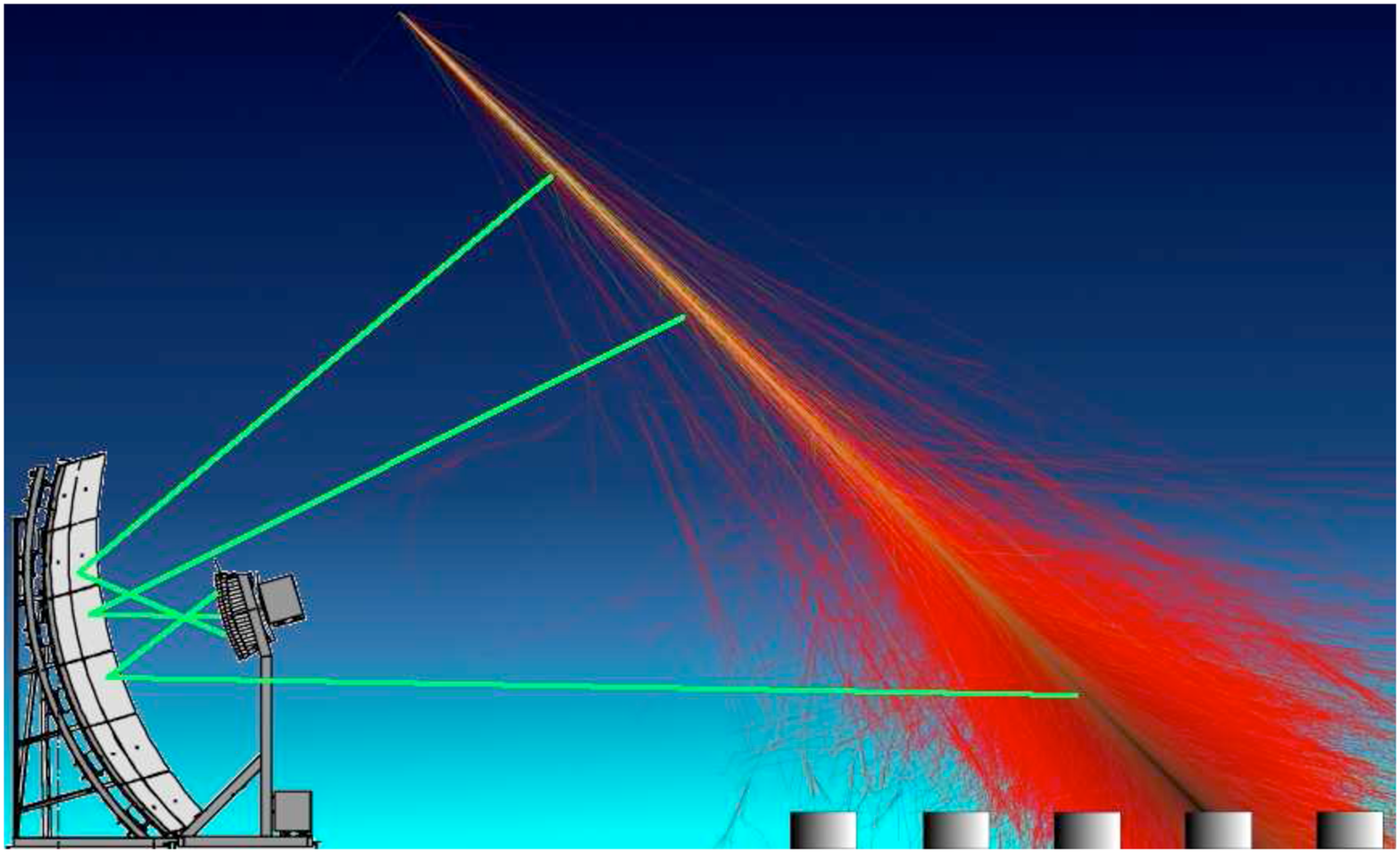} 
   \vspace*{-0.3cm} \caption{Sketch of the detection principles of
 a fluorescence detector. The fluorescence light emitted by the air
 shower is collected on a large mirror and focussed onto a camera
 composed of photo-multipliers (Auger collaboration).}
   \label{fig:FD}
\end{figure}

 The first fluorescence detector assembled for UHECR detection
 was laid down by Greisen and his team in the mid
 60's (\cite{Bunner-1967a, Bunner-1967b}).
 Small mirrors and the atmospheric conditions did not allow to
 record signals from EAS.
 Detectors were built in the late 70's by a group of
 the University of Utah and tested at the Volcano Ranch ground
 array (\cite{bergeson77}) while the first detection of fluorescence light from UHECR
 was made by Tanahashi and his collaborators (\cite{Hara-1970}). Later on, 
 a fully functional detector  was installed at Dugway (Utah) under the  name of {\em Fly's Eye}
 (\cite{Baltrusaitis}).
It took data from 1981 until
 1993 and fully demonstrated the extraordinary potential of the technique . The highest
 energy shower ever detected (320~EeV) was observed by 
 this detector. An updated version of this instrument, the
 High-Resolution Fly's Eye, or HiRes (\cite{Hires}),
 ran on this same site from 1997 until 2006.

 The fluorescence yield is 4 photons per electron per meter at ground
 level pressure. Under clear moonless night conditions, using square-meter
 scale telescopes and sensitive photodetectors, the UV emission from the
 highest energy air showers can be observed at distances in excess of
 20 km from the shower axis. This represents about two attenuation lengths 
 in a standard desert atmosphere  at ground level. 
 Such a large aperture, instrumented from a single site,
 made this technique a very
 attractive alternative to ground arrays despite a duty cycle of about 10\%.  

 Fluorescence photons reach the telescopes in a direct line from their source.
 Thus the collected image reflects exactly the development of the EM cascade (see figure~\ref{fig:FD}).
 From the fluorescence profile it is in principle straightforward to obtain
 the position of the shower maximum and a calorimetric estimate of the primary
 energy. In practice a number of corrections must be made to account for
 the scattering and the absorption of the fluorescence light. Also pollution
 from other sources such as the Cherenkov component which can be emitted
 directly, or diffused by the atmosphere into the telescope,  must be
 carefully evaluated and accounted for. A constant monitoring of the
 atmosphere and of its optical quality is necessary together with a
 precise knowledge of the shower geometry for a careful account for those
 corrections.    

\begin{figure}[thb]
\centerline{\includegraphics[width=7.5cm]{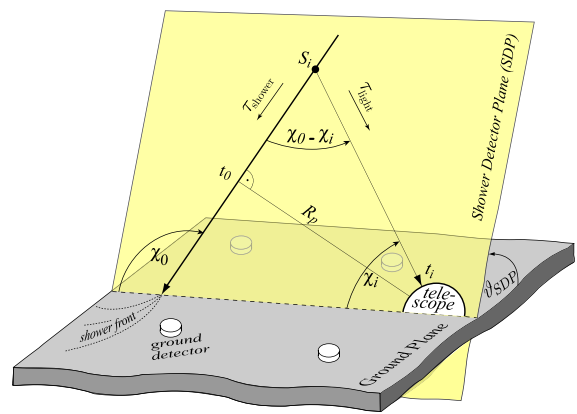}}
\caption{Geometry of the detection of an air shower by a fluorescence telescope, from~\cite{Kuempel-2008}.
}
\label{fig:FDGeom}
\end{figure}

 The shower geometry as viewed from a fluorescence telescope is depicted
 in Fig.~\ref{fig:FDGeom}. It is defined by the shower detector plane
 (SDP), the distance of closest approach $R_p$, the time $t_0$ along the
 shower axis at the distance of closest approach and the angle $\chi_0$
 within the SDP between the ground plane and the shower axis.
 This geometry is usually reconstructed in two steps. 
 First, the shower-detector-plane (SDP) is determined. Next,
 the arrival time $t_i$ of the signal in each pixel
 in the direction of SDP $\chi_i$ 
 is used to determine $\chi_0$, $R_p$ and $t_0$
 from the equation (\cite{Baltrusaitis}).
\begin{equation}
\label{eq:FDGeom}
t_i=t_0+\frac{R_p}{c}\tan{\left(\frac{\chi_0-\chi_i}{2}\right)}\; .
\end{equation}

 One important property of this equation is that unless the angular
 velocity in the camera and its rate of change can be measured there
 is a degeneracy between the impact parameter $R_P$ and the angle
 $\chi_0$. This degeneracy leads to poor pointing resolution - 
 the  three parameters defining the shower geometry cannot be
 constrained accurately (\cite{Sommers-1995}).
 The situation can be improved using fast electronics to achieve 
 a good precision on $t_i$  and for those showers with sufficient track
 length in the camera (over about 10$^\circ$). This was first used by
 the HiRes collaboration with the HiRes-II detector (\cite{HiresFADC-2005}).
 Alternatively the HiRes collaboration also developed a profile
 constrained time fit (PCF) for the part of its detector not equipped
 with fast electronic (\cite{HiresFADC-2005}). Nevertheless, in both
 cases the geometrical resolution remains at a few degrees. The best
 option to resolve this ambiguity  is to improve the measurements.
 This can be done in two ways:\\
A) Using a second telescope viewing the shower from a different position.
 The intersection of the two SPD will constrain the geometry of the
 shower axis to within a fraction of a degree.  This is called a stereo
 reconstruction and is the technique used by the HiRes detector.\\
B) Constrain the $t_0$ parameter by a direct measurement of the time of
 arrival of the shower at the ground. This is the hybrid technique used by
 the Auger detector. Again the geometry can then be constrained to within
 a fraction of a degree.

 Once the geometry has been determined, the fluorescence technique is the
 most appropriate way to measure the energy of the incident cosmic ray.
 The amount of fluorescence light emitted along the shower axis is
 proportional to the number of electrons in the shower. The EAS has
 a longitudinal development usually parameterized by the
 4 parameter Gaisser-Hillas function giving the size $N_e$ of the shower
 as a function of the atmospheric depth $X$ (\cite{Gaisser-Hillas}):
\begin{equation}
N_e(X)=N_{\text{max}}\left(\frac{X-X_0}{X_{\text{max}}-X_0}
\right)^{(X_{\text{max}}-X_0)/\lambda}e^{(X_{\text{max}}-X)/\lambda}
\label{GH}
\end{equation}

 The total energy of the shower is proportional to the integral of this
 function, knowing that the average energy loss per particle is
 2.2~MeV/g~cm$^{-2}$.

 The Pierre Auger fluorescence reconstruction uses this formula while
 the HiRes group has used both the Gaisser-Hillas form and a
 three-parameter Gaussian in age (\cite{Abu-Zayyad-2001}).
 Alternatively,  analytic shower theory led to yet another form
 popularized by~\cite{Greisen-1956}.
 In a  recent study (\cite{Matthews-2010})
 it was shown that the introduction of the profile full width
 half-maximum ($fwhm$) and its asymmetry (defined by the ratio
 of the left-width at maximum to the $fwhm$) could unify the
 parameterization of all three profile functions. 
 Greisen and Gaisser-Hillas profiles are shown to be essentially
 identical while gaussian in age profile only differ at the very
 early and very late development stages of the cascade.    

 Beside the corrections arising from the experimental conditions
 discussed above, the energy transported by the neutral particles
 (neutrinos), the hadrons interacting with nuclei (whose energy is
 not converted into fluorescence) and penetrating muons, 
 whose energy is mostly dumped into the Earth, must also be accounted
 for to estimate properly the primary CR energy. This missing energy
 correction is calculated using detailed simulations and varies 
 with energy, composition and the interaction model used. It is  about
 20\% at 10$^{18}$~eV for iron (10\% for proton) and about 12\% at
 10$^{20}$~eV (6\% for protons). Variations from one model 
 to another are of about 50\% (\cite{Pierog-2007}). 
 
 Despite the fact that fluorescence measurements give direct experimental
 access to the position of $X_{max}$ the separation of hadronic
 primaries according to their mass cannot be done on a shower by shower
 basis because of the intrinsic fluctuation of this parameter. 
 One must look for statistical means of studying the chemical composition
 and/or use additional information such as the muon content that can
 be provided by particle detectors as in the hybrid detection system. 

 Statistical methods relying on the measured fluorescence profile only
 are based on the elongation rate plot, or the RMS($X_{max}$) plot.
 In the former,  one calculates the average value of $X_{max}$ from a
 set of showers of the same energy and plot it as a function of 
 energy. In the latter, it is the width of the distribution of
 $X_{max}$ of shower of the same energy that is plotted against energy. 
 Those measurements are discussed in section~\ref{sec:chem}.

\section{GALACTIC COSMIC RAYS}
\label{sec:knee}

\subsection{Origin of the galactic cosmic rays}
 Galactic cosmic rays are believed to be accelerated
 at supernova remnants.
 This idea was justified by~\cite{GSbook}
 through simple and powerful
 arguments based on the energetics of supernova remnants. If only
 5\% to 10\% of the kinetic energy of supernova remnants is
 converted to accelerated cosmic rays this would  provide the
 energy of all galactic cosmic rays.

  Supernova remnants are attractive candidates for cosmic ray
 acceleration because they have higher magnetic fields than the
 average interstellar medium. They are also large and live long
 enough to carry the acceleration process to high energy. The
 acceleration mechanism is believed to be stochastic acceleration
 at supernova blast shocks.

 The idea of stochastic particle acceleration was first developed by
 E.~Fermi who
 proposed (\cite{Fermi_acc}) to use the charged particle
 interactions with interstellar clouds to accelerate cosmic rays.

 The shock ahead of  the expanding supernova remnant is formed because
 the expansion velocity of the remnant is much higher than the
 sound velocity of the interstellar medium.
 Shock  acceleration  is much faster than the original Fermi
 acceleration mechanism. The energy gain is proportional to $\beta$
 (first-order acceleration) rather than to $\beta^2$ (second-order
 (Fermi) acceleration) where $\beta$ is the velocity of the magnetic
 cloud or the blast shock velocity
 in terms of $c$. In addition, the supernova shock velocity 
 is much higher than the average velocity of molecular clouds. As a
 result shock acceleration is orders of magnitude more efficient, and
 correspondingly much faster. The shock acceleration scenario was
 suggested in the late 1970s (\cite{ALS77,Bell78,Krym77})
 and is under continuous development (\cite{Drury-1983,
 Blandford-1987,Jok87,Jones-1991}).
 The prediction is for a flat, E$^{-2}$ cosmic ray spectrum in
 acceleration in non relativistic shocks and for a steeper
 E$^{2.2-2.3}$ spectrum at acceleration in highly relativistiv shocks
 (\cite{Achterberg2001}).

  The maximum energy that a charged particle could achieve is then
 expressed as a function of the shock velocity and extension and
 the value of the average magnetic field as
\begin{equation}
 E_{max} \, = \, \beta Z e B r_S,
\label{E_max}
\end{equation}
 where $\beta$ is the shock velocity in terms of the speed of light,
 $r_S$ is the shock radius, and $Ze$ is the particle charge.
 Equation (\ref{E_max}) is
 valid during the period of the free expansion of the
 supernova remnant when the shock velocity is constant.
 During the Taylor--Sedov phase, when the shock
 has collected enough interstellar matter to start slowing
 down, the
 maximum energy starts decreasing as the radius is only proportional
 to the time to the power of 2/5. Detailed more recent
 calculations (\cite{Berezh})
 derive maximum energy values close to 5 $\times$ 10$^5$ GeV
 and even higher in some cases (\cite{VSP2010}).
 An important component of the
 expression for $E_{max}$ is its dependence  on the particle
 charge $Z$. It means that a fully ionized heavy nucleus of
 charge $Z$ could achieve $Z$ times higher energy than a proton.

 Since cosmic rays scatter in the galactic magnetic fields
 we cannot observe them coming from particular sources.
 The only way we can study their acceleration sites is by
 observing the neutral particles, gamma-rays and neutrinos,
 generated by their interactions during acceleration.
 There are two epochs in supernova remnant evolution when one
 can expect $\gamma$-ray and neutrino emission. One of them is 
 shortly after the supernova explosion, when the density of the
 expanding supernova envelope is very high and thus contains enough
 of a target for hadronic interactions.
 The emission will continue for about 2 to 10 years (depending on the
 mass distribution and expansion velocity of the SNR) until
 proton energy loss on inelastic interactions becomes dominated
 by the adiabatic loss due to the SNR expansion. The $\gamma$-ray
 emission will fade for a long time, until the SNR reaches the
 Sedov phase, when most of the galactic cosmic rays are 
 accelerated. Since this phase lasts for more than 1,000 years
 there should be many supernova remnants that are gamma-ray sources.

  The modern expectations of the $\gamma$-ray emission of mature
 supernova remnants was developed by~\cite{DAV}.
 The assumption is that cosmic rays
 at the source have a much flatter spectrum than the one observed
 at Earth as acceleration models suggest.
 As an example of the expectations from a concrete SNR~\cite{DAV}
 apply the calculation to the Tycho (1572) supernova remnant
 which should be close to the Sedov phase. One can take the
 average supernova energy and density from different estimates
 $E_{\rm SN} = 4.5 \pm 2.5 \times 10^{50}$
 ergs, matter density $n_1 = 0.7 \pm 0.4$ and estimate the
 $\gamma$-ray flux for conversion efficiency $\theta$ = 0.2 and
 distance $d = 2.25 \pm 0.25$ kpc. 
 The expected flux is 
\begin{equation}
 F(>E_\gamma) \; = \; \simeq 1.2 \times 10^{-12}
 \left( \frac {E_\gamma}{\rm TeV} \right) ^{-1.1} \; \;   
 {\rm cm}^{-2} \, {\rm s}^{-1} \; .
\label{DAV_Tycho}
\end{equation}
 The detection of such a flux is easily within the capabilities of the
 last generation of $\gamma$-ray Cherenkov telescopes.

\begin{figure*}[thb]
\centerline{\includegraphics[width=0.8\textwidth]{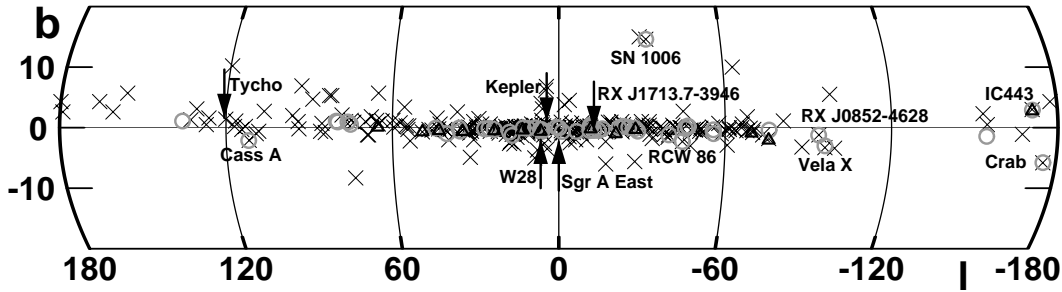}}
\caption{Comparison of the positions of supernova remnants
 (x's) with GeV (Fermi/LAT: triangles) and TeV (circles) gamma-ray sources. 
 }
\label{dagreen3}
\end{figure*}

  Figure~\ref{dagreen3} compares the positions of supernova 
 remnants from the~\cite{green} catalog
  with the 
 positions of GeV gamma-ray sources from the
 Fermi/LAT observations (\cite{Bright})
  and of TeV sources from
 the TeVCat catalog~[{\em http://tevcat.uchicago.edu}].
  One can clearly see several coincidences. There are others that
 are more difficult to find by eye because there are so many
 supernova remnants close to the galactic center at very low
 galactic latitude.
 Many of the
 TeV sources come from the HESS survey of the galactic
 plane of galactic longitudes from --30$^o$ to 30$^o$ and 
 latitudes below 3$^o$ (\cite{HESS_survey}).
 The names of some of the SNR that emit 
 TeV $\gamma$-rays are indicated in the figure whenever possible.
 There are though no gamma-rays coming from
 the Tycho supernova remnant.

  The number of direct coincidences of the supernova remnants 
 locations with the directions of the gamma-ray sources is
 relatively small. 
 What is the fraction of the supernova remnants
 that are $\gamma$-ray sources and thus are cosmic ray
 accelerators? The small 
 fraction of $\gamma$-ray producing SNRs creates doubts that galactic
 cosmic rays are generated at these objects. This may be true, but the
 HESS group put together an alternative explanation of this effect
 in their study of the galactic ridge (\cite{ridge}).
  Hadronic gamma-ray production is only possible when the matter
 density of the medium is much higher than 1 cm$^{-3}$.
 A very likely gamma-ray production site is the location of dense
 clouds of matter close to an
 acceleration site of cosmic rays. HESS observed that the peaks 
 of the $\gamma$-ray emission from the region of the galactic center
 ridge, after subtraction of known sources,
 coincide with the positions of molecular clouds with a matter
 density of hundreds per cm$^3$.
 The total amount of mass in these clouds
 is 2--4 $\times$ 10$^7$ solar masses. In addition, the energy spectrum
 of the $\gamma$-rays is about E$_\gamma^{-2.3}$ which is likely to
 happen close to the cosmic rays acceleration site.
 This observation may explain the fact that many sources of TeV
 $\gamma$-rays do not exactly coincide with the positions of SNR
 where the cosmic rays that produce them are accelerated, rather
 with close by molecular clouds.
 Higher energy cosmic rays
 diffuse faster away from their sources. For this 
 reason it is possible that a molecular cloud could be a source 
 of TeV $\gamma$-rays before it becomes a strong source of 
 GeV gamma-rays. 
 
 TeV $\gamma$-rays have been detected from the Crab nebula
 and the supernova remnants SN1006, Cas A, RX~J1713.7-39466,
 RX~J0852.0-4622, W28, W48, RCW~86 and others.
 The Crab nebula is the standard candle
 in TeV $\gamma$-ray astronomy; it has a steady flux
 which is used to measure the fluxes of other sources.
 The models that explain best its gamma-ray emission do not
 involve hadronic interactions. They are electromagnetic 
 models that rely on electron acceleration and the inverse Compton
 process.

\subsection{Energy spectrum and composition
 at the knee}

 The energy range in which the cosmic ray spectrum changes its slope is called
 `the knee'. Its existence was first suggested by the Moscow State
 University group (\cite {MSU}) on the basis of their air shower data.
 Many groups
 have studied the knee region and the change of the cosmic ray spectrum is
 well established. Up to an energy of 10$^{6}$ GeV the spectrum of all cosmic
 ray nuclei is a power law with  differential spectral index $\alpha$ of
 2.70--2.75. The spectral index increases by $\Delta \alpha$ of about 0.3
 above the knee.
 A flattening of the spectral index has been
 detected (\cite{Ahn-2010,Panov-2011}) just before the knee.

 There is no lack of theoretical ideas about the origin of the knee. 
 \cite{Peters59}
 suggested that the knee is
 a rigidity-dependent effect. Rigidity $R$ is the ratio of 
 the particle momentum to its charge.
 It could be related to the maximum rigidity
 that can be achieved in acceleration processes
 or to rigidity-dependent escape
 of the cosmic rays from the Galaxy.
 Rigidity-dependent effect is an attractive idea.
 We know that heavy charged nuclei can achieve $Z$ times
 higher energy at acceleration. So a natural assumption could
 be that at the approach to the knee cosmic ray sources can
 not accelerate protons to higher energy. Then the next nucleus,
 He, takes over and the process continues in order of charge
 until at some higher energy galactic cosmic rays contain 
 only iron nuclei. Mostly the common nuclei of H, He, C, O, Si,
 Mg and Fe are represented in the cosmic ray spectrum.

\begin{figure}[thb]
\centerline{\includegraphics[width=8.5cm]{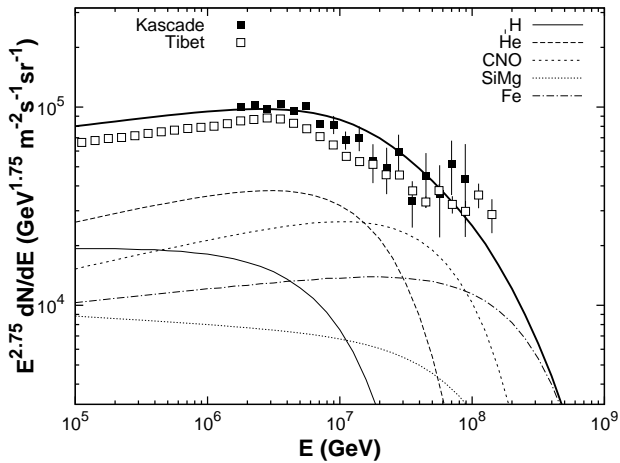}}
\caption{ 
 A simple model of the knee which extends the low energy
 spectrum and composition to high energy with an exponential
 cutoff at 10\protect$^7$ GeV. The model is normalized to the
 all particle flux measured by the Kascade experiment (\cite{Kascade})
 which is shown with full squares. The measurements of
 the Tibet III experiment (\cite{Tibet})
 is shown with empty squares.
}
\label{spec_mod}
\end{figure}

 Figure \ref{spec_mod} shows a very simple 
 flux model with rigidity-dependence. It uses the spectrum and
 composition at low energy and extends it to high energy
 with exponential cutoff in rigidity at 10$^7$ GV.
 The thin lines show the contribution of different nuclear groups
 to the all-particle spectrum. The proton spectrum turns over
 at 10$^7$ GeV and those of heavier nuclei turn over at energies
 of $Z \times {\rm 10}^7$ GeV. At energies above 10$^8$ GeV there 
 are only heavy, high $Z$, nuclei in the cosmic ray flux. 
 The end of the modeled spectrum is where the Fe component 
 is also exponentially cut off.

  This simple model agrees pretty well with the measurements of
 the Kascade (\cite{Kascade}) and Tibet\ III (\cite{Tibet}) air
 shower arrays. The normalization of these experiments is
 slightly different which affects both the magnitude of the flux
 when it is multiplied by $E^{2.75}$ and the position of the knee. 
 The analysis of air shower data depends on the hadronic interaction
 models used in the simulations. The dependence is stronger 
 for the Kascade experiment which is located much lower in the
 atmosphere. Tibet~III is close to the depth of shower maximum
 $X_{max}$ where the ratio between shower electrons and muons 
 is at its maximum. This ratio decreases with the 
 atmospheric depth but predictions depend strongly on the hadronic
 interaction model. The normalization of both spectra, however,
 depends on the cosmic ray composition.

\begin{figure}[thb]
\centerline{\includegraphics[width=8.5cm]{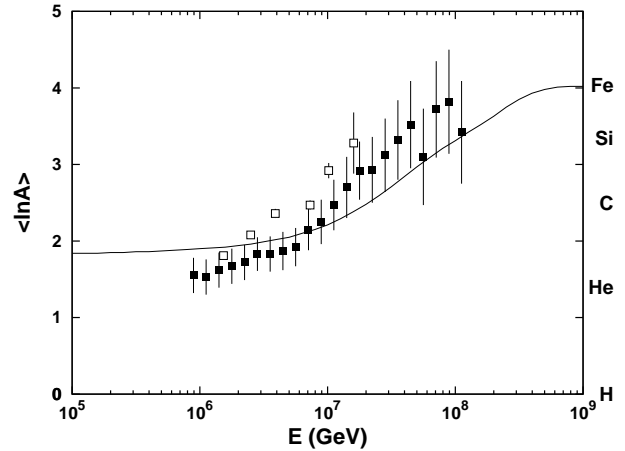}}
\caption{ 
 Results from studies of the cosmic ray composition in the
 region of the knee compared to the predictions of the
 simple model presented in Fig.~\ref{spec_mod}.  
}
\label{comp}
\end{figure}

 The cosmic ray composition estimated from air shower data is
 usually presented as the average value
 of the logarithm of the primary particle mass $\langle \ln{A} \rangle$.
 Different composition estimates are not in very good agreement.
 As an illustration Fig.~\ref{comp}
 presents the results from the analyses of data from the
 Kascade (\cite{Kascade}) and EAS-TOP (\cite{EASTOP_c})
 experiments.  
 Both composition results come from the ratio of the shower
 muon density at predefined distances from the shower core
 as a function of the total
 number of electrons in the shower, $N_e$. These two 
 measurements are in a fairly good agreement.

 Figure~\ref{comp} shows that the composition
 becomes significantly heavier with increasing energy. It is
 fully consistent with the rigidity dependent idea.
 The simple composition model, however, does not describe the data
 well. It predicts heavier composition at 10$^6$ GeV and lighter
 composition at 10$^8$ GeV. A better model would require a
 different low energy composition and possibly lower maximum
 rigidity.
  
 Although the majority of the experiments measure a cosmic ray
 composition
 that becomes heavier between 5$\times$10$^6$ and 10$^7$ GeV,
 it is difficult to draw a definite conclusion about the exact
 changes of the cosmic ray spectrum and composition at the knee.
 All experiments agree that the cosmic ray spectrum steepens
 above 10$^6$ GeV. The exact position of the spectral change 
 and the width of the transition region are not yet well
 determined. The composition studies, both with surface air
 shower arrays and with optical detectors, indicate a change
 in the average mass of the cosmic ray nuclei after the
 steepening of the spectrum, once again with large uncertainty
 in the energy range and shape. All these numerous data sets
 are consistent with rigidity dependent effects, either in the cosmic
 ray acceleration or in their propagation. This second scenario
 assumes that lower rigidity nuclei are contained in the
 Galaxy longer. 

\section{ORIGIN OF COSMIC RAYS UP TO 10$^{20}$ eV}
\label{sec:acc}


 The question of how to accelerate cosmic rays up to 10$^{20}$~eV has
 been pending since their very first observation in the 1960's.
 More than thirty years later, in the mid 90's, the data collected by
 the AGASA and HiRes experiments generated a profusion of ideas.
 Some of them  aimed at an explaination of the possible absence of the GZK
 cutoff and the lack of visible astrophysical sources.
 All ideas tried to find a
 solution to the basic problem of how to transfer efficiently a
 macroscopic amount of energy, of the order of 20 Joules,
 to a microscopic particle.

 To circumvent this difficulty  one of the main axis of research was
 the mere suppression of the accelerator itself. Particles are not
 accelerated as such but directly produced, via the decay of some
 supermassive relic of the Big Bang, or by the collapse of topological
 defects, with energies in excess of 100 EeV. While attractive from
 theoretical point of view, these models had the disadvantage of
 replacing the acceleration problem with the question of the nature
 and existence of such top-down sources.

 With the observational facts collected by HiRes and Auger
 in the past decade the situation has been greatly
 clarified.  A cutoff in the high energy end of the spectrum is
 clearly visible and the limits on the fraction of photons and
 on the flux of high energy neutrinos have strongly reduced the
 interest in the top-down models.  On the other hand the possibility
 of a dominant iron component in the very end of the energy spectrum
 decreases by a factor 26 the hard conditions placed on "standard"
 bottom-up cosmic accelerators to reach the 100 EeV barrier.

 Nevertheless, after many decades of investigation the problem has
 not been solved, and even the extragalactic nature of the sources
 above 3~EeV has been challenged (\cite{Wick-2003},\cite{Calvez-2010}).
 In the following we briefly reviews some of the necessary conditions
 for the acceleration of UHECR at astrophysical sites and enumerate
 some possible candidates. We also briefly review the main
 characteristics of the top-down models. 
 More details on this subject can be found in the recent review of
 \cite{Olinto-2010}, now in press.

\subsection{Possible acceleration sites}

 Acceleration at astrophysical sites may occur principally through
 two distinct mechanisms: diffusive shock acceleration, based on the
 Fermi mechanism and one shot acceleration in very high electric
 field generated by rapidly rotating compact magnetized objects such
 as young neutron stars.

 Diffusive acceleration takes place near shock waves and rely on the
 repeated scattering of charged particles
 on magnetic irregularities back and forth across the shock.
 In the case of non relativistic shock velocities the energy gain
 at each crossing is of the order of $\Delta E \sim E$.
 To reach energies above 1 EeV large acceleration regions and/or
 highly relativistic blast waves are necessary.
 In the case of relativistic shock the energy gain reaches
 $\Gamma_s^2$E where $\Gamma_S$ is the shock bulk Lorentz
 factor. Such gain appears, however, to be limited to the first
 crossing (\cite{Achterberg-2001}).

 One of the principal advantages of the diffusive shock acceleration
 mechanism is that it naturally provides a power low spectrum whose
 predicted index $\gamma$ is within the range of the experimental
 measurements. Depending on the exact geometry of the shock and on
 its relativistic nature, the combination of the energy gain per
 crossing and of the escape probability leads to a power law index
 of exactly 2  for the case of a strong
 non relativistic shock in an ideal gas and to indexes
 between 2.1 and 2.4 for relativistic shocks.

\begin{figure}[thb]
\centerline{\includegraphics[width=0.45\textwidth]{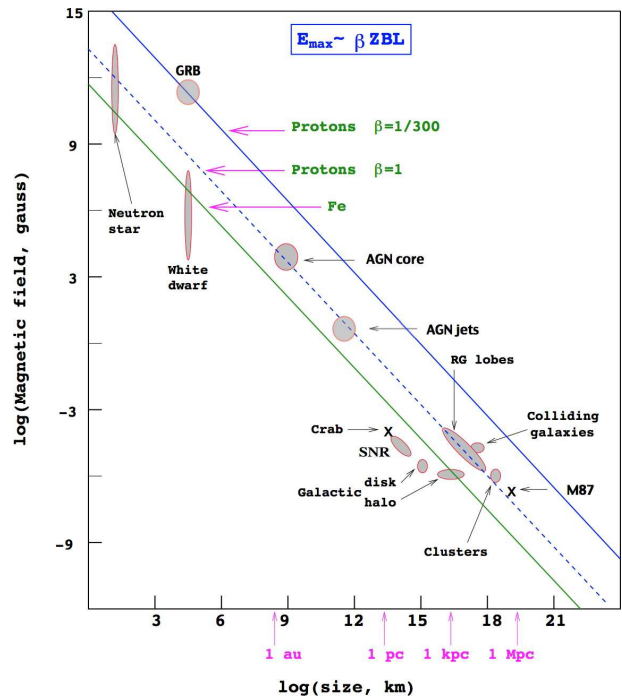}}
\caption{Hillas plot for candidate acceleration sites, relating their
 size and magnetic field strength. To accelerate a given particle
 species above 100 EeV objects must lie above the corresponding lines.}
\label{fig:hillas-plot}
\end{figure}

  \cite{Hillas-1984} summarized the conditions
 on potential acceleration sites using a relation between the maximum
 energy of a particle of charge Ze and the size and strength of the
 magnetic field of the site:
$$
E_{max}=\beta Ze\left (\frac{B}{1\mu G}\right)\left(\frac{R}{1kpc}\right)\mbox{ EeV}
$$
 where $\beta$ represents the velocity of the accelerating shock wave
 or the efficiency of the accelerator\footnote{In the case of a relativistic
 shock the bulk Lorentz factor $\Gamma_s$ enters the right hand side of
 this equation (\cite{Achterberg-2001})}. We show in
 Fig.~\ref{fig:hillas-plot} the now famous "Hillas plot" illustrating
 this condition.

 Looking at the Hillas diagram one sees that only a few astrophysical
 sources satisfy this necessary, but not sufficient, condition. 
 Among the possible candidates are neutron stars and other similar
 compact objects, large-scale shocks due to merging galaxies or
 clusters of galaxies, the core  and jets of Active Galactic Nuclei
 (AGN), hot spots of Fanaroff-Riley class II (FR-II) radio galaxies
 and processes associated with Gamma Ray Bursts (GRB).

 AGN have long been considered as potential sites where energetic
 particle production might take place (\cite{GSbook,Hillas-1984}).
 AGN jets have dimensions of the order of a fraction of a parsec
 with magnetic field of the order of a few Gauss (\cite{Halzen-1997}).
 These parameters could in principle lead to a maximum energy for
 protons of a few tens of EeV. Similarly, AGN cores with a magnetic
 field of order 10$^3$~G and size of a few 10$^{-5}$~parsec can reach
 about the same energy. However those maxima, already marginally
 consistent with acceleration up to 100 EeV, are unlikely to be
 achieved under realistic conditions.
 The high radiation field around the central engine of an AGN is
 likely to interact with the accelerated protons
 while energy losses due to synchrotron radiation,
 Compton processes, and adiabatic losses will also take place.
 The situation is
 worse for nuclei that will photodisintegrate even faster.
 Such processes may lead
 to a maximum energy of only a small fraction of EeV
 (\cite{Bhattacharjee-2000}).
 To get around this problem, the acceleration site must be away
 from the active center and in a region with a lower radiation density
 as in the terminal shock sites of the jets, a requirement possibly
 fulfilled by FR-II galaxies.

 The link with GRB and UHECR acceleration was initially made by
 \cite{Waxman-1995}, \cite{Vietri1995}, and \cite{MU1995},
 who pointed out that the observed at the time
 cosmic ray  flux beyond 100 EeV (now estimated to be lower by a
 factor 3 to 10 after the measurements of Auger and HiRes)
 is consistent with a scenario in
 which these particle are produced in GRB's provided that each burst
 produces similar energies in gamma rays and in high energy cosmic rays.
 From a phenomenological point of view, based on the gamma-ray observations,
 bursts can be described by the product of the dissipation of the
 kinetic energy of  a relativistic expanding fireball.
 The time variability of the phenomena and the compact nature of the
 source suggest that the expanding wind has a bulk Lorentz factor of
 a few hundreds, a condition in principle sufficient to accelerate
 charged particles up to 100 EeV.  In a more recent analysis taking
 into account the cosmological nature of the GRB distribution
 similar conclusion has been drawn, placing GRB's as one of the
 prominent sites of cosmic ray acceleration.
 Note that in such a scenario, UHECR sources are not visible since
 the detected cosmic rays come from various bursts and reach the Earth
 long after ($10^3$ to $10^7$ years) the gamma ray burst
 itself (\cite{Waxman-2004}).

 Direct observation of radio galaxies gives us their main characteristics
 in term of their radio luminosity (10$^{39} $ $-$ 10$^{44}$ ergs/sec),
 their size (10$^3$ $-$ 10$^6$ parsecs), their brightness morphology
 and the polarization level of the radio emission. From these parameters 
 one can indirectly infer their mean magnetic field (of the order of
 10 $-$ 10$^3$ $\mu$G) and kinetic power (10$^{42}$ $-$ 10$^{47}$ ergs/sec).
 However the exact characteristics of the jets and in particular their
 Lorentz factor, density and composition are still under
 debate (\cite{Massaglia-2008}).  Among radio loud galaxies  the
 Fanaroff-Riley radio galaxies of class II are of particular interest
 because they combine a very powerful engine and relativistic blast
 wave (with Lorentz factor of the order 2 $-$ 10) together with a
 relatively scarce environment. Hence, in the associated hot spots
 where the relativistic jet terminates, they not only satisfy the
 acceleration criterion but also the requirement that the
 accelerated particle does not lose all of its energy via radiation
 or interactions on its way out of the
 source (\cite{Rachen-93}). Finally,
 invoking the sheared jet mechanisms where inductive acceleration
 can take place at the interface of the central spine and outer
 flow of the jet, acceleration of UHECR can take place in the
 jets themselves (\cite{Lyutikov-2007}).

\subsection{Exotic top-down models}


 One way to overcome the many problems related to the acceleration of
 UHECR is to introduce the existence of a new unstable or meta-stable
 super-massive particle. Its decay should produce
 quarks and leptons, which will result in a large cascade of energetic
 photons, neutrinos, and light leptons with a small
 fraction of protons and neutrons. In such a model no acceleration is
 required and cosmic rays are emerging directly  with ultra high energy
 from the decay cascade. Hence their name of top-down models.

 For this scenario to produce  observable particles above 50 EeV three
 conditions must be met:\\
\ \ \ \ $\bullet$ The decay must occur in recent time, i.e.
 at distances less than about 100~Mpc\\
\ \ \ \ $\bullet$ The mass of this new particle must be well above the
 observed highest energy (100~EeV range), a hypothesis well satisfied by
 Grand Unification Theories (GUT) whose scale is around $10^{6}-10^{7}$~EeV.\\
\ \ \ \ $\bullet$ The ratio of the volume density of this particle to
 its decay time must be compatible with the observed flux.\\
 Two distinct mechanisms may produce such energy release.\\
\ \ \ \ $\bullet$ Radiation, interaction or collapse of Topological
 Defects (TD), producing GUT particles that
 decay instantly. In those models the TD are leftovers
 from the GUT symmetry breaking phase transition in the very early universe.
 However very little is known about the phase transition itself and
 about the
 TD density that survives a possible inflationary phase, and quantitative
 predictions are usually quite difficult to rely on.\\
\ \ \ \ $\bullet$ Super-massive metastable relic particles from some primordial
 quantum field, produced after the  inflationary stage of our Universe.
 Lifetime of those relics should be of the order of the age of the
 Universe and must be guaranteed by some almost conserved protecting
 symmetry.
 It is worth noting that in some of those scenarios the relic particles
 may also act as non-thermal Dark Matter.

 In the case of TD the flux of UHECR is related to their number density
 and their radiation, collapse or interaction rate,  while in the case of
 massive relics the flux is driven by the ratio of the density of the
 relics over their lifetime.

 The very wide variety of topological defect models together with their
 large number of parameters makes them difficult to review in detail.
 Many authors have addressed this field. Among them let us mention
 \cite{Shellard-1994} and
 \cite{Vachaspati-1997,Vachaspati1998} for a review on TD formation
 and interaction. For a review on experimental signatures see
 \cite{Bhattacharjee-1998},
 \cite{Bhattacharjee-2000} and
 \cite{Berezinsky-1998b}

 Basic principles ruling the formation of TD in the early universe
 derive from the current picture on the evolution of the Universe.
 Several symmetry breaking phase transitions such as
 $GUT\Longrightarrow H~...\Longrightarrow SU(3)\times
 SU(2)\times U(1)$ occurred during the cooling. For those ``spontaneous''
 symmetry breakings to occur, some scalar field (similar to the
 Higgs field generating masses to elementary particles) must acquire
 a non vanishing expectation value in the new vacuum (ground) state.
 Quanta associated to those fields have energies of the order of the
 symmetry breaking scale, e.g. $10^{15}-10^{16}$~GeV for the
 Grand Unification scale.

 During the phase transition process non causal regions may evolve
 towards different states in such a way that at the different domain
 borders, the Higgs field keeps a null expectation value.
 Energy is then trapped in a TD whose properties depend on the
 topology of the manifold where the Higgs potential reaches its
 minimum. Possible TDs are classified
 according to their dimensions: magnetic monopoles (0-dimensional, point-like);
 cosmic strings (1-dimensional); a sub-variety of the previous which
 carries current and is supra-conducting; domain walls (2-dimensional);
 textures (3-dimensional). Among those, only monopoles
 and cosmic strings are of interest as possible UHECR sources.

 Supermassive relic particles may be another possible source of
 UHECR (\cite{Berezinsky-1998a}, \cite{Bhattacharjee-2000}).
 Their mass should be larger than $10^{12}$~GeV
 and their lifetime of the order of the age of the Universe since
 these relics must decay now (close by)
 in order to explain the UHECR flux. Unlike strings and monopoles,
 relics aggregate under the effect of gravity like ordinary matter
 and act as a (non thermal) cold dark matter component.
 The distribution of such relics should consequently be biased
 towards galaxies and galaxy clusters.

 Regardless of the details and dynamic of the topological collapse
 or of the massive particle decay the cascade that is produced will
 contain, possibly among many other things,  quarks, gluons and leptons.
 Those particles will in turn  produce far more photons and neutrinos
 than any type of nucleons. Hence, in all conceivable top-down scenarios,
 photons and neutrinos dominate at the end of the hadronic cascade.
 This is \emph{the} important distinction from the conventional
 acceleration mechanisms. The spectra of  photons and neutrinos can
 be derived from the charged and neutral pion densities in the jets as:
$$\Phi_{\gamma}^{\pi^0}(E,t) \simeq 2 \int_E^{E_{\text{jet}}}\Phi_{\pi^0}(\varepsilon,t)d\varepsilon/\varepsilon$$
$$\Phi_{\nu}^{\pi^\pm}(E,t) \simeq 2.34 \int_{2.34E}^{E_{\text{jet}}}\Phi_{\pi^\pm}(\varepsilon,t)d\varepsilon
/\varepsilon$$
 where $E_{\text{jet}}$ is the total energy of the jet
 (or equivalently the initial parton energy). Since
$\Phi_{\pi^\pm}(\varepsilon,t)\simeq 2\Phi_{\pi^0}(\varepsilon,t)$,
 photons and neutrinos should have very similar spectra.
 These injection spectra must then be convoluted with the transport
 phenomena to obtain the corresponding flux on Earth. In particular
 the photon  transport equation strongly depends on its energy and on
 the poorly known universal radio background and extragalactic magnetic
 fields (\cite{Stanev-book}).
 Nevertheless,  if top-down scenario dominates the UHECR
 production above  a certain energy the photon fraction should become
 very large. However, the recent observations, in particular of
 the Auger observatory, showed the contrary. This has considerably
 reduced the possibility for such models to be the source of UHECR.

\section{UHECR DETECTORS}
\label{sec:det}

\subsection{Older experiments - AGASA}\label{sec:Agasa}

 The AGASA (Akeno Giant Air-Shower Array) is the largest air shower
 array of the previous generation of detectors. 
 AGASA covered 100 km$^2$. It consisted of 111 scintillator counters
 of area 2.2 m$^2$ at an average distance of 1 km from each other.
 Initially AGASA was divided in four {\em branches} that operated
 individually (\cite{Chiba-1992}). In 1995 the data acquisition system
 was improved (\cite{Ohoka-1997}) and the four branches were unified in
 a single detector. This increased the effective detector area by a
 factor of 1.7 to reach 100 km$^2$.

  Each AGASA station was viewed by a 125mm photomultiplier and  had a
 detector control unit (DCU) that controled the high voltage of the PMT,
 adjusted the gain and recorded the timing and pulse hight of 
 every signal. The stations were connected by two optical fibers. One of
 them was used to send commands to the DCU. The other reported to
 the center triggers, shower data, and monitor data. AGASA operated
 for more than 12 years. 

  Muon detectors of sizes from 2.4 m$^2$ to 10 m$^2$ were installed 
 at 27 of the detector stations. In the Southeast corner of AGASA was
 the Akeno 1 km$^2$ array which has been in operation since 1979.
 It was a densely packed array with detector separations from 3 to 120 m.
 Akeno has studied the cosmic ray energy spectrum from 3$\times$10$^{14}$ eV
 to 3$\times$10$^{18}$ eV (\cite{Nagano-1984}).

\subsection{HiRes}\label{sec:Hires}

 The Hires observatory was a much improved follow-up of the pioneer and
 very successful  fluorescence detector Fly's Eye (\cite{Baltrusaitis}).
                                                          
 Also constructed by the University of Utah, the observatory was comprised of
 two air fluorescence detector sites separated by 12.6 km (\cite{HiRes_prl1,
 Abbasi-2005}). 
 It was located at the U.S. Army Dugway Proving Ground in the state of Utah
 at 40.00$^o$N, 113$^o$W, at atmospheric depth of 870 g/cm$^2$.
 The two detectors, referred to as HiRes-I and Hires-II operated on clear
 moonless nights with a effective duty cycle for physics data of about
 10\% typical for fluorescence detectors.   

 The HiRes-I site (\cite{Abu-Zayyad-1999})
 consisted of 21 telescope units, each equipped with a 5 m$^2$ spherical mirror
 and 256 phototube pixels at its focal plane. Each telescope covered an
 elevation range of 14$^\circ$ between 3$^\circ$ and 17$^\circ$ and
 360$^\circ$ in azimuth. The phototubes were equipped with sample and hold
 electronics which integrated the fluorescence signal within a 5.6 $\mu$s
 window. This was enough to contain 
 the shower signal but also because of the limited elevation range of
 the detectors did not allow to extract the shower geometry
 from Eq.~\ref{eq:FDGeom} alone. HiRes-I was in operation from
 June 1997 up until April 2006. 
\begin{figure}[thb]
\centerline{\includegraphics[width=8cm]{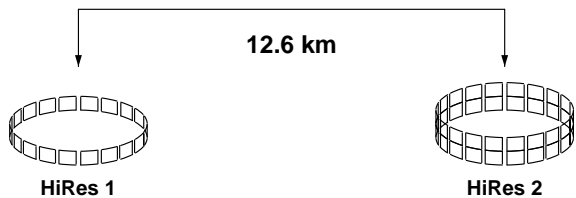}}
\caption{ Sketch of the HiRes fluorescent experiment. Each rectangle represents a fluorescence telescope
including a mirror and a camera. Each site of the HiRes detectors has a nearly full azimuth coverage and site 2 
which consists of two rings of mirrors covers elevation from $3^\circ$ to $30^\circ$  while site one,
 with a single ring, covers elevation from $3^\circ$ to $16^\circ$.  
}
\label{fig:HiRes}
\end{figure}

 The HiRes-II site was completed at the end of 1999. Detectors were similar
 to those of HiRes-I but with twice as many mirrors organized in two rings
 covering elevation from 3$^\circ$ to 31$^\circ$ and still 360$^\circ$ in
 azimuth. Moreover the HiRes-II phototubes were equipped with fast FADC
 electronics which sampled the shower signal every 100~ns.  This allowed
 the reconstruction of the shower geometry from timing alone
 (Eq.~\ref{eq:FDGeom})
 with a precision of about 5 degrees (\cite{HiRes_stereo}).

 Although the two detectors of HiRes could trigger and reconstruct events
 independently, HiRes was designed to measure the fluorescence light
 stereoscopically.  Stereoscopic mode allows the reconstruction of the shower
 geometry with a precision of 0.4$^\circ$ and provides very valuable
 information and cross checks about the atmospheric conditions at
 the time of the event. HiRes-I and II took data until April 2006 for
 an accumulated exposure in stereoscopic mode of 3460~hours
 (\cite{HiRes_stereo}).
  On the other hand, the "monocular" mode had better statistical power
 and covered a much wider energy range. 

 In monocular mode, the geometry of the HiRes-I events was
 calculated using an expected form of the shower development in addition
 to Eq.~\ref{eq:FDGeom} (the PCF technique) . The shower profile was
 assumed to be described by the Gaisser-Hillas parameterization  which is
 in good agreement with HiRes measurements  and detailed
 simulations (\cite{Abu-Zayyad-2001, Song-2000, Kalmykov-1997}).
%
%
%
 Significant contamination from the forward-beamed direct Cherenkov light
 degraded its reliability and tracks with $\chi_0 > 120^\circ$ or with
 large Cherenkov fraction (as estimated from Monte Carlo simulation)
 were rejected. Monte Carlo studies showed that  the RMS energy
 resolution for this method was better than 20\% only at the highest
 energies (above $10^{19.5}$~eV). 

 For monocular reconstruction, from either HiRes-I or Hires-II, the
 aperture is energy and composition dependent and must be evaluated
 by Monte-Carlo simulations. The HiRes collaboration made extensive
 and detailed simulation of both the atmospheric cascade and their
 detector and studied the systematics uncertainty in the estimation of
 the monocular aperture (\cite{Abbasi-2007}).The stereo aperture
 was determined by the requirement that the Monte Carlo
 events trigger both telescopes. Because of the better reconstruction
 of stereo events the quality cuts for them were not as strict
 and the stereo aperture is somewhat higher above energy of 
 10$^{19.7}$ eV (\cite{HiRes_stereo}). It is smaller for events 
 below 10$^{18.5}$ eV.

 To determine the correct shower energies, the air fluorescence technique
 requires accurate measurement and monitoring of the absolute gain of the
 telescope. In HiRes, two methods of calibration were used. One provided
 nightly relative calibration and used a YAG laser connected to two mirrors,
 the other relied on a stable and standard light source and provided monthly
 absolute calibration. The pulses
 from a YAG laser were distributed to 2 mirrors via optical fibers. They
 provided a nightly relative calibration.  Relative phototube gains were
 stable to within 3.5\% and the absolute gains were known to
 ±10\% (\cite{Abu-Zayyad-2005}).
 Fluorescence light from air showers is also attenuated by molecular
 diffusion (Rayleigh) and aerosol scattering. While the former is
 approximately constant, the aerosol concentration varies rapidly with
 time. At HiRes (likewise at the Auger observatory) the aerosol
 content was  measured by observing scattered light from steerable
 laser systems.

 The fluorescence yield has been measured by~\cite{Kakimoto-1996}, 
by~\cite{Nagano-2004},
 and more recently by the AIRFLY collaboration (\cite{AIRFLY-2008}). 
 A review of those measurements is available in~\cite{Arqueros-2008}.
 The HiRes collaboration used the fluorescence spectrum compiled by
 \cite{Bunner-1967a} and normalized it to
 the yield of \cite{Kakimoto-1996}. 

 For both HiRes-I and HiRes-II events, the photo-electron count was
 converted to a shower size at each atmospheric depth, using the known
 geometry of the shower, and correcting for atmospheric attenuation.
 The reconstructed profile was integrated over the atmospheric depth. The
 integral was then multiplied by the average energy loss per particle
 to give the visible shower energy. A correction (about 10\%) for the
 invisible energy, carried off by non-observable particles,
 was applied to give the total shower energy.

 The HiRes data contains two events at or above 10$^{20}$~eV, measured at
 1.0 and 1.5$\times$10$^{20}$~eV. 
 Assuming a purely molecular atmosphere a lower energy limit of
 0.9 and 1.2$\times$ 10$^{20}$~eV was obtained for these events.
 The flux values were on average 13\% lower than the stereo
 spectrum reported by Fly's Eye collaboration~\cite{Bird-1993}. 
 This difference can be explained by a 7\% offset in the energy
 calibration, well within the uncertainty of the two experiments.

\subsection{Auger}
 The Pierre Auger Observatory is the largest operating cosmic ray
 observatory ever build. It is based on the hybrid concept
 where both fluorescence and surface array detection techniques
 are used and combined.  

\begin{figure}[thb]
\centerline{\includegraphics[width=6.5cm]{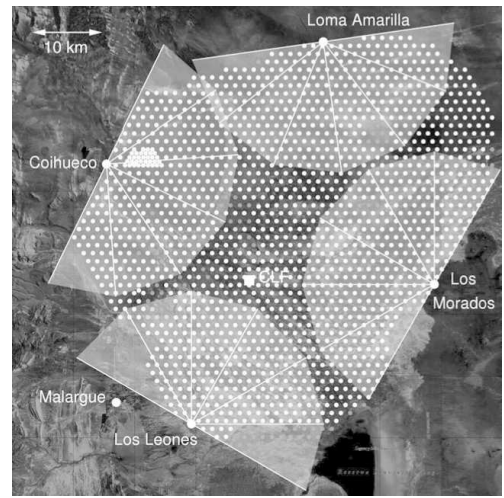}}
\caption{ The Pierre Auger observatory at the end of March 2009. Individual white dots represent Cherenkov tanks, while gray ones are un-equiped positions. A denser (infill) area is visible in the upper left. Big white dots at the periphery of the array are fluorescence detector sites with the field of view of individual telescope given by the radial white line. Also shown is the Central Laser Facility (CLF) used for FD calibration and atmospheric monitoring purpose. 
}
\label{fig:AugerDet}
\end{figure}

 The Southern site of the Auger observatory is located in the "Pampa Amarilla"
 region (35.1$^\circ$-35.5$^\circ$~S, 69.0$^\circ$-69.6$^\circ$~W and
 1300-1400~m a.s.l.) of the province of Mendoza, Argentina (\cite{EA}). 
 Construction was completed in 2008 but stable data taking started as
early as  the beginning of 2004 when Auger already had 100 detectors, covering and area  in excess of 150 km$^2$,
installed in the field . 
The arrangement of the detectors is shown in
 Fig.~\ref{fig:AugerDet}. 

\subsubsection{Surface Array}
 The surface array (SD) of Auger South is composed of 1600 water Cherenkov
 tanks, distributed on a triangular grid of 1500~m. It covers a total
 surface area of 3000~km$^2$.  Each tank is equipped with three
 photomultiplier tubes to measure the Cherenkov light, a data acquisition
 (DAQ) and front-end electronic (FE) card for control and trigger,
 a solar panel and two batteries for power, a GPS receiver for the time
 tagging, and a custom radio emitter/receiver for trigger and data
 transfer (\cite{SDpaper}).
 A central site, located on the Southwest corner of the array hosts the
 central DAQ system (CDAS), including the central trigger processors and
 the permanent data storage area.

 The SD has a 100\% duty cycle, and a well defined purely geometrical 
 aperture ($\propto$ $\cos\theta$) above trigger saturation at
 3$\times$10$^{18}$~eV. The coverage is largely uniform in right ascension.
 Modulation in the event rate due to the atmospheric conditions are
 at the level of 2\% for daily modulation and about 10\% for seasonal
 ones. Those effects have been carefully studied and can be corrected
 for (\cite{Weather}).  

 The water tanks are 1.2~m in height and are mainly sensitive to muons,
 electrons, positrons and photons. 
 A vertical GeV muon
 hitting the tank deposits an energy of about 240~MeV, to be
 compared to few tens of MeV for an average electron. The unit for
 the shower signal is a vertical equivalent muon (VEM).
 This allows
 for an in situ calibration of the PMT gain based on the rate
 of atmospheric muons. The gain is adjusted so that a single
 ADC count corresponds to about 1.5~ MeV. Local triggers are
 adjusted to a rate of about 20 Hz for a simple threshold
 trigger and a few Hz for a more sophisticated time over threshold
 (counting time bins over a certain threshold within a given time
 window) (\cite{SdAperture}).
 Local triggers are sent to CDAS where space-time coincidences of
 at least three tanks are required to trigger the  upload and permanent
 storage of the full event data. 
 
 The large sensitivity to muons and the height of the individual tanks
 allows the Auger array to have excellent sensitivity to horizontal
 showers be them from hadronic origin or from neutrinos. 

The shower signal is sampled at a rate of 40 MHz. The analysis of
 its time structure allows for example to identify the presence of an electromagnetic component
 in the ground signal, at appropriate distances from the core to identify the short high pulse from the individual muons and to count them and to calculate signal shape parameters such as the rise time. Those
 information allow to efficiently distinguish neutrinos from the hadronic background 
 in nearly horizontal showers and photons.
Additionally they also allow  to construct hadronic mass sensitive parameters.
 Timing information is obtained from GPS receiver functioning in
 position hold mode. The absolute time resolution is about 10~ns,
 combined with the sampling of the shower front and of the FE
 electronics which allows for an angular resolution of better than
 1$^\circ$ above 10$^{19}$~eV (\cite{SdAr,SdTime}). 
 The aperture of the Southern Auger Observatory is 
 energy independent when
 the surface array triggers and is determined by the area of 
 the SD and the maximum shower zenith angle (60$^o$)used in the analysis.
%

 The lateral distribution function of the Auger tank signals 
 is fitted to a NKG (\cite{Greisen60} and \cite{NK58}) function:
\begin{equation}
f(r)_{NKG}= S_{1000}\left (\frac{r}{1000\,\mbox{m}}\right)^\beta\left (\frac{700\,\mbox{m}+r}{1000\,\mbox{m}+700\,\mbox{m}}\right)^{\beta+\gamma}
\end{equation}
 where $S_{1000}$ is the adjusted normalization and where the exponent
 $\beta$ is adjusted to the data using a second order polynomial in
 $\sec{\theta}$ whose coefficient is a linear function of $S_{1000}$
 in VEM(e.g. $a = a_0+a_1\log_{10}({S_{1000}/\mbox{VEM})}$.
 The exponent $\gamma$ is very close to zero.

\subsubsection{Fluorescence detector}
 The fluorescence detector of Auger South (FD) is composed of 24 telescopes
 distributed in four sites installed at the periphery of the surface array
 and looking inward. Each telescope has a field of view of
 30$^\circ$x30$^\circ$ in elevation and azimuth. A set of six telescopes in
 each site covers 180$^\circ$ in azimuth and observes the atmosphere above
 the ground array. This geometrical arrangement ensures full detection
 efficiency for showers in excess of 10$^{19}$ eV over the entire surface
 of the array (\cite{AugerFD-2010}).

 In each telescope the optical system is composed of an entrance
 filter selecting the UV light, an aperture and corrector ring
 maintaining a large aperture while reducing  spherical and eliminating
 coma aberrations, and a 3.6~m diameter mirror illuminating a camera
 composed of  440 PMT tubes. Each tube has a field of view of
 1.5$^\circ$x1.5$^\circ$.  

 Triggering is done at the hardware level of each camera for the first
 (pixel) and second (alignment) levels. A third level trigger (TLT) is
 implemented in software mainly to reject lightning events and random
 alignments.  Each TLT is then processed at the fluorescence site level
 to merge all the telescope information and to send via CDAS, a preliminary
 shower direction and ground impact time to the surface array.
 The information from the tanks closest to the shower core is retrieved for
 showers that do not independently trigger three tanks. Together with
 additional fiducial cuts this hybrid trigger is fully efficient above
 10$^{18}$~eV. Above this energy,  the FD trigger is always accompanied
 by at least one station, independent of the mass and direction of the
 incoming primary particle (\cite{Auger_pl}).

 Event reconstruction proceeds in two steps. First the shower geometry is
 found by combining information from the shower image and timing measured
 with the FD with the trigger time of the surface detector station that has
 the largest signal (\cite{Mostafa-2007}). From this timing information  it
 is possible to break the degeneracy intrinsic to equation~\ref{eq:FDGeom}.
 Therefore the hybrid approach to shower observation enables the shower
 geometry and consequently the energy of the primary particle to be
 determined accurately. The Auger collaboration uses a fluorescence yield
 in air at 293 K and 1013 h Pa from the 337 nm band of 5.05$\pm$0.71~photons/MeV 
 of energy deposited taken from the measurements of Nagano and collaborators
 (\cite{Nagano-2004}). 
 The wavelength and pressure dependence of the yield adopted follows the
 measurements of 
 the AIRFLY collaboration (\cite{AIRFLY-2008}). Note that the fluorescence
 yield used by the 
 HiRes collaboration (\cite{Kakimoto-1996}) in the same conditions is 5.4
 photons/MeV (\cite{Arqueros-2009}) so very close
 to the value used by Auger.

 In the second step light attenuation from the shower to the telescope is
 estimated and all contributing light sources are
 disentangled (\cite{Unger-2008}). A great deal of effort is spent by the
 Auger collaboration to accurately monitor the atmospheric transparency
 and maintain the absolute calibration of the telescopes. An extensive
 set of instruments is installed and operated at the Auger site for
 this sole purpose (\cite{AugerStudy-2010}). Finally the profile of energy
 deposition of the shower is reconstructed using a Gaisser-Hillas
 functional form (\cite{AugerFD-2010}).

 The reconstruction accuracy of hybrid events is much better than what
 can be achieved using SD or FD data independently. For example, the
 angular and energy resolution of hybrid measurements at 1~EeV is better than
 0.5$^\circ$ and 6\% respectively compared with
 about $2.5^\circ$ and 20\% for the surface detector alone.

\section{COSMIC RAY ENERGY SPECTRUM}

\subsection{The end of the cosmic ray spectrum}
 In 2008 the HiRes Collaboration published a paper
 (\cite{HiRes_prl2}) with a title emphasizing 
 the experimental proof of the GZK suppression. Soon after that
 the Auger Collaboration confirmed the observation of the end
 of the cosmic ray spectrum (\cite{Auger_prl}. 
 In 1966~\cite{Greisen66} and
 independently~\cite{ZK66} predicted that the cosmic ray spectrum
 will end at several times 10$^{19}$ eV because of the interactions
 of the UHECR with the microwave background (CMB). Although
 the energy of the CMB photons is very low, the center of mass
 energy of these interactions is enough to produce pions and
 cause high energy loss for these particles that decreases their
 flux.

 The importance of these observation is that the previously
 biggest air shower array, AGASA, has observed 11 events above
 10$^{20}$ eV and no decrease above the predicted cutoff
 (\cite{Agasa_prl}). A re-evaluation of the energy assignment of
 AGASA based on a larger data set was published later in~\cite{Agasa2}. The energy determination
 of AGASA was tied up to the particle signal at 600 meters from
 the shower axis $S_0(600)$. The Monte Carlo calculations 
 suggested that the primary energy is
 $$ E \; = \; 2.17\times 10^{1.03} E_0(600)\; {\rm eV}. $$
 Since the detectors of AGASA were scintillator counters the
 signal is produced by the shower electromagnetic component 
 with a small contribution of the shower muons. The ultrahigh energy
 events published by AGASA provided the inspiration for the 
 exotic `top-down' models of these particles.

 The HiRes energy spectrum was based on monocular observations 
 with the two fluorescent telescopes HiRes-I and II that were operated
 respectively from 1997 to 2005 and from 1999 to 2004. Some 
 corrections were made on the previous HiRes spectrum
 release (\cite{HiRes_prl1}).
 A special attention was paid to the detector calibration 
 and the atmospheric conditions which were studied by standard meteorological
 methods and by the observation and analysis of laser shots from 
 different locations surrounding the two detectors.
 Most of the
 highest energy events were observed with HiRes-I. The exact 
 shape of the spectrum depends strongly on the calculation of the
 aperture for the two telescopes. The HiRes found two breaks in the 
 cosmic ray spectrum - one at energy 10$^{18.65 \pm 0.05}$ (the cosmic
 ray ankle) and another at 10$^{19.75 \pm 0.04}$ at the GZK cutoff.
 The spectral index between the two breaks is 2.81$\pm$0.03 and after
 the cutoff is 5.1$\pm$0.7.
 The spectrum is  consistent with various models and in particular
 the model of~\cite{BerGG05} with pure proton composition.
 An important parameter is $E_{1/2}$ 
 where the cosmic ray flux is one half of what it should have been
 without the GZK effect. $E_{1/2}$ was predicted by~\cite{BerGri}
 to be 10$^{19.76}$ and HiRes measured 10$^{19.73 \pm 0.07}$. 
\begin{figure}[thb]
\centerline{\includegraphics[width=8cm]{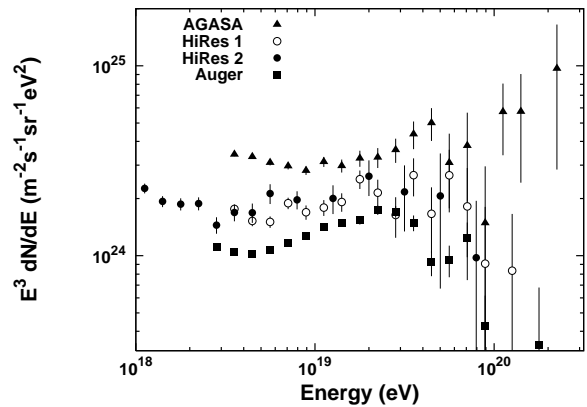}}
\caption{Cosmic ray spectrum as presented by AGASA, HiRes and Auger. 
 }
\label{spe_300}
\end{figure}
 The statistical significance of the cutoff is more than 5$\sigma$.

 The energy spectrum derived by the Auger collaboration (\cite{Auger_prl})
 is shown with
 those of HiRes and AGASA in Fig.~\ref{spe_300}. As well as HiRes,
 Auger observes the shower profile with its fluorescent detectors.
 They, however, have a live time of 13\% compared to that of the surface
 detector. Taking  full advantage of its hybrid design
 the Auger collaboration decided to 
 correlate the energy derived from the fluorescence observations
 to the shower signal at 1,000 meters from the shower core (S$_{1000}$),
 which is the least sensitive to the cosmic ray composition.
 To account for the angular dependence of this quantity it was
 corrected to the median angle of 38$^o$, S$_{1000}^{38}$ using 
 the constant intensity curve (\cite{CIC}) observed by the
 surface array. The constant intensity curve is a study of the
 change with angle of the signal threshold above which cosmic 
 rays arrive with constant rate, i.e. study of the shower
 absorption in the atmosphere. 
 
 The correlation between FD and S$_{1000}$ was
 studied in high quality hybrid events, that
 were seen in both the surface and the fluorescent detectors.
 The correlation showed that 
\begin{equation}
 E_{FD}= 1.49\pm 0.06\pm 0.12 ({\rm syst}) \times  S_{1000}^{1.08\pm 0.01 \pm 0.04 } \times 10^{17}\; {\rm eV}.
\end{equation}
 The Auger collaboration then used  the surface detector statistics to produce
 the energy spectrum. The uncertainty in the energy reconstruction by
 the fluorescent telescopes was estimated to 22\% and the width
 of the observed correlations was consistent with the statistical
 uncertainty of both measurements.
 The surface detector exposure for this publication was
 twice that of HiRes and four times higher than AGASA.
 Since the fluorescent energy measurement does not depend on the 
 hadronic interaction model used in the analysis, such an estimation of the spectrum  
 was considered to be model independent.

 The Auger spectrum has a slightly different shape in addition 
 to the energy assignment. From 4$\times$10$^{18}$ eV to 4$\times$10$^{19}$
 eV the slope of the spectrum is 2.69$\pm$0.02$\pm$0.06(syst) and above it
 is 4.2$\pm$0.4$\pm$0.06(syst). A single power law for the whole data set
 is rejected at 6$\sigma$ level.

\begin{figure}[thb]
\centerline{\includegraphics[width=8cm]{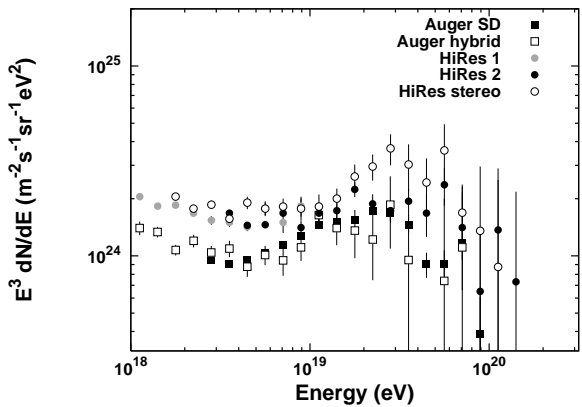}}
\caption{Cosmic ray spectrum as measured by HiRes and Auger. 
 }
\label{spe_301}
\end{figure}

 The measurements of the cosmic ray spectrum were extended by using
 stereo events in HiRes (\cite{HiRes_stereo}) and hybrid events
 in Auger (\cite{Auger_pl})). Stereo events are reconstructed
 much more precisely
 than monocular ones and they confirmed the previously measured spectrum.
 Auger used hybrid events to extend the spectrum to lower energy.
 The Auger exposure at the time of 
 this last publication was 12,790 km$^2$yr.sr. 
 All measured spectra from 
 HiRes and Auger are shown in Fig.~\ref{spe_301}.
 
 The new Auger spectrum is a bit flatter than the older one with
 an index of 2.59$\pm$0.02 between the breaks and 4.3$\pm$0.2
 above that. $E_{1/2}$ value is 10$^{19.61 \pm 0.03}$.
 The differences in the interpretation of the HiRes
 and Auger spectra are significant. 
 The Auger spectrum can be 
 explained by several different models some of which include
 mixed chemical composition at acceleration in the sources. 
 The end of the cosmic ray spectrum measured by Auger is 
 consistent with the GZK effect.
   
\subsection{Cosmic ray energy loss in propagation}

 In addition to the adiabatic energy loss because of the expansion
 of the Universe, there are two important energy loss processes for
 protons: pion photo-production interactions and $e^+e^-$ pair production (BH)
 interactions identical to the pair production interactions
 of $\gamma$--rays in the nuclear field.
 The average interaction length $\lambda_{ph}$ for interactions with
 the CMB is the inverse of the product of the interaction cross section
 $\sigma_{ph}$ and the photon density $n$. For $\sigma_{ph}$ =
 10$^{-28}$~cm$^2$ and $n$ = 400~cm$^{-3}$, $\lambda_{ph}$ = 8.3~Mpc. 

 Heavy nuclei lose energy in photo-disintegration (spallation) processes
 when the center of mass energy exceeds the giant dipole resonance.
 Since less energy is required in the center of mass, the cross section
 is higher but the energy loss depends on the mass of the nucleus that
 loses one or two nucleons. 
 The photo-production energy loss follows the
 same energy dependence as for protons but in the Lorentz factor
 space, i.e. in E/A units. The pair production cross section 
 is a quadratic function of the charge of the nucleus $Z$. 

 In the case of $\gamma$-rays the energy loss is due to the
 $\gamma \gamma \rightarrow e^+e^-$ process.
 
 A photo-production interaction is possible when the center of mass
 energy of the interaction $\sqrt{s}$ is higher than the sum of a
 proton mass $m_p$ and a pion mass $m_\pi$. In the Lab system
 the square of the center of mass energy is
\begin{equation}
 s \; = \; m_p^2 + 2 E_p \epsilon (1 - \cos{\theta})\; ,
\end{equation}
 where $\epsilon$ is the photon energy and $\theta$ is the angle between
 the proton and the photon. In a head on collision ($\cos{\theta} = -1$)
 with a photon of the average CMB energy (6.3$\times$10$^{-4}$~eV) the
 minimum proton energy is
\begin{equation}
E_p \; = \; \frac{m_\pi}{4 \epsilon}(2 m_p + m_\pi) \; \simeq \;
 {\rm 10^{\rm 20}\; eV.}
\end{equation}
 There are many CMB photons with higher energy and the threshold 
 proton energy is actually lower, about 3$\times$10$^{19}$ eV.

 The cross section for pion photo-production was very well studied at accelerators.
 The highest cross section 
 is at the mass of the $\Delta^+$ resonance (1232 MeV). 
 At the peak of the resonance the cross section 
 is about 500 $\mu$b. The cross section decreases
 to about 100 $\mu$b and then increases logarithmically. The neutron 
 interaction cross section is very similar to the proton one.

 The CMB spectrum and density are also very well known, so the proton 
 interaction length can be calculated exactly. Since protons lose only
 a fraction of their energy $(K_{inel})$, another quantity - the energy loss
 length $L_{loss} \; = \; - \frac{1}{E} \frac{dE}{dx}$ becomes important.
 The energy loss length is longer than the interaction length by 
 $1/K_{inel}$, by about a factor of 5 at threshold. At higher energy 
 $K_{inel}$ grows and this factor is about 2. 

 In the case of $e^+e^-$ pair production (\cite{BerGri})
 the addition of two electron masses to the center of mass energy $\sqrt{s}$
 requires much lower proton energy and the process has lower threshold.
 The cross section is higher than $\sigma_{ph}$,
 but the fractional energy loss is of order of $m_e/m_p$.
 The energy loss length has a minimum
 around 2$\times$10$^{19}$~eV and is always longer than 1,000 Mpc.

 The last proton energy loss process is the redshift due to the expansion
 of the Universe. The energy loss length to redshift is the
 ratio of the velocity of light to the Hubble constant ($c/H_0$) and is
 4,000 Mpc for $H_0$ = 75 km.s$^{-1}$Mpc$^{-1}$.  
     
\begin{figure}[thb]
\centerline{\includegraphics[width=8truecm]{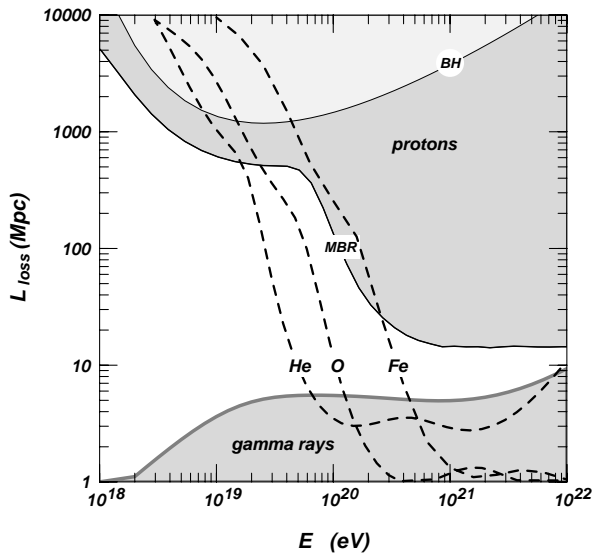}}
\caption{Energy loss length for protons, nuclei and gamma rays.
 The heavier shading points at the proton and gamma-ray L$_{loss}$
 and the light one shows the contribuion of the e$^+$e$^-$
 pair production. The adiabatic energy loss is not included.}
\label{xloss}
\end{figure}

  The energy loss length of protons in the CMB is shown in 
 Fig.~\ref{xloss}.

 The energy loss length for several nuclei is also shown in
 Fig.~\ref{xloss} as calculated by~\cite{APKGO05}. The minimum
 value of $L_{loss}$ is significantly lower than that of protons
 but is achieved at higher energy: $A\times E_p$. Since only iron
 has similar $L_{loss}$ to protons around 10$^{20}$ eV it is considered
 the only nucleus that can compete with protons in the 
 chemical composition of UHECR.
 The effect of propagation on the accelerated UHECR can not
 be calculated directly from the energy loss lengths shown in
 Fig.~\ref{xloss} because an accelerated nucleus changes its mass
 after the first photodisintegration.
 A code treating the propagation of nuclei should account
 for the energy loss of all nuclei and isotopes
 lighter than the injected nucleus. 

  The process $\gamma \gamma \rightarrow e^+e^-$ has a
 resonant character and the cross section peaks at $E_\gamma \epsilon \; = \;
 2 m_e^2$ where $\epsilon$ is the ambient photon energy.
 For CMB this corresponds to $E_\gamma$ of
 8$\times$10$^{14}$~eV and the mean free path decreases with increasing
 $E_\gamma$. For gamma rays of energy 10$^{20}$~eV the relevant seed
 photon frequency is about 1 MHz - in the radio band. 
 This creates some uncertainty in the estimates of the UHE $\gamma$-ray
 energy loss length because the density of the radio background at such
 frequencies is not well known.
 
 A different source of uncertainty in the $\gamma$-ray propagation
 is the strength of the extragalactic magnetic fields. If they are
 negligible the electrons have inverse Compton interactions,
 whose interaction length is similar to that of the pair production,
 and generate a second generation of very high energy $\gamma$-rays.
 If, however,
 the magnetic fields are significant, electrons lose energy very fast on
 synchrotron radiation and the created $\gamma$-rays are in the
 MeV-GeV energy range.
 The energy loss distance on synchrotron
 radiation is 2.6$E_{18}^{-1} B_{-9}^{-2}$~Mpc, where $E_{18}$ is the
 electron energy in units of 10$^{18}$~eV and $B_{-9}$ is the strength
 of magnetic field in $n$Gauss. 

\subsection{Formation of the cosmic ray energy spectrum in propagation}

 Predictions of the shape of the cosmic ray spectrum requires much
 more than the energy loss in propagation. The necessary astrophysical
 input includes at least the following items:\\
 \ \ \ \ $\bullet$ UHECR source distribution\\
 \ \ \ \ $\bullet$ cosmic ray source emissivity\\
 \ \ \ \ $\bullet$ cosmic ray injection (acceleration) spectrum\\
 \ \ \ \ $\bullet$ maximum acceleration energy E$_{max}$\\
 \ \ \ \ $\bullet$ cosmic ray chemical composition\\
 \ \ \ \ $\bullet$ cosmic ray source cosmological evolution\\
 that are not independent of each other. As an example we will
 discuss the formation of the proton spectrum in propagation.
    
\begin{figure}[thb]
\centerline{\includegraphics[width=9truecm]{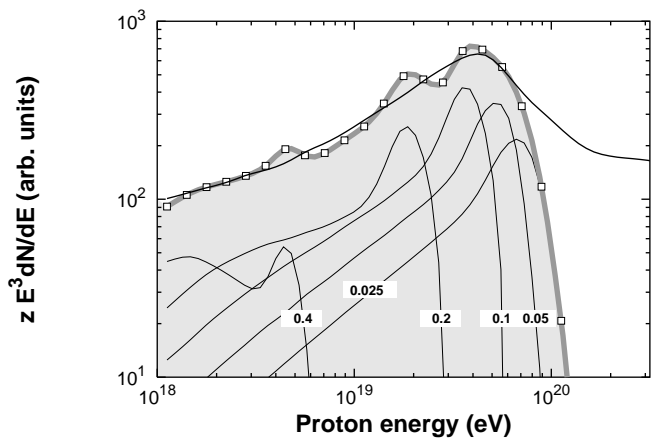}}
\caption{Contribution of different 
 redshifts to the arrival spectrum for $E^{-2}$ injection spectrum
 with no cosmological evolution. The thick gray line shows the sum
 of the contributions from these five redshifts while the black
 line is result of a full integration.
 }
\label{prop1} 
\end{figure}
 
 Figure~\ref{prop1} shows the contribution 
 to the observed UHE cosmic ray proton flux by sources located at
 different redshifts that inject protons on a E$^{-2}$ spectrum
 with an exponential cutoff at 10$^{21.5}$ eV. One can see how the
 energy loss increases the contribution to the 2-6$\times$10$^{19}$ eV
 energy range after propagation to $z$=0.1. In models with
 cosmological evolution of the sources the effect is stronger and
 proportional to the strength of the source evolution.

 A simplification in such calculation is the assumption that
 sources are  isotropically and homogeneously distributed in
 the Universe
 and the contribution of all sources are identical.
 In such a case the cosmic ray flux at Earth can be determined
 by an integration of the fluxes from different redshifts
 shown in Fig.~\ref{prop1}.
 In the case of cosmological evolution of the sources the integral is
\begin{equation}
 N(E) \; = \; \int^{z_{max}}_0 \int_E^{E_0} L(z) N_0(E_0) P(E_0, E', z) 
 \frac{dt}{dz} dE' dz \; ,
\end{equation}
 where $L(z)$ is the cosmic ray source emissivity as a function of redshift
 and $N_0(E_0)$ reflects the injection spectrum. $P(E_0,E',z)$ is the
 probability for a proton injected with energy $E_0$ at redshift $z$ to
 reach us with energy $E'$.
 The derivative $dt/dz$ depends on the cosmological model
 and is
 $$ {{dt} \over{dz}} \; = \; {{1} \over {H_o (1 + z)}}
 [ \Omega_M (1+z)^3 + \Omega_\Lambda ] ^{-1/2}$$
 and is simplified to $(1+z)^{-5/2}/(H_o(1 + z))$ for the
 Einstein-deSitter Universe.
 
 It is important to note that the contribution of different redshifts
 depends not only on the cosmological evolution but also on the 
 injection spectral index as the photoproduction energy loss is
 a strong function of the injection energy. Since in steep 
 injection spectra a larger fraction of the observed flux 
 comes from lower primary energy (that do not change as much
 on propagation) the contribution of higher redshifts is larger.
 
 One can see in Fig.~\ref{prop1} that even $z$=0.05 does
 contribute to UHECR above 6$\times$10$^{19}$ eV
 where the GZK cutoff is already present. Another explanation  is
 that the cutoff is just the end of the acceleration
 power of the sources that does not much exceed
 10$^{20}$ eV (\cite{Watson-2007,Aloisio-2009}).
 The extragalactic magnetic fields can also be involved in the
 explanation. If they are high UHECR would scatter
 often and their real pathlength would be considerably larger 
 than the distance to the sources as in~\cite{Stanevetal00}.
\section{CHEMICAL COMPOSITION OF UHECR}
\label{sec:chem}

 One has to use the properties of the extensive air showers
 to identify the type of the primary particle whose interaction
 in the atmosphere has initiated the shower. Because of the
 high level of fluctuations in the shower development it is
quite difficult to distinguish showers originating from different hadronic primaries
 on  an event by event basis - it can only be done
 on a statistically significant set of showers. At lower energy,
 around the knee, the main parameter in composition studies is
 the ratio of the shower muon to electron components which 
 increases with the primary nucleus mass.

 Another way is to study the shower longitudinal development.
 This is usually done by observing the depth of shower maximum
 $X_{max}$ with fluorescent detectors that can determine the
 atmospheric depth where the shower particles emit the highest
 amount of fluorescent light. The shower longitudinal development,
 as we will see in this section, can also be studied by the 
 surface detectors. 

\subsection{Limits on the flux of neutrinos}

 Possible shower neutrino primaries may be the easiest to
 identify (\cite{Capelle-1998}).
 The reason is the many orders of magnitude difference between
 the hadronic and neutrino cross sections. If neutrinos interact 
 in the atmosphere at all, they would interact very deep.
 It is more likely that they interact in the rock of the Earth.
 This was used by the Auger collaboration in order to set a limit 
 on the flux of $\tau$ neutrinos. Setting such a limit is equivalent
 to a limit on the total neutrino flux. Although $\nu_\tau$'s are rarely
 produced in particle interactions, cosmic neutrinos oscillate in
 propagation to Earth. While at production the neutrino flavor ratio
 ($\nu_e$:$\nu_\mu$:$\nu_\tau$) is close to 1:2:0 after propagation
 it is close to 1:1:1.

 The tau neutrino detection idea (\cite{Bertou-2002})
 is that in a small fraction of the solid angle, at
 zenith angles $\theta$ between 90.1$^o$ and 95.9$^o$ tau neutrinos
 will graze the Earth, possibly interact and after escaping the earth the tau decay will 
 generate a shower that
 can be seen by the shower array. The neutrino identification is based
 on the different quality of vertical and almost horizontal showers.
 Vertical showers are young: they exhibit long, $\sim\mu$s waveforms
 in the surface detectors. Old showers, after penetrating about two
 atmospheric depths, consist mostly of muons. The waveforms they
 generate in the surface detectors are much shorter, of order of
 100 $n$s. If one detects an almost horizontal shower that has the
 waveforms of young showers it would mean that the primary particle
 has interacted near to the detector and is most likely a neutrino.

 Atmospheric interaction of tau neutrinos are also 
 especially interesting because they should
 develop two showers (\cite{double-bang}), one when the $\tau$ neutrino
 interacts and the second one when the $\tau$ lepton decays.
 Since the $\tau$ energy loss is much lower than that of a muon, 
 most of the neutrino energy is released through the $\tau$ decay.
 The exact parameters for $\nu_\tau$ shower identification are
 found with Monte Carlo calculations.

 The Auger collaboration is using two related parameters. The first
 one is the shower shape in the surface array. Because of the large
 zenith angle it obviously should
 be elongated. The collaboration defined shower {\em length} and
 shower {\em width}. The Monte Carlo calculations showed that there
 is no chance for a nucleus or a $\gamma$-ray to generate a shower
 with length/width ratio higher than 5.

 The second parameter is the shower {\em ground speed}. If the shower
 is indeed horizontal it has to move with velocity equal to
 the speed of light. So they looked for showers with velocity
 between 0.29 m/$n$s and 0.31 m/$n$s. Only showers with RMS (ground speed)
 better than 0.08 m/$n$s are included in the sample. These requirements,
 together with the general requirement that the tank with the maximum VEM
 signal is surrounded by six active tanks significantly decrease 
 the size of the sample. Anyway, no such showers were found in the
 Auger statistics. The Auger exposure as a function of energy was
 determined by Monte Carlo calculations. 

 Using the statistics between January 2004 and August 2007 the Auger
 collaboration
 set an integral limit on the $\nu_\tau$ flux between 2$\times$10$^{17}$ eV
 and 2$\times$10$^{19}$ eV of $E_\tau^2 dN/dE_\tau$ of 1.3$\times$10$^{-7}$ GeV.
 This limit assumes that the $\nu_\tau$ energy spectrum is E$^{-2}$.
 In the same publication (\cite{Aug-tau1}) the collaboration uses the
 exposure as a function of the neutrino energy to also give a
 differential limit. Extending the statistics
 to April 2008 the Auger collaboration decreased the integral limit
 of $E_\tau^2 dN/dE_\tau$ of  to about 6$\times$10$^{-8}$
 GeV.cm$^{-2}$s$^{-1}$sr$^{-1}$ (\cite{Aug-tau2})
 for the same flat $\nu_\tau$ spectrum. This limit is shown
 with a gray line in Fig.~\ref{up_down}. The integrated Auger limit
 is competitive with the limits set by the neutrino telescope
 AMANDA (\cite{AMANDA}) at lower energy.

 The HiRes Collaboration has also set limits on the fluxes of tau
 and electron neutrinos (\cite{HiRes-Nu1,HiRes-Nu2}).
 The first limit is based on the same assumptions as the Auger one.
 The HiRes collaboration has simulated $\tau$-neutrino induced showers
 hitting the Earth with elevations between 10$^o$ and -10$^o$ with
 an account of the topography of the detector.
 After detection simulation they obtained 6,699 monocular triggers
 and 870 stereo ones. Then the collaboration analyzed with some quality 
 cuts simulated and real data events events with reconstructed zenith
 angles between 88.8$^o$ and 95.5$^o$. The data sample yielded
 a total of 134 events that happened to be laser events, which 
 passed the cuts because of the light scattering near the ground.
 Thus they were left with no neutrino candidates.

 For the limit on electron neutrino events only the HiRes~2 detector
 data were used because of its superior reconstruction. HiRes 
 looked at upward going showers with zenith angles above 105$^o$.
 Lower zenith angles do not yield more events because of the
 neutrino absorption in the Earth. The basis of the search is
 the fact that high energy electrons have much lower energy
 loss at high energy and especially in dense materials 
 because of the LPM effect (\cite{LP,M}). After electron neutrino
 charge current interactions in the ground the generated electrons
 would not lose much energy and may produce upward going 
 air showers. The simulations were done for neutrinos of energy
 exceeding 10$^{18}$ eV that generate horizontal and upward going
 air showers with more than 10$^7$ particles at maximum. These showers were
 then treated with the HiRes detector simulations and compared 
 to experimental data. No neutrino candidates were observed.
 The combination of the two searches reduced the integral neutrino flux
 limits in the three decades above 10$^{18}$ eV to 3.8$\times$10$^{-7}$,
 9.7$\times$10$^{-6}$, and 4.7$\times$10$^{-6}$ GeV cm$^{-2}$sr$^{-1}$s$^{-1}$.

\subsection{ Limits on the fraction of gamma-rays}

 The limit on the fluxes of ultrahigh energy neutrinos is
 astrophysically important because it is related to the
 dynamics of the systems where UHE cosmic rays are accelerated
 and the importance of hadronic interactions in such objects.
 The limit on the fraction of photons in UHECR  determines the
 general origin of the highest energy
 particles in the Universe. All top-down models of the UHECR
 origin predict that 90\% of these particles are $\gamma$-rays and
 neutrinos. There have been previous limits from several UHECR
 air shower arrays for energies above 10$^{19}$ eV but none of them 
 was less than 10\%.

 Auger has the advantage of being a hybrid array. A fluorescent 
 detector trigger with a single surface detector trigger improves
 significantly the shower reconstruction and lowers the detection
 threshold. The Auger collaboration set limits on the fraction
 of $\gamma$-rays at energies above 2, 3, 5, and 10$\times$10$^{18}$ eV
 (2, 3, 5, and 10 EeV) (\cite{Aug-phot0,Aug-phot2}). The limit is based on 
 the measurement of the shower depth of maximum $X_{max}$.
 Because of the low secondary multiplicity in electromagnetic interactions
 UHE $\gamma$-ray induced showers reach maximum much later than
 proton showers. In addition at energies above 10$^{19}$ eV the LPM
 (Landau-Pomerntchuk-Migdal) effect, which significantly decreases
 the pair-production cross section, starts becoming important and 
 increases $X_{max}$ even more. The key in such a measurement is
 to make certain that the event selection is not biased versus showers
 with deep $X_{max}$. 

 There are many cuts that are applied to the detected showers
 that exclude the possible biases. Auger requires that the tank 
 with the largest signal is less than 1.5 km from the reconstructed
 shower axis and the time difference between the fluorescent and
 tank signals is small. Another requirement is that the shower $X_{max}$
 is observed in the telescope field of view. The minimum angle between
 a fluorescent pixel and the shower direction has to be above 10$^o$ 
 to exclude Cherenkov light contamination. The shower zenith angle has
 to be higher than 35$^o$ and the distance of the shower core to the 
 telescope less than 24 km.

\begin{figure}[thb]
\centerline{\includegraphics[width=8cm]{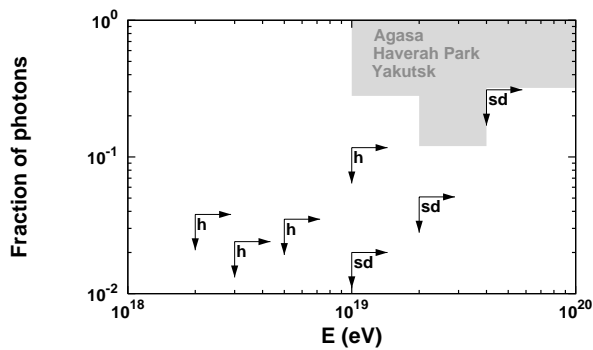}}
\caption{Fraction of photon showers above certain energy determined by
 the hybrid analysis (labeled {\em h}) and the surface detector ({\em sd}).
 The shaded area shows limits set by previous experiments. 
 }
\label{Aug-phot}
\end{figure}

 All these different cuts, together with the requirement for a good
 reconstruction in the fluorescent detector, decrease the total
 statistics above 2 EeV to 2063 events. Eight of these events have
 $X_{max}$ consistent with possible photon showers. Using the 95\%
 confidence value for the number of photon candidate events and 
 the systematic uncertainty of 22\% in the energy estimate and 11\%
 in $X_{max}$ the Auger collaboration arrives at an 95\% upper limit
 on the fraction of photon showers above 2 EeV of 3.8\%. At higher 
 energy bins the number of photon candidate showers is 1, 0, and 0, but
 the statistics are lower, 1021, 436, and 131 events respectively.
 This leads to 95\% proton fraction estimates of 2.4\%, 3.5\%, and
 11.7\% as shown in Fig.~\ref{Aug-phot}.

 At higher energy the photon fractions are calculated using the surface
 detector results (\cite{Risse-2007}). This analysis (\cite{Aug-phot1})
 is very interesting
 because it uses surface detector shower characteristics rather than
 the direct $X_{max}$ measurement. The idea of such analysis was
 first developed for the inclined showers detected by the Haverah Park
 array (\cite{HP-phot}). When the air shower particles hit the surface
 array detectors
 the shower front is not flat, it has a certain curvature, i.e., the shower
 particles away from the shower axis arrive later than those close
 to it. If the shower curvature is assumed to be spherical (which is an
 oversimplification) the delay is proportional to $r^2/H$ where $H$ is
 the altitude of the particle production and $r$ is the distance from
 the shower axis. This means that in early developing showers the shower
 curvature is smaller than in later developing ones. The radius of shower
 curvature $R_c$ is the first parameter that can be measured by the 
 surface array. 

 The second parameter is the width of the shower front, i.e. the time
 that it takes the shower particles to arrive at the surface array.
 The spread of the arrival time at certain distance $r$ from the core
 also increases with the depth of $X_{max}$. This could be measured 
 by the surface array as the arrival time at a fixed distance from the
 shower core $\tau_{1/2}$, which is defined by Auger as the time in which
 10\% to 50\% of the signal arrives at a detector. Since photon showers 
 develop deeper in the atmosphere than nuclear showers, one can identify
 them by their large $R_c$ and $\tau_{1/2}$. For photon showers one 
 can use Monte Carlo calculations that have the advantage to depend very
 little on the hadronic interaction models used in nuclear shower 
 calculations. The only remaining problem is the energy assignment of
 the photon showers - the fluorescent detector value is used for
 hadronic showers. Auger developed an estimate that gave them 25\% accuracy
 for photon showers.

 The Monte Carlo calculations of photon showers generated values of these two 
 parameters for all zenith angles that were very different from those
 of the detected air showers from 2004 to the end of 2006. The data set
 used includes 2761, 1329, and 372 showers above 10, 20, and 40 EeV.
 There were no $\gamma$-ray candidates in either bin, while 570, 145, and
 21 showers are certainly nuclear showers. Since zero events per bin
 corresponds to less than 3 events at 95\% confidence level, the fraction
 of photon showers was calculated to 2.0\%, 5.1\% and 31\% in the three
 bins.  

 These results demonstrate that the top-down models are not responsible 
 for the production of the majority of UHECR. There is still a little 
 space remaining at the highest energies but in 5 years of operation
 Auger South will be able to bring down these limits if the current
 trend continues.
 
 It is indeed true that the differences between different hadronic 
 interaction models do not play much of a role in the photon showers
 Monte Carlo and these analyses are mostly independent of the 
 details of those interactions.
  
\subsection{Depth of maximum data and their interpretation}
  The application of Heitler's toy shower model to the 
 shower longitudinal development demonstrates its dependence
 on the mass of the primary particle. With the advent of
 fluorescence detectors the measurement of the shower depth
 of maximum, $X_{max}$, quickly became a major component of
 the cosmic rays composition studies. An important parameter
 is the shower elongation rate $D_{10}$, the relation of which
 to the changes in the cosmic ray composition is discussed first
 in~\cite{LW1981}.


  The first analysis of the $X_{max}$ energy dependence with
 fluorescent detector data was done with the Fly's Eye.
 This analysis (\cite{Gaisseretal93}),
 which used only two chemical components - H and Fe, showed a
 trend of increasing the proton fraction in cosmic rays of
 energy above 10$^{18}$ eV.  

  In 2005 the HiRes Collaboration published an analysis of the
 UHECR composition from $X_{max}$ measurements~ (\cite{HiResApJ}). 
 The data sample
 included 553 events of energy above 10$^{18}$ detected in stereo
 by both fluorescent detectors
 during 20 months from 1999 to 2001. The sample is relatively small
 because of the different cuts made on the total event sample.
 The first set of cuts are related to the atmospheric conditions.
 About 3/4 of the total sample had hourly data on the vertical
 aerosol optical depth obtained with laser shots from the location
 of both detectors. 
 The cuts on the remaining events 
 used the average atmospheric conditions and the records made during
 the measurement. 
 The rest of the cuts address the reconstruction
 quality. They include a minimum viewing angle of more than 20$^o$ 
 in both detectors, more than 5$^o$ difference in the shower detector
 plane (to decrease the effect of scattered Cherenkov light),
 $\chi^2$ of the global fit of less than 15 p.d.f and bracketing 
 of $X_{max}$ within the observed tracks. The application of the
 same cuts to a Monte Carlo data set gives a $X_{max}$ resolution of
 30 g/cm$^2$ and energy resolution of 13\%. 

  This analysis was presented in a combination with an earlier 
 result (\cite{HiResMIA}) obtained by the HiRes prototype 
 working in coincidence with the MIA air shower array.
 The elongation rate between 10$^{17}$ eV and 10$^{18.5}$ eV
 was measured to be 93$\pm$8.5 g/cm$^2$ with a systematic 
 uncertainty of 10.5 g/cm$^2$. This result suggested a quick
 transition from heavy to light cosmic ray composition.  
  
\begin{figure}[thb]
\centerline{\includegraphics[width=8cm]{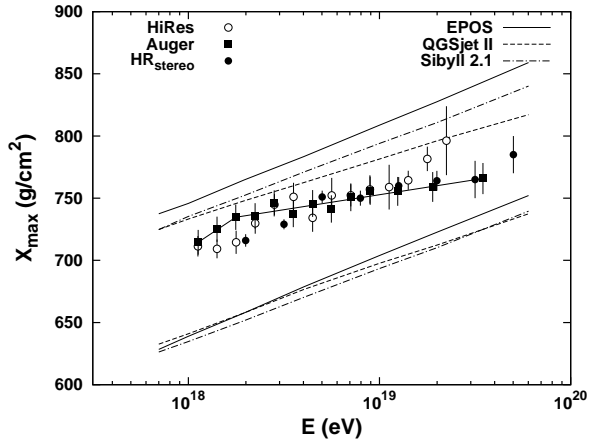}}
\caption{ Depth of maximum measurements of UHECR by the HiRes collaboration
 (2005) analysis shown with empty circles and 2010 analysis with full
 circles and the Auger collaboration - full squares are compared with
 the predictions of three different interaction models for H and Fe.    
 }
\label{compos_A}
\end{figure}

  The HiRes 2005 paper measured $D_{10}$ = 54.5$\pm$6.5 g/cm$^2$
 consistent with the values for constant composition from
 different hadronic interaction models. The data points above
 10$^{18}$ eV agreed within the errors with the results from
 the HiRes-MIA coincidence experiment. The HiRes points,
 which are derived from fitting the shower profile with
 the Gaisser-Hillas formula, are 
 shown with empty circles in Fig.~\ref{compos_A}. The lines
 in the same figure show the expectations from three different
 hadronic interaction models: EPOS~1.99, QGSjet~II, and SIBYLL~2.1.
 The fraction of protons is different for these interaction models.
 QGSjet~II shows almost pure proton composition with a possible
 small He contamination. In the case of SIBYLL~2.1 the fraction 
 of protons is smaller but still significant.
  QGSjet~II has a moderate cross section energy dependence and
  fast multiplicity increase. SIBYLL~2.1 has a fast cross section
  growth and a relatively low multiplicity. 
  The comparison with EPOS~1.99, which has the largest $D_{10}$,
 may even point at large, but decreasing with energy, fraction of protons.

  The interpretation of the $X_{max}$ data will become better after the LHC
  results are accounted for in the hadronic interaction models used
  in the analysis. The current models do not disagree with each other
  in the energy range studied in accelerators. After the normalization
  of these models to the LHC data the differences in the interpretation
  of the experimental results will significantly decrease. An extensive
  description of the current hadronic interaction models used for
  air shower analysis can be found in~\cite{Engel-2011}.

  HiRes also studied the width of the $X_{max}$ distributions in
 different energy bins. Proton showers do have a wider distribution
 while iron showers have widths lower by at least a factor of two.
 The comparisons with simulations using QGSjet~01 were consistent
 with a proton fraction of 80\% and those with SIBYLL~2.1 suggested
 a proton fraction of about 60\%. EPOS~1.99 did not exist at that
 time. The general conclusion of the HiRes $X_{max}$ study is that
 the cosmic ray composition was heavy at 10$^{17}$ eV, progressed
 to light one in one order of magnitude in energy and stayed light
 with a proton fraction from 60 to 80\% above 10$^{18}$ eV.

  Although the Auger Collaboration has presented its $X_{max}$
 measurements at different conferences the first journal paper
 on this topic appeared in 2010 (\cite{Auger_comp_prl}). 
 The results of this study are unfortunately very different from
 those of HiRes. This analysis is done using only hybrid events,
 i.e. events detected by one or more fluorescent telescopes plus
 at least one surface detector. Using the timing of the surface
 detector vastly improves the quality of the reconstruction. 
 The light collected by the fluorescent detector is corrected
 for attenuation using the atmospheric monitoring devices. 
 The fitting is done using the Gaisser-Hillas function. 
 Events with light emission angle less than 20$^o$ are not used.
 Neither are events where $X_{max}$ uncertainty due to shower
 geometry and atmospheric conditions is more than 40 g/cm$^2$.
 The limit on the reconstruction $\chi^2$ is set to less than 
 2.5 p.d.f. The resulting $X_{max}$ resolution above several
 EeV is about 20 g/cm$^2$. This number is consistent with the
 checks with stereo fluorescent detector observations.

  Using data taken between 2004 and March 2009 there are 3754 events
 above 10$^{18}$ passing all cuts. The highest energy event is of
 energy 59$\pm$8$\times$10$^{19}$ eV. The measured $X_{max}$ in 10
 logarithmic bins per decade are shown in Fig.~\ref{compos_A} with 
 full squares. The elongation rate of the three points below
 10$^{18.25}$ eV is 106$^{+35}_{-21}$ g/cm$^2$ and that above this point
 is 24$\pm$3 g/cm$^2$. Both these values are determined with good
 $\chi^2$ fits. Systematic uncertainty is around 10 g/cm$^2$.
 In absolute value this data set does not appear extremely different
 from the HiRes 2005 analysis (\cite{HiResApJ}) but its interpretation
 is. Instead of a constant elongation rate of 54.5 g/cm$^2$ we have 
 a large one, maybe similar to that of HiRes-MIA, in the lower energy part and
 a short one at higher energy. The cosmic ray composition thus 
 have to become lighter up to 10$^{18.25}$ eV and then consistently 
 heavier up to the highest energy measured. The Auger Collaboration is 
 careful enough to state that such interpretation is reasonable if
 there are no drastic changes in the hadronic interactions in this
 energy interval. 

\begin{figure}[thb]
\centerline{\includegraphics[width=8cm]{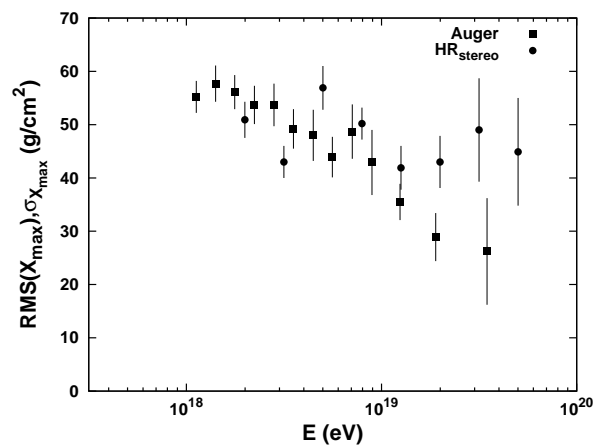}}
\caption{ Width of the $X_{max}$ distributions as measured by
 HiRes (full circles) and Auger (full squares). Note that the width
 is presented in different ways (see text) and the points cannot be 
 compared directly.    
 }
\label{RMSs}
\end{figure}

  The Auger collaboration also studied the width of the $X_{max}$
 distributions (RMS) in the same energy bins. While the RMS's
 in the first five bins look consistent with a light composition
 at higher energies there is a steep decrease of RMS($X_{max}$) 
 consistent with heavier and heavier composition
 as shown in Fig.~\ref{RMSs}. The RMS values
 decrease from about 55 g/cm$^2$ to 26 g/cm$^2$ in the last bin.
 This distribution width is consistent with cosmic ray composition
 dominated by iron. The interaction model predictions for proton showers
 give 60 g/cm$^2$ with a slight energy dependence and these for iron
 showers give about 22 g/cm$^2$. The linear decrease of RMS($X_{max}$)
 is not, however, consistent with a simple change of the cosmic ray 
 composition from pure proton to pure iron.

  It is worth noting that the Auger Collaboration has also attempted
 to use the surface detector data for studies similar to those
 of $X_{max}$ (\cite{Auger_SD_comp}).  Since these results are 
 preliminary we will only give the
 general idea. The width of the shower front depends on the 
 depth of the shower maximum. One can study the shower front width
 by measuring the rise-time of the surface detector signals. The
 attempt to do that is fully consistent with the more detailed
 fluorescent detector analysis.

  The final analysis of the stereo measurements of HiRes in the period
 1999-2006 was published in~\cite{HiRes_comp_2}. The cuts on the data
 are more stringent
 than in the previous analysis. Apart from the good weather requirement,
 they limit the chance of noise coincidence to less than 1\% and the
 longitudinal development fit $\chi^2$ to less than 4 p.d.f. The final
 data set of 815 events includes only events with zenith angle uncertainty
 of less than 2$^o$, $X_{max}$ uncertainty of less than 40 g/cm$^2$, 
 zenith angle less than 70$^o$ and distance to HiRes~II more than 10 km.
 The measured $X_{max}$ should be bracketed by the HiRes~II field of 
 view and have shower detector plane between 40$^o$ and 130$^o$. 
 The application of the vertical aerosol optical depth hourly measurements
 to the amount of light received by the detectors requires a mean
 upward correction of $\sim$15\% to shower energy for an event 25 km
 distant from the observatory. Shower segments with emission angles
 of less than 5$^o$ of a bin pointing direction are not used in the
 analysis.

  The measured light profile of the shower is fit to a Gaussian function
 of the age parameter $s$ to determine the shower energy and $X_{max}$.
 The claim is that the use of the Gaisser-Hillas function does not 
 change the results within the errors. Showers of energy between
 1.6$\times$10$^{18}$ eV and 6.3$\times$10$^{19}$ eV are included
 in the analysis.

  All uncertainties in the $X_{max}$ measurement come from the
 treatment of simulated showers after the detector is accounted for.
 Comparisons of the reconstructed $X_{max}$ with the original one
 showed that the selection and reconstruction results in $X_{max}$ 
 shallower by about 15 g/cm$^2$ than the original one. For this reason
 for the interpretation of the measurements the predictions are
 appropriatly scaled.  
 The Monte Carlo measured uncertainty of $X_{max}$ is better than 25 g/cm$^2$
 over most of the energy range. 
    
  This analysis finds a constant elongation rate of 47.9$\pm$6 g/cm$^2$
 with fit $\chi^2$ of 0.86 p.d.f 
 over the whole range with systematic uncertainty of 3.2 g/cm$^2$.
 Most of the systematic uncertainty is due to the event selection cuts.

  HiRes also presents the energy dependence of the $X_{max}$ fluctuations
 in the same energy bins. These numbers are obtained in a different
 way from those of the Auger Collaboration. Since these $X_{max}$
 distributions are wide and asymmetric, the HiRes analysis fits them
 to Gaussian distributions truncated at 2$\times$RMS. The distributions
 are still wide as shown in Fig.~\ref{RMSs}. 
 
  The heavy cosmic ray composition derived from the Auger data
 suggests that the strong decline of the cosmic ray flux may be
 caused by exceeding the maximum acceleration energy at the cosmic
 ray sources. In such a case only iron nuclei could be accelerated
 to energies exceeding 10$^{20}$ eV.

\subsection{Transition from galactic to extragalactic cosmic rays}
 
\begin{figure}[thb]
\centerline{\includegraphics[width=8cm]{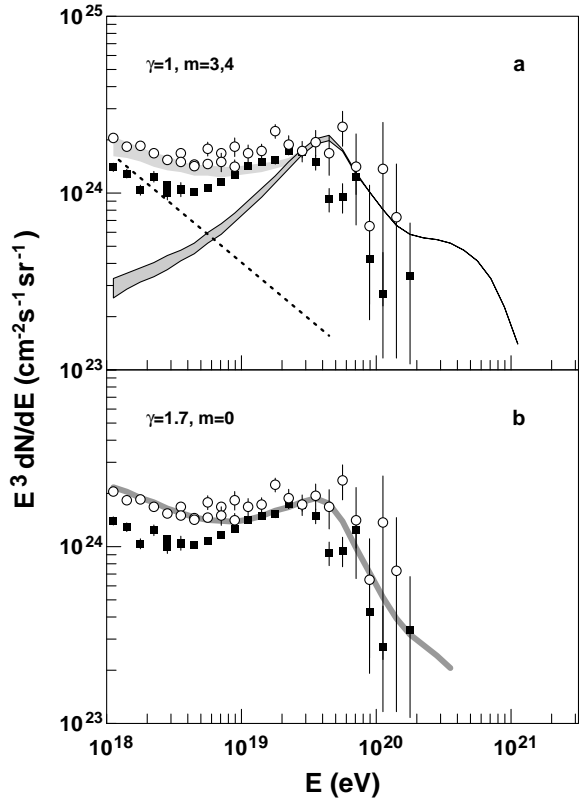}}
\caption{The models of \cite{BahWax03} (a) and \cite{BerGG05}
(b) compared to the more recent data of Auger (full squares) and HiRes
( circles).}
\label{e_spec}
\end{figure}

 One of the reasons for identifying different features at the end of
 the cosmic ray spectrum is to study the transition between the galactic
 and extragalactic components. The common opinion is that most of the
 cosmic rays above 10$^{19}$ eV are of extragalactic origin and the 
 GZK feature supports that. The main question was (and is) the
 origin of the {\em dip} at around 3$\times$10$^{18}$ eV. The prevailing
 school of thought was that the dip is at the intersection of the
 galactic and extragalactic components as explained in \cite{Hillas-1984} and ~\cite{BahWax03}.
 In this model the extragalactic cosmic rays have a flat E$^{-2}$
 spectrum and the galactic ones have a steep E$^{-3.5}$ spectrum as 
 shown in the upper panel of Fig.~\ref{e_spec} with two values (3\&4)
 of the parameter $m$ in the evolution described as $(1 + z)^m$ up to
 redshift of about 2. As is seen from the figure the cosmological 
 evolution does not affect much the predicted spectra because the observed UHECR
 have to be local. The galactic cosmic rays, although with
 a small contribution, extend well above 10$^{19}$ eV as shown with
 a dashed line in the figure. It became soon obvious that extragalactic
 cosmic rays cannot have such a flat injection spectrum subsequent models
 deal with E$^{-2.3}$ or slightly steeper spectra.

  Soon after that~\cite {BerGG05} suggested a totally different model.
 The dip is caused by the pair production interaction of the extragalactic
 protons with CMB as predicted by~\cite{BerGri}. The model underwent
 some development later~(\cite{Aloietal06}) to convey some of its
 details. The shocking part of this model, illustrated in the lower panel
 of Fig.~\ref{e_spec} is that the cosmic ray injection spectrum was
 a steep E$^{-2.7}$ rather than the expected flat one.
 At about 10$^{18}$ eV the injection spectrum has to become much flatter
 for the flux not to exceed the measured cosmic ray spectrum.
 The transition from galactic to extragalactic cosmic rays should
 then happen below 10$^{18}$ eV.
 There is no need for
 cosmological evolution of the cosmic ray sources in the model.
 Extragalactic cosmic rays had to be almost exclusively protons.

  At about the same time a third model for the transition became
 available~(\cite{APO07}) following the calculations of the 
 heavy nuclei propagation in extragalactic space and their
 interactions with the CMB and other photon fields
 (\cite{APKGO05};\cite{HST07}). Since the extragalactic cosmic rays
 in such a model may have at least five chemical components,
 i.e. many more parameters, the model
 is much more complicated, although
 it could be made to fit the cosmic ray spectra as well as the first
 two. These propagation calculations also showed
 that protons and iron nuclei have approximately
 equal energy loss lengths, while all intermediate nuclei would
 disintegrate at much shorter distances.

  Since the spectrum shape alone cannot answer the
 questions about the transition, the answer could only come from
 accompanying composition
 study. The chemical composition derived by the HiRes experiment
 together with the measurement of ~\cite{HiResMIA} would claim that
 the galactic cosmic ray spectrum does not extend above 10$^{18}$ eV
 and higher energies contain only protons with a small admixture of
 light nuclei. This admixture may be different in the 2005 and 2010
 analyses but it seems to be constant and belong to the same population.
 Such interpretation may not be consistent with a proton cosmic rays
 knee at 3$\times$10$^{15}$ eV or lower as derived by the Kascade
 experiment since in such a case the iron knee would start at 10$^{17}$
 eV and would leave about one order of magnitude of energy unexplained
 (\cite{Hillas-2005}).

 A similar simple interpretation of the Auger $X_{max}$ and RMS
 result is impossible. The elongation rate derived from the first
 three points seems to show a quick transition from heavy to 
 light nuclei followed by a slower transition to heavy nuclear
 composition.
 
 We first have to understand if at least 
 a part of the $X_{max}$ behavior is not due to a sudden  change
 of hadronic interactions at $\sqrt{s}$ close to 50 TeV, well above the
 LHC maximum energy. Then we have to relate the change of
 the injection composition to the shape of the energy spectrum
 that is in this range quite different from that of HiRes.
 These are not simple problems to solve and in our opinion
 they will take years. The main hope is that Auger and HiRes
 would examine each others $X_{max}$ analysis techniques and
 will come up with similar, if not identical, $X_{max}$ values
 as a function of the energy. The lower energy extensions of
 the UHECR arrays described in Section~\ref{sec:fut} and the
 use of different composition related  
 parameters (muon to electron density ratios) may also be
 of big help.  

\section{SEARCH FOR THE SOURCES OF UHECR}

 Before describing the searches for the sources of UHECR we will
 briefly introduce some of the ideas about the strengths of 
 the magnetic fields in the Universe. These are important parameters
 because particle scattering in the magnetic fields can hide
 the sources. A nice review of the investigations and results of the
 studies of astrophysical magnetic fields can be found in \cite{Beck01}.  

\subsection{Galactic magnetic fields}

 Galactic magnetic fields are very important for the scattering
 of UHECR because they definitely have a large scale structure.
 This means that cosmic rays coming from the same direction will
 scatter in a similar way and the scattering will shift the 
 arrival direction away from the true source. 

 The regular magnetic field strength is measured mostly by studies of
 the Faraday rotation of the radio emission of pulsars. The rotation
 measure $RM$, measured in rad/m$^2$ is proportional to 
 $\int_0^d n_e B_\parallel dl$ where $d$ is the distance to the source and
 $n_e$ is the electron density. The integral over the measured or
 assumed $n_e$ is used to extract the magnetic field strength as
 described in a recent review of the galactic magnetic field
 (\cite{Hanetal06}). In principle the magnetic field strength
 is proportional to the matter density in the Galaxy and is
 decreasing with the galactocentric distance. The field decrease 
 in the galactic plane is best described with an exponential
 function $e^{-R_{GC}/8.5}$
 where the distance from the galactic center $R_{GC}$ is in kpc.
 The local regular field in the vicinity of the Solar system
 has a strength of about 2 $\mu$Gauss and points counterclockwise
 close to the direction of the Carina-Sagittarius arm.

 This expression is valid for $R_{GC}$ bigger than 3 kpc
 because the field inside that circle is difficult to study and is
 not well known. In regions near the galactic center {\em m}Gauss fields
 have been observed pointing almost perpendicular to the galactic
 plane. This led to suggestions that there is a strong magnetic
 dipole in the galactic center. At the solar system the dipole 
 field strength is about 0.3 $\mu$Gauss and points North in 
 galactic coordinates. 

 The more general estimates of the total field strength at our
 location give values of 6$\mu$Gauss which results in random 
 field strength twice as big as the regular field. There are
 also ideas that the random field reaches maximum inside the
 galactic arms (because of the stellar fields pointing in
 different directions) and the regular field reaches maximum
 in the inter-arm space. The random field is not very important
 for UHECR scattering because its scale size is only 50-100 pc.

 A very important question which is far from solved is the galactic
 magnetic halo, i.e. the extension of the magnetic field above
 and below the galactic plane. More recent measurements tend
 to show an extended halo that can contribute a lot
 to the cosmic ray scattering angle (\cite{Jiang2010}).
 A standard way to study
 UHECR scattering angle is to inject negatively charged nuclei
 in a magnetic field model and follow their trajectories until 
 they leave the Galaxy (\cite{Stanev97}).
 Such exercises with different {\em toy}
 galactic field models give scattering angles at 100~EeV between 2$^o$ and
 4$^o$ depending on the cosmic ray direction. Some other estimates,
 however, give much higher values, up to 10$^o$ ({\em R.~Beck, private
 communication}).  

\subsection{Extragalactic magnetic fields}

 Our knowledge of the extragalactic magnetic fields is much smaller
 and still on the basic level of the excellent review of~\cite{Kronberg94}.
 Although $\mu$Gauss magnetic fields have been observed in clusters
 of galaxies, such objects enclose a small fraction of the
 Universe (10$^{-6}$ or less) and the upper limit of the average
 magnetic fields is 10$^{-9}$ Gauss~=~1~$n$Gauss if the correlation
 length of the field $L_c$ is 1 Mpc, the average distance between
 galaxies. But even such
 small fields can affect the propagation of UHECR.

 The angular deflection due to random walk $\theta$ would then be
 \begin{equation}
 \theta \, = \, {\rm 2.5}^o \, E_{20}^{-1} \, B_{-9}
 \, d_{100} \, L_{C}^{1/2} \;,
 \end{equation}
 where $E_{20}$ is the energy in units of 10$^{20}$ eV, $B_{-9}$
 is the magnetic field strength in $n$Gauss,
 $d_{100}$ is the source distance in units
 of 100 Mpc and $L_C$ is the correlation length in Mpc. The random
 walk causes a propagation path length $\Delta d$ that is larger than
 the distance to the source and causes increased energy loss.
 It depends on the square of the parameters above and is
 \begin{equation}
 \Delta d = 0.047 \, E_{20}^{-2} \, B_{-9}^2 \,
 d_{100}^2 \, L_C \; {\rm Mpc.}
\end{equation}
 The increased propagation distance causes a corresponding time delay.
 In case UHECR are generated in a GRB, or in an active state of an
 AGN we may not be able to correlate these events with the resulting
 cosmic rays.

 The situation changes drastically if the UHECR encounters an extended 
 region with organized magnetic field. In principle this should be a
 rare occasion except close to a powerful astrophysical system where
 such fields have been observed. Depending on the field strength,
 its direction toward us, and structure of the field, the angular
 deflection could be  much larger.
 
\subsection{Correlation of the arrival directions of UHECR  with astrophysical objects.}

  The first attempt to correlate the arrival direction of UHECR with
 known astrophysical objects was published by~\cite{Stanevetal95}.
 The authors used 143 events of energy more than 2$\times$10$^{19}$ eV
 detected by the Haverah Park array, together with the statistics of the
 Vulcano Ranch, Yakutsk and the preliminary data of AGASA.
 The authors studied the angular distance between the UHECR events and
 the super-galactic plane (SGP), which is the plane of weight of almost all
 extragalactic objects within redshifts below 0.04 (\cite{DeVau}). 
 The conclusion was that at energy above 4$\times$10$^{19}$ eV the average
 and RMS distances of UHECR to SGP are much closer than would be expected
 from an isotropic distribution of the UHECR sources.

  With the increase of the AGASA statistics that started to dominate
 in the late 1990s the correlation with the SGP decreased. Other effects 
 were claimed by that experiment: a large scale isotropy and small scale
 anisotropy. The anisotropy was defined by the fact that three pairs
 and a triple of events coming within 2.5$^o$ of each other were
 found (\cite{Takeda99}) among 47 events of energy above 4$\times$10$^{19}$ eV.
 The chance probability of this happening from isotropic distribution
 was less than 1 \%. Soon after this clustering analysis was extended
 to include the previously detected events (\cite{Uchihori00}).
 The conclusions from that analysis were slightly different. Since the
 angular resolution of the older experiments was worse the clustering 
 was analyzed in angular distances of 3, 4, and 5$^o$. Twelve doubles
 and 2 triples were found within 3$^o$. The emphasis, though, was on
 the fact that 8 of the doubles and the 2 triples lie within 
 10$^o$ of the super-galactic plane, which had a chance probability of 
 0.1 and 0.2 \% for an isotropic distribution of the sources.
   
\subsubsection{Correlation of the Auger events with AGN}

\begin{figure*}[thb]
\begin{center}
\includegraphics[width=0.75\textwidth]{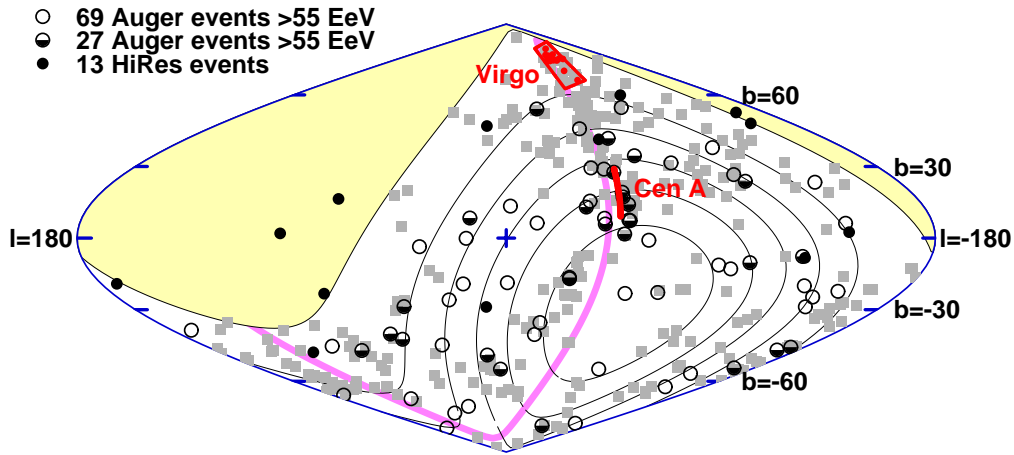}
\end{center}
\caption{Correlation of the arrival directions of UHECR with AGN from
 the VCV catalog. The shaded part of the sky is not visible by Auger.
 The gray squares are the AGN within $z$ less than 0.018. The Auger 
 events are shown with circles. The first 27 events are half 
 filled  . The 13 HiRes events are shown with black dots. The thin lines show
 the six regions of the sky to which Auger has equal exposure.
 The wide gray line is the supergalactic plane.  
 }
\label{Auger69}
\end{figure*}

  After collecting and exposure of 4,390 km$^2$ sr yr the Auger 
 collaboration noticed that many of their higher energy events are
 close to the active galactic nuclei from the VCV (\cite{VCV}) catalog.
 They did a scanning analysis varying the angular distance,
 event energy and the source distances. The best correlation appeared 
 for angular distance of 3.1$^o$, event energy above 5.7$\times$10$^{19}$ eV
 (57 EeV) and distance of 75 Mpc (redshift less than
 0.018). The same analysis was repeated when the exposure more than
 doubled and the total number of high energy events reached 27.
 Twenty of these events were within 3.1$^o$ of the AGN from the
 VCV catalog when only 7.4 were expected for isotropic sources.
 The chance probability for this happening was 
 1.7$\times$10$^{-3}$ (\cite{Auger_corr1,Auger_corr2}).
 A significant number of
 events were close to the nearby (distance of 3.8 Mpc) radio galaxy
 Cen A. There were no events close to the powerful AGN M87.
 When the events with galactic latitude less than $\pm$12$^o$, where
 the catalog coverage is smaller and UHECR scattering in the galactic
 magnetic field is supposed to be stronger, were excluded the strength
 of the correlation increased and 19 out of 21 events
 correlated with at least one AGN as shown in Fig.~\ref{Auger69}.
 This strong correlation was surprising because of several reasons.
 First of all the VCV optical catalog includes many low power objects
 that are not likely to accelerate particles to such high energy.
 Secondly, the 0.018 redshift does not correspond to the GZK 
 horizon (the distance up to which cosmic ray sources contribute significantly to the flux {\em observed} above a certain energy) for energy of 57 EeV and the question arose if the Auger 
 energy scale was low by about 25\% which would bring the two distances
 close to each other. 

  The Auger collaboration did not claim that the AGN from the VCV
 catalog are the actual sources, which may have a sky distribution 
 similar to that of the correlating AGN. The anisotropy of the UHECR
 sources was emphasized in the papers.

  The analysis was repeated by the HiRes experiment (\cite{HiRes_corr}) 
 as close to the original as possible. There were only two out of
 13 events with similar energy that correlated with the same AGNs and the
 conclusion was the opposite. The HiRes field of view is not the same
 as that of Auger and the VCV catalog has different coverage of the 
 corresponding fields of view. Still the results from the two
 analyses appeared to be controversial since HiRes sees one half 
 of the Auger field of view.

  Recently the Auger collaboration presented the correlation results 
 from an exposure of 20,370 km$^2$.sr.yr (\cite{Auger_corr3})
 which contains 69 events of 
 energy above 55 EeV (corresponding in the contemporary energy assignment
 to 57 EeV in 2007). The complete catalog of the 69 Auger events published to date are shown in Fig.~\ref{Auger69}.
 The correlation is now weaker - 42\% of all
 events (29/69) correlate, compared with 74\% in 2007.
 The event reconstruction is constantly improving and because
 of that a couple of events move from one group (above 55 EeV or not) 
 to the other.
 If the events participating in the initial parameter scan are excluded
 the corresponding fractions of correlated events are  69\% and 38\% respectively. 
 This does not mean, though, that the observed UHECR are isotropically 
 distributed as the expected fraction of correlated events  is 21\%.
  
 It should be noted that the scattering in the extragalactic 
 magnetic fields should be much stronger according to some
 calculations. \cite{Ryu-2010} predict an average scattering 
 angle of 15$^o$ for cosmic ray protons above 60 EeV.

\subsubsection{Correlation with sources from other catalogs}

  The most difficult part of the search for UHECR sources is that
 we have no idea what is the best proxy for cosmic ray acceleration
 to the highest energy - is it the optical/UV luminosity, or the
 X-ray one, or still higher energy $\gamma$-ray emissivity?
 The last paper on the Auger events anisotropy explores the
 correlations with two more catalogs: the 2MRS catalog (\cite{2MRS}),
 which contains the brightest galaxies from the 2 MASS catalog,
 and the Palermo Swift-BAT hard X-ray catalog (\cite{Swift-BAT}).
 2MRS contains 13,000 galaxies within 100 Mpc and 22,000 galaxies
 within  200 Mpc. The Swift-BAT catalog has the advantage to cover
 well the region of the galactic plane. It contains 133 
 extragalactic sources within 100 Mpc and 267 within 200 Mpc.
 The correlation of the Auger UHECR arrival directions with
 the positions of the objects in both catalogs is much better than
 an isotropic source distribution would suggest. 

  To fully understand the correlations with catalogs containing
 different number of objects and to estimate the statistical 
 significance of these correlations the Auger collaboration 
 used a different approach. The catalogs were used to create maps
 of possible sources using the object densities per unit area of sky
 where each object position was extended by several degrees. 
 These extensions are supposed to account for the particle scattering
 in magnetic fields and the angular sensitivity of the experiment.
 The UHECR luminosity of the sources were scaled with the distance
 and with the observed source luminosity at different wavelengths.
 With the use of simulations the events were then separated in
 {\em source} and {\em isotropic} fractions with different
 confidence levels. The isotropic fraction became on the average
 0.64 for the 2MRS catalog and 0.62 for Swift-BAT with huge
 error bars even at 1$\sigma$ level. In a way this analysis 
 produced similar results to the contemporary correlation with
 the VCV catalog. 

  The last test of isotropy was made with studies of self 
 correlation - a comparison of the number of event pairs
 as a function between the angular distance of the pair compared
 to that of isotropic source distribution. The number of 
 experimental pair events is consistently above the expectations.
 The biggest deviation is at angular distance of 11$^o$, where 
 the experimental events show 51 pairs while 34.8 pairs are
 expected for an isotropic distribution. At angular distances
 higher than 45$^o$ the number of pairs is consistent with
 isotropy but below 30$^o$ it is not.  
 
\subsubsection{Events coming from specific objects}
        
  Ever since the publication of the first analysis of the 
 correlation of the Auger events with extragalactic objects
 the question was why there are so many events coming from directions
 close to Cen A and there are none from the Virgo cluster
 and M87. With the increased contemporary statistics Auger was
 able to analyze better this fact. There are still no events
 coming from less than 18$^o$ from M87. And there are now 13
 events coming from less than 18$^o$ from Cen A and two events
 very close to it. Cen A is close to
 the direction of the Centaurus cluster but is not a part of it. 

  M87 is almost 5 times more distant than Cen A,
 which is at a distance of 3.8 Mpc. It is also in a region
 where the Auger exposure is three times less as shown in
 Fig.~\ref{Auger69}. Using these two rough numbers
 one expects 75 times less events from M87 than from
 Cen A. In other words one expects 13/75 events coming from
 M87 if it has the same CR luminosity as Cen A. The lack of events
 then is not a problem.

\begin{figure}[thb]
\begin{center}
\includegraphics[width=7cm]{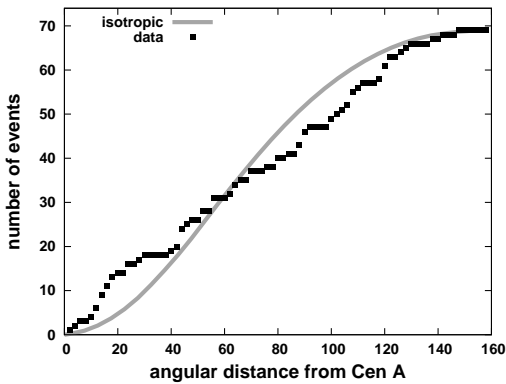}
\end{center}
\caption{ Number events as a function of the angular
 distance to Cen A. The thick gray line shows the expectation from
 isotropic distribution of the cosmic ray sources. 
 }
\label{cenA}
\end{figure}

  The 13 events coming from directions close to Cen A are mostly
 responsible for the excess of self correlation discussed above.
 The events coming from this direction have 28 pairs coming 
 with separation less than 11$^o$. For an isotropic distribution
 one expects 3.2 events rather than 13, while the map based
 on the 2MRS catalog predicts 9.2 and that based on Swift-BAT
 catalog predicts 20.6.    

  Figure~\ref{cenA} shows the comparison of the number of events
 coming at different distances from this object compared to the
 expectations from an isotropic distribution. A Kolmogorov-Smirnov
 test of this distributions establishes 96\% significance or
 about 2$\sigma$ deviation from an isotropic distribution. 
 The question then is if at least a part of these events come 
 from the Centaurus cluster rather than from Cen A. This does not
 appear likely because the Centaurus cluster is
 further away than Virgo and one would expect a small fraction
 of events coming from there for equal CR luminosities, which is 
 of course not guaranteed. 
 
\section{ULTRAHIGH ENERGY NEUTRINOS}
\label{sec:neut}

 The relationship between UHECR and ultrahigh energy neutrinos 
 was first noted by~\cite{BerZat69}. Later on, an important  relation between the observed UHECR flux and the 
 flux of diffuse neutrino was derived~\cite{WaxBah98}. The authors  took a simple basic
 approach to the problem and using the measured cosmic ray flux 
 at 10$^{19}$ eV. Then they assumed a flat, $\gamma$=2 injection
 spectrum and calculated the emissivity of UHECR in the Universe
 which came to 10$^{44}$ erg/(Mpc${^3}$.yr) in the range 10$^{19}$ -
 10$^{21}$ eV. The next observation is
 that a fraction of these cosmic rays would have photo-production
 interaction at their sources and a fraction of their energy loss
 $\epsilon$ would go to neutrinos. The upper bound of the ultrahigh
 energy muon neutrinos and antineutrinos would be if UHECR lose
 all of their energy to neutrino production. Using the average 
 energy loss they arrived at an upper bound of
 \begin{equation}
 E_\nu^2 dN_\nu/dE_\nu \; = \; {\rm 1.5} \times {\rm 10}^{-8}
 {\rm GeV} ({\rm cm^2. s. sr})^{-1}
 \end{equation}
 which after accounting for the cosmological evolution of the sources as
 $(1 + z)^3$ is increased by a factor of 3.
 The limit was criticized by~\cite{MPR} who derived
 a more realistic limit that only touched the limit of Waxman\&Bahcall
 at 10$^{18}$ eV. Both limits are shown with thick gray lines in 
 Fig.~\ref{WB_ESS}. 

\subsection{Cosmogenic neutrinos}

\begin{figure}[thb]
\centerline{\includegraphics[width=8cm]{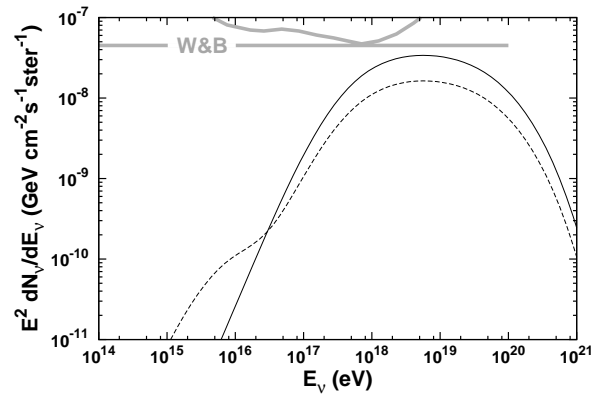}}
\caption{ The upper limits on the ultrahigh energy neutrino fluxes
 derived by~\cite{WaxBah98} (labelled) and~\cite{MPR} are compared to
 a calculation of cosmogenic neutrinos produced by UHE protons
 by~\cite{ESS}. Electron neutrinos and antineutrino fluxes are plotted with
 a dashed line. One can see the contribution of neutron decay
 at lower energy.   
 }
\label{WB_ESS}
\end{figure}

 Cosmogenic neutrinos were first suggested by~\cite{BerZat69}.
 These are neutrinos that are produced by UHECR in photo-production
 interactions in the CMB and other photon fields in propagation.
 The original paper did not produce much interest since the
 contemporary experiments could not detect high energy neutrinos.
 The shapes of the cosmogenic neutrino spectra are very different 
 from those of the Waxman\&Bahcall limit. Muon neutrino and 
 antineutrino spectra peak at about 10$^{18}$ eV and significantly
 decline at both lower and higher energy. These spectra are shown
 in Fig~\ref{WB_ESS} together with the two limits
 assuming the same astrophysical input. Even after
 the multiplication by E$_\nu^2$ the electron neutrino and antineutrino
 spectra show an extension to lower energy which is due to $\bar{\nu}_e$
 from neutron decay. 

 The magnitude of the cosmogenic neutrino spectra depends on the cosmic
 ray injection spectrum and composition, on the distribution of
 UHECR sources, and very strongly on the cosmological evolution of
 these sources. For flatter injection spectra  more UHECR can
 undergo photo-production interactions and hence generate more neutrinos.
 Cosmological evolution importance has a simple explanation --
 UHECR can arrive to us only
 from very low redshifts (less than 0.05) while neutrinos can travel without
 energy loss (except adiabatic) from the whole Universe. If the
 cosmological evolution of the UHECR sources peaks at $z$=2, as it does
 in many models, the cosmogenic neutrino production would peak close
 to $z$=3, when the source emission is much stronger. 

\begin{figure}[thb]
\centerline{\includegraphics[width=8cm]{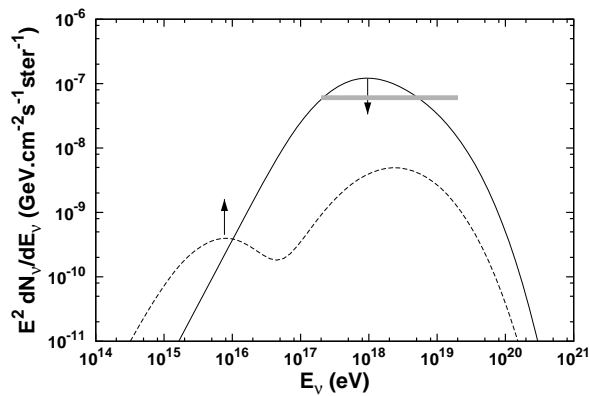}}
\caption{ Neutrino fluxes calculated from the Auger energy
 spectrum interpretation that produces the larger amount of
 cosmogenic neutrinos. The solid line shows the sum of $\nu_\mu$ +
 $\bar{\nu}_\mu$ + $\nu_e$. The dash-dot line shows the flux of $\bar{\nu}_e$.
 Any heavier nuclei contribution to the UHECR flux can only raise
 the dash-dot line and decrease the solid one. The thick gray line
 shows the median (halfway between optimistic and pessimistic) limit
 set by Auger on $\nu_\tau$.    
 }
\label{up_down}
\end{figure}

 The influence of the cosmic ray composition on the cosmogenic
 neutrino flux is even stronger, although
 more difficult to evaluate. Fig.~\ref{up_down} shows the fluxes of
 cosmogenic neutrinos calculated for UHE protons. The solid line 
 shows the sum of muon neutrinos and antineutrinos and electron neutrinos
 and the dash dot line shows the flux of electron antineutrinos 
 from neutron decay.
 The input parameters come from the Auger energy spectrum fit that
 produces the larger amount of cosmogenic neutrinos (protons, $\gamma$=1.3,
 cosmological evolution $(1+z)^5$).  
 If  UHECR are not all protons, the solid line should come down
 keeping all other parameters stable. For 20\% protons in UHECR
 the flux would be lower by a factor of 5. At the same time the flux
 of cosmogenic $\bar{\nu}_e$ would rise since heavy nuclei
 photo-disintegrate and emit neutrons that decay. The estimate of
 the increase of the $\bar{\nu}_e$ flux is more complicated 
 but it would increase roughly also by a factor of 5. 
 The cosmogenic neutrino flux would then be dominated at production
 by this  neutrino flavor. After propagation it would be shared 
 equally by all three neutrino flavors.

 The conclusion is that a measurement of the cosmogenic neutrino flux
 is a complimentary measurement to that of the UHECR spectrum and
 composition. Even the detection of a few events will considerably
 help the analysis of the UHECR features and origin. The problem here 
 is that the current limits on the UHE neutrino fluxes are generally
 above the predictions shown in Fig.~\ref{up_down}. The measurement of the 
 cosmogenic $\bar{\nu}_e$ spectra and their oscillations is even more 
 difficult because of the energy dependence of the neutrino-nucleon
 cross section.

\section{REMAINING PROBLEMS AND EXPECTATIONS FROM FUTURE EXPERIMENTS}
\label{sec:fut}

\subsection{Remaining problems}

 It is obvious from the controversial results on the chemical
 composition of UHECR that this is the main unsolved problem.
 The uncertainty of the chemical composition also affects the
 interpretation of the end of the cosmic ray energy spectrum.
 If the UHECR composition is dominated by protons the most likely
 explanation is the GZK effect. If, however, the composition is 
 increasingly heavier to the highest energies it could be a
 result of reaching the maximum acceleration rigidity at the
 UHECR sources.

 At the highest energies (above a few tens of EeV) two important
 observables need to be measured with better precision, the composition
 and the anisotropies. Both will tell us about the sources and their
 distribution as well as about the mechanisms at play in accelerating
 particles up to 100 EeV or more. 

 Currently the available data from Auger and HiRes appear contradictory
 and no model is able to explain in a coherent way all the observations. 
 Moreover, in the Auger data Centaurus A is today the sole possible 
 source candidate that may have been seen in the sky. Can this possibility
 be confirmed, is CenA the only source visible from the Southern hemisphere?
 From both hemispheres?
 
 At the highest energies the effort needs to be pursued along at
 least three lines: covering the whole sky, increasing the statistics
 by instrumenting larger surfaces or volumes, and improving the measurements
 adding new detector components.  To make definite progress, the
 next generation of detectors  should be able to measure independently,
 and if possible redundantly, all EAS components. This includes in particular, 
 electromagnetic shower profile with a maximum of a few tens of g/cm$^2$
 resolution, as well as the muonic and electromagnetic components at
 ground to better constrain hadronic model and the first interaction 
 dynamics. 

 At the EeV scale, the expected transition from galactic to extra-galactic
 origin in the cosmic ray spectrum has not been confirmed. Several features in
 the energy spectrum need attention. Is there a second knee around 0.1 EeV ?
 or at almost 1~EeV as measured by the Akeno array~\cite{Nagano-1992} ?
 How pronounced is the ankle ? What is its origin? Today
 the interpretations in terms of a pure proton composition undergoing
 $e^+e^-$ pair creations, or in terms of the galactic to extra galactic
 transition of a mixed composition seem equally (in)valid.
 What is the level of anisotropy in this energy range, can the above
 models accommodate or predict the  low values already
 reported (\cite{Bonino-2009}).
 Again the only hope for light can come from more accurate measurements
 of this regions, both  in terms of statistics and in terms of multi-parametric observations. 

\subsection{Extensions of Auger South}
  
 The High Elevation Auger Telescope (HEAT, \cite{Kleifges-2009}) and
 the Auger Muon and Infill for the Ground Array (AMIGA, \cite{Platino-2009})
 have been added to the original design of the Pierre
 Auger Observatory. Improving the efficiency of the observatory in
 the 0.1 to 1 EeV range, these extensions will efficiently
 test the various models for the acceleration and transport of galactic
 and extragalactic  cosmic rays in the transition region.
 
 Studies of this region require not only a better collection efficiency
 to improve the statistics but also powerful mass discrimination
 capabilities.
 While very high energy showers can be efficiently
 measured by fluorescent telescopes
 from distances up to several tens of kilometers, 
 lower energy ones do not emit sufficient light to bee seen further
 than a few kilometers away.  For the same position of $X_{max}$
 closer showers appear  at higher elevation than distant ones.
 Since low energy showers reach their maximum of development
 faster, coverage above the  30$^\circ$ limit of the original Auger
 telescopes is required.  HEAT is composed of three fluorescence
 telescopes of the same basic design as the original Auger telescope
 and is  installed  at the western fluorescence detector site
 (Coihueco) of the observatory. They can operate in two positions.
 Horizontally they share the same field of view as the original telescopes.
 This position is used for laser and drum calibration of the instruments
 as well as for inter-calibration using shower data. Tilted upward by
 29$^\circ$, this is the normal operation mode in which the nearby
 upper part of the atmosphere is observed. Construction took place in
 2008-2009 and first light was seen from one of those telescopes in
 January 2009.

\begin{figure}[thb]
\centerline{\includegraphics[width=0.40\textwidth]{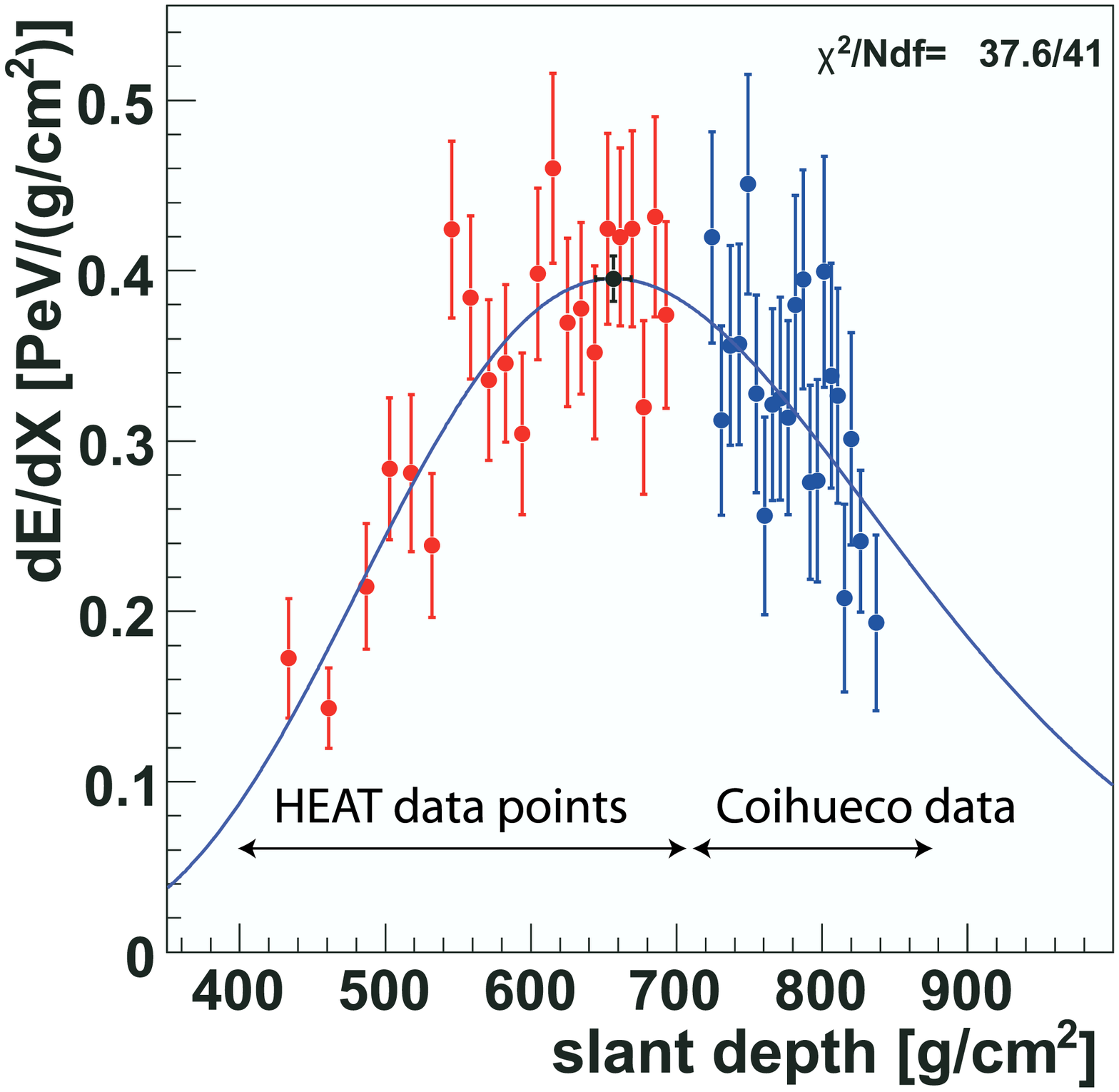}}
\vspace*{-2cm}
\caption{Longitudinal profile of a 0.2$\pm$0.02~EeV shower recoded in coincidence by the HEAT and Coihueco fluorescence telescope of the Auger observatory. }
\label{fig:heat}
\end{figure}

 Routine observation with the HEAT telescope began in 2010.
 Figure~\ref{fig:heat} shows the longitudinal shower profile
 of an event recorded in coincidence with the Coihueco telescope.
 The reconstruction of this event gives a shower energy of
 0.2$\pm$0.02~EeV and a distance of 2.83$\pm$0.06~km from Coihueco.
 It is clear from the plot that the data points provided by the
 HEAT telescope are mandatory to properly reconstruct the
 shower development profile.

 The Auger observatory reconstruction is based on the hybrid technique.
 To provide the HEAT telescope with adequate information from the surface
 array it was necessary to also increase the surface detector density at
 the foot of the telescope. An infill array of 85 detectors is deployed
 on two grids of one half (750~m) and one fourth (433~m) of
 the regular Auger surface array grid.  
 Measuring the muon densities on the ground together with the
 electromagnetic component provides important information on the
 cosmic ray composition in
 addition to the longitudinal shower development.
 Such a multi parametric measurement allows to study
 independently the evolution of $X_{max}$ and of the muon
 densities which are linked in a similar way in all interaction models.   
 The AMIGA extension aims to provide such information
 by measuring the shower muons with buried muon counters.
 Each counter is made of a segmented plastic scintillator
 read out by wave shifting fibers connected to a 64 channels
 multi anode PMT.  

 The muon lateral distribution function is adjusted to the counter
 data to provide the number of muons at 600 m from the shower axis. 
 Realistic Monte Carlo analysis together with an improved
 reconstruction showed a relative precision on the
 estimated muon density better than 20\% in the energy range of
 0.4 to 3 ~EeV accessible to the 750~m infill alone (\cite{Supanitsky-2008}).

 At the time of writing nearly all of the 750~m infill grid is
 completed and is operating while a  muon counter has been buried
 and successfully tested. Completion of this effort is expected to
 take place in 2011-2012. 

 Finally, co-located with the infill array, the Auger collaboration
 is currently  installing the first phase of the Auger Engineering
 Radio Array (AERA, \cite{VanDenBerg-2009}).
 The base line parameters for AERA comprise about 150 radio detection
 stations distributed over an area of  20 km$^2$. The main scientific
 goals of the project are the thorough investigation of the radio
 emission from an air shower at the highest energies, the exploration
 of the capability of the radio-detection technique, and the provision of
 additional observables (calorimetric energy and  shower profile determination with 100\% duty cycle)
 for the composition measurements between
 10$^{17.4}$ and 10$^{18.7}$~eV 

 In order to increase the amount of data on the shower longitudinal
 development, today severely limited by the 10\% duty cycle of
 the fluorescence detector, the Auger collaboration is pursuing
 several R\&D programs aiming at measuring the shower longitudinal
 development using microwave radio techniques (\cite{AMBER, MIDAS, EASIER}).  

\subsection{Telescope Array}
  The Telescope Array (TA) is a new hybrid detector that started collecting
 data in 2009 in Utah, USA, at 39$^o$N, 120$^o$W and altitude of
 1500 m a.s.l. Its surface array (SD) currently consists of 607 
 scintillator counters on a square grid with dimension of 1.2 km.
 Each scintillator detector consists of two layers of thickness
 1.2 cm and area of 3 m$^2$. The phototube of each layer is connected
 to the scintillator via 96 wavelength shifting fibers which make the
 response of the scintillator more uniform. Each such station is
 powered by a solar panel that charges a lead-acid battery.
 The total area of the surface array is 762 km$^2$. The surface array is
 divided in three parts that communicate with three control towers 
 where the waveforms are digitized and triggers are produced.
 Each second the tower collects the recorded signals from all
 stations and a trigger is produced when three adjacent stations
 coincide within 8 $\mu$sec.The SD reaches a full efficiency 
 at 10$^{18.7}$ eV for showers with zenith angle less than
 45$^o$ (\cite{Nonaka-2009}). This angle corresponds to SD 
 acceptance of 1,600 km$^2$sr.
\begin{figure}[thb]
\centerline{\includegraphics[width=4.5cm]{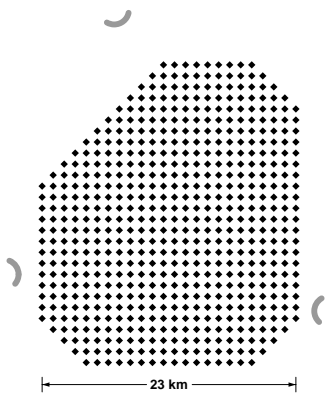}}
\caption{Sketch of the Telescope array geometry. The surface detectors are
 indicated with full diamonds and the telescope stations with arcs.    
 }
\label{ta}
\end{figure}

  The fluorescence detector (FD) consists of three fluorescence stations
 as shown in Fig.~\ref{ta}. Two of them are new and consists of
 12 telescopes with field of view from elevations of 3$^o$ to 31$^o$.
 The total horizontal field of view of each station is 108$^o$.
 Each telescope has a camera consisting of 256 PMT with field of
 view 1$^o \times$1$^o$. The signals are digitized by 40 MHZ sample 
 FADC and the waveforms are recorded when signals are found
 in 5 adjacent PMTs.
 The third station has 14 telescopes that use
 cameras and electronics from HiRes-I and
 mirrors from HiRes-II. The fluorescent telescopes are calibrated 
 with N$_2$ lasers, Xe flashers, and an electron linear
 accelerator (\cite{Tokuno-2009}).

 The atmosphere is monitored for clouds by IR cameras and with the
 use of the central laser facility which is in the center of the
 array at 20.85 km from each station. The fluorescent stations 
 are positioned in such a way that they cover the whole area of
 the surface detector. The mono acceptance of the FD is 1,830 km$^2$sr
 and the stereo one is 1040 km$^2$sr. The total energy resolution is 25\%
 and the $X_{max}$ resolution is 17 g/cm$^2$.

 TALE is the lower energy extension of the Telescope Array which will be 
 deployed to study cosmic rays of energy 10$^{16.5}$ to 10$^{18}$ eV.
 It consists of an infill array and a fourth fluorescent station
 inside TA. The field of view of this station will be elevations 
 of 33$^o$ to 71$^o$ so it will be able to see $X_{max}$ of the lower 
 energy showers. The infill will consist of 100 scintillator counters 
 on a square grid of 400 m and muon counters. See its description at
 {\em http://wow.telescope array.org}.
  
\subsection{Auger North}
  


 Based on the same detection principle as the southern observatory the
 design of the northern site of Auger, or Auger North for short, is
 focussed on collecting significantly larger statistics
 (\cite{Bluemer-AN-2010}).
 Its target energy lies above the 60 EeV threshold where anisotropy in 
 the distribution of cosmic ray sources within the GZK sphere has been
 detected. Motivation for such a detector are plenty, they principally
 concern the determination of the cosmic ray composition up to at least
 100 EeV.
 This would lead, on one hand, to the identification of the trans-GZK
 cosmic ray sources and their acceleration mechanisms and, on the other hand,
 to the study of particle interactions at center of mass energies far
 beyond any man made accelerators. 

 Covering 20,000 km$^2$  this new facility is to  be deployed in the state
 of Colorado (USA) in the northern hemisphere to provide the Auger
 collaboration with full sky coverage. Composed of a particle array
 of 4000 Cherenkov tanks principally on a $\sqrt{2}$ miles grid and 39
 fluorescence telescope overlooking the atmosphere over the whole array.
 This configuration should reach 90\% efficiency above 30 EeV for proton
 primaries.

 The expected performances of this detector are similar to its southern
 counterpart but at a higher energy due to the larger spacing.
 For example the angular resolution is expected to be better than
 2.2$^\circ$ above 50 EeV.  The statistics above 60~EeV is expected
 to be of the order of 150 events per year. Out of those, of
 order of 10 per year should have an appropriate profile
 reconstruction from the FD telescopes for mass composition measurements.

 Construction of this facility was planned for 2011 and should last
 5 years. However as of the end of 2010, the funding situation and
 prioritization in the USA does not allow for such construction to
 start in the short term. 

\subsection{EUSO - JEM-EUSO}
  The Extreme Universe Space Observatory (EUSO) is an UV telescope
 mounted on an external facility of the International Space Station (ISS)
 which observes the atmosphere to detect light signals from 
 UHE cosmic rays and neutrinos. It is a monocular 
 telescope that measures the air shower fluorescence light
 and the Cherenkov light diffusively reflected from the surface of the Earth.
 The initial idea of such experiment was
 proposed and developed by the pioneer of the UHECR
 research John~\cite{JL98}. He was later joined by  Livio Scarsi and
 a group of scientists from the University of Palermo.
 Initially EUSO was approved by the European Space
 Agency for a Phase A conceptual study (\cite{Catalano}).
 It was not approved to be mounted on the European research module 
 of ISS and was taken over by the Japanese Space Agency. It is
 now known as JEM-EUSO.

  EUSO is a wide angle ($\pm$30$^\circ$) camera with a diameter of
 2.5 m. The UV light is imaged through Fresnel lens optics and
 detected by a segmented focal surface detector using multi-anod PMT.
 The aim is to have 1 km$^2$ resolution on the surface of the Earth,
 which provides an angular resolution of 2.5$^\circ$. The surface
 area covered on Earth is about 160,000 km$^2$. The duty cycle of EUSO
 will be similar to that of surface fluorescent telescopes, of
 order of 10\%. The plan is to have JEM-EUSO looking straight down
 and also in a tilted mode that will increase the viewing area by a
 factor up to 5 but decrease its resolution.
 EUSO will also be equipped with devices that measure
 the  transparency of the atmosphere and the existence of clouds. Clouds
 are not always bad for such a detector because some shower signals
 could be reflected by them and detected better by the instrument.
 
  The motivation for EUSO is the study of cosmic rays
 of energy above 5$\times$10$^{19}$ eV as well as very high energy
 neutrinos (\cite{Ebisuzaki}).

  Currently JEM-EUSO is the second experiment to be launched to the
 Japanese Experimental Module of ISS in 2015. The launch will be 
 provided by the Japanese transfer vehicle HTV.   
 
\section*{Acknowledgments}
 We acknowledge extended discussions with D. Allard, P. Billoir,
 R. Engel, T.K. Gaisser,  
 A. Olinto, P. Sokolsky, M. Unger, A.A. Watson and others.
 Special thanks to C.~Dobrigkeit, P.~Ghia, and A.A.~Watson for
 the careful reading of the manuscript and their comments. 
 ALS thanks the whole Auger Collaboration. TS is grateful to
 LPNHE/UPMC-CNRS/IN2P3
 for their support of his visit to Paris University
 Pierre et Marie Curie in 2009.
 The work of TS is supported in part by the US Department of Energy
 grant UD-FG02-91ER40626. 

\bibliographystyle{apsrmp4-1}
\bibliography{RMP}


\end{document}